\documentclass[fleqn,usenatbib]{mnras}

\usepackage[T1]{fontenc}
\usepackage[usenames,dvipsnames]{xcolor}
\usepackage{academicons}
\usepackage{xcolor}
\usepackage[normalem]{ulem}
\usepackage{tikz,xcolor,hyperref}

\DeclareRobustCommand{\VAN}[3]{#2}
\let\VANthebibliography\thebibliography
\def\thebibliography{\DeclareRobustCommand{\VAN}[3]{##3}\VANthebibliography}

\usepackage{graphicx}	% Including figure files
\usepackage{amsmath}	% Advanced maths commands
\usepackage{amssymb}	% Extra maths symbols
\graphicspath{{./}{figures/}}
\usepackage{newtxtext,newtxmath}

\definecolor{lime}{HTML}{A6CE39}
\DeclareRobustCommand{\orcidicon}{%
    \begin{tikzpicture}
    \draw[lime, fill=lime] (0,0) 
    circle [radius=0.16] 
    node[white] {{\fontfamily{qag}\selectfont \tiny ID}};
    \draw[white, fill=white] (-0.0625,0.095) 
    circle [radius=0.007];
    \end{tikzpicture}
    \hspace{-2mm}
}

\newcommand{\orcidBoris}{\href{https://orcid.org/0000-0002-1857-2088}{\orcidicon}}
\newcommand{\orcidKalina}{\href{https://orcid.org/0000-0001-5294-8002}{\orcidicon}}
\newcommand{\orcidDanilo}{\href{https://orcid.org/0000-0001-9002-3502}{\orcidicon}}
\newcommand{\orcidAnton}{\href{https://orcid.org/0000-0002-6610-2048}{\orcidicon}}
\newcommand{\orcidMarc}{\href{https://orcid.org/0000-0002-9946-4731}{\orcidicon}}
\newcommand{\orcidArjen}{\href{https://orcid.org/0000-0002-5027-0135}{\orcidicon}}
\newcommand{\orcidAdina}{\href{https://orcid.org/0000-0002-9464-8101}{\orcidicon}}
\newcommand{\orcidAdam}{\href{https://orcid.org/0000-0002-9330-9108}{\orcidicon}}
\newcommand{\orcidKate}{\href{https://orcid.org/0000-0001-7160-3632}{\orcidicon}}
\newcommand{\orcidHeath}{\href{https://orcid.org/0009-0007-1787-2306}{\orcidicon}}
\newcommand{\orcidJeyhan}{\href{https://orcid.org/0000-0001-9187-3605}{\orcidicon}}
\newcommand{\orcidGabe}{\href{https://orcid.org/0000-0003-2680-005X}{\orcidicon}}
\newcommand{\orcidRos}{\href{https://orcid.org/0000-0001-7393-3336}{\orcidicon}}
\newcommand{\orcidNick}{\href{https://orcid.org/0000-0003-3243-9969}{\orcidicon}}
\newcommand{\orcidEvelyn}{\href{https://orcid.org/0000-0002-2368-6469}{\orcidicon}}

\newcommand{\orcidMauro}{\href{https://orcid.org/0000-0001-7768-5309}{\orcidicon}}
\title[Galaxy Component Properties]{Bulge+disc decomposition of HFF and CANDELS galaxies: UVJ diagrams and stellar mass--size relations of galaxy components at 0.2 $\leq \boldsymbol{z} \leq$ 1.5 }

\author[K.~V.~Nedkova et al.]{
Kalina V.~Nedkova\orcidKalina$^{1,2,3}$\thanks{E-mail: knedkov1@jhu.edu},
Boris Häußler\orcidBoris$^{4}$,
Danilo Marchesini\orcidDanilo$^{3}$,
Gabriel B.~Brammer\orcidGabe$^{5,6}$, \newauthor 
\ Adina D.~Feinstein\orcidAdina$^{7\thanks{NHFP Sagan Fellow}}$,
Evelyn J.~Johnston\orcidEvelyn$^{8}$,
Jeyhan~S.~Kartaltepe\orcidJeyhan$^{9}$, 
Anton M.~Koekemoer\orcidAnton$^{2}$,\newauthor
\ Nicholas S.~Martis\orcidNick$^{10}$, 
Adam Muzzin\orcidAdam$^{11}$,
Marc Rafelski\orcidMarc$^{2,1}$,
Heath V.~Shipley\orcidHeath$^{12}$,\newauthor
\ Rosalind E.~Skelton\orcidRos$^{13}$, 
Mauro Stefanon\orcidMauro$^{14}$,
Arjen van der Wel\orcidArjen$^{15}$, and Katherine E.~Whitaker\orcidKate$^{16,6}$
\\ \\
$^{1}$Department of Physics and Astronomy, Johns Hopkins University, 3400 North Charles Street, Baltimore, MD 21218, USA\\
$^{2}$Space Telescope Science Institute, 3700 San Martin Drive, Baltimore, MD 21218, USA\\
$^{3}$Department of Physics and Astronomy, Tufts University, 574 Boston Avenue Suites 304, MA 02155, USA\\
$^{4}$European Southern Observatory, Alonso de Cordova 3107, Casilla 19001, Santiago, Chile\\
$^{5}$Niels Bohr Institute, University of Copenhagen, Jagtvej 128, København N, DK-2200, Denmark\\
$^{6}$Cosmic Dawn Center (DAWN), Copenhagen, Denmark\\
$^{7}$Laboratory for Atmospheric and Space Physics, University of Colorado Boulder, UCB 600, Boulder, CO 80309\\
$^{8}$Instituto de Estudios Astrof\'isicos, Facultad de Ingenier\'ia y Ciencias, Universidad Diego Portales, Av. Ej\'ercito Libertador 441, Santiago, Chile \\
$^{9}$Laboratory for Multiwavelength Astrophysics, School of Physics and Astronomy, Rochester Institute of Technology, 84 Lomb Memorial Drive, Rochester,\\ \ NY 14623, USA\\
$^{10}$Department of Mathematics and Physics, University of Ljubljana, Jadranska ulica 19, SI-1000 Ljubljana, Slovenia\\
$^{11}$Department of Physics and Astronomy, York University, 4700 Keele Street, Toronto, ON, M3J 1P3, Canada\\
$^{12}$Department of Physics, Texas State University, San Marcos, TX 78666, USA\\
$^{13}$South African Astronomical Observatory, P.O. Box 9, Observatory, Cape Town, 7935, South Africa\\
$^{14}$Leiden Observatory, Leiden University, NL-2300 RA Leiden, The Netherlands\\
$^{15}$Sterrenkundig Observatorium, Universiteit Gent, Krijgslaan 281 S9, 9000 Gent, Belgium\\
$^{16}$Department of Astronomy, University of Massachusetts Amherst, 710 N Pleasant Street, Amherst, MA 01003, USA
}

\date{Accepted XXX. Received YYY; in original form 2023 December 12}

\pubyear{2024}

\begin{document}
\label{firstpage}
\pagerange{\pageref{firstpage}--\pageref{lastpage}}
\maketitle

\begin{abstract}
Using deep imaging from the CANDELS and HFF surveys, we present bulge+disc decompositions with \textsc{GalfitM} for $\sim$17,000 galaxies over $0.2 \leq z\leq 1.5$. We use various model parameters to select reliable samples of discs and bulges, and derive their stellar masses using an empirically calibrated relation between mass-to-light ratio and colour. Across our entire redshift range, we show that discs follow stellar mass--size relations that are consistent with those of star-forming galaxies, suggesting that discs primarily evolve via star formation. In contrast, the stellar mass-size relations of bulges are mass-independent. Our novel dataset further enables us to separate components into star-forming and quiescent based on their specific star formation rates.
We find that both star-forming discs and star-forming bulges lie on stellar mass-size relations that are similar to those of star-forming galaxies, while quiescent discs are typically smaller than star-forming discs and lie on steeper relations, implying distinct evolutionary mechanisms. Similar to quiescent galaxies, quiescent bulges show a flattening in the stellar mass-size relation at $\sim$10$^{10}$M$_\odot$, below which they show little mass dependence. However, their best-fitting relations have lower normalisations, indicating that at a given mass, bulges are smaller than quiescent galaxies. 
Finally, we obtain rest-frame colours for individual components, showing that bulges typically have redder colours than discs, as expected. We visually derive UVJ criteria to separate star-forming and quiescent components and show that this separation agrees well with component colour. HFF bulge+disc decomposition catalogues used for these analyses are publicly released with this paper.

\end{abstract}


%%%%%%%%%%%%%%%%%%%%%%%%%%%%%%%%%%%%%%%%%%%%%%%%%%

%%%%%%%%%%%%%%%%% BODY OF PAPER %%%%%%%%%%%%%%%%%%

\section{Introduction}

 While galaxies are typically treated as single component objects, most galaxies are complex systems that can consist of several components, such as a star-forming disc, a central spheroidal bulge, a bar, and other substructures. The presence or absence of these components as well as their characteristics provide valuable clues about the formation histories of the galaxies which host them. 
 However, properly decomposing galaxies into components, especially at high redshift, is difficult. Despite this, many studies \citep[e.g.][]{Allen2006MNRAS, Simard2011, LacknerGunn2012MNRAS, Mendel2014ApJS, Vika2014, Kennedy2016MNRAS, Lange2016, Dimauro2018MNRAS, Kim2018ApJ, Bottrell2019MNRAS, MegaMorph2, Casura2022, Jegatheesan2024} have shown that, to first order, most galaxies are well represented by a central bulge and/or a disc and that modelling galaxies as bulge+disc systems can help place strong constraints on evolutionary models of galaxies. In this paper, we set out to perform bulge+disc decompositions for galaxies from the Hubble Frontier Fields (HFF, \citealt[]{Lotz2017}) and  Cosmic Assembly Near-infrared Deep Extragalactic Legacy Survey (CANDELS, \citealt[]{Grogin2011ApJS, Koekemoer2011ApJS}), probing the bulge and disc properties of less massive galaxies than previously studied with large photometric surveys over $0.2\leq z\leq1.5$.

It is crucial that bulges and discs are modelled and treated separately because they have fundamentally distinct properties.
Classical bulges, which are spheroidal in structure, generally have old stellar populations, leading to redder colours, higher metallicities, and higher $\alpha$ abundances \citep[e.g.][]{Moorthy2006MNRAS, Morelli2008MNRAS.389..341M, Morelli2016MNRAS.463.4396M, Coccato2018MNRAS.477.1958C, Johnston2022MNRAS.514.6141J}. 
 Galaxy discs, on the other hand, are usually sites of ongoing star formation. Hence, they contain younger stars and are bluer in colour and more metal-poor than bulge components.
 Bulges and discs are also structurally different as a result of stellar kinematics. 
 The orbits of stars in classical bulges are randomly oriented while the stars within discs rotate around the centre of the galaxy in a plane, exhibiting little random motion. 
 Thus, bulges have light profiles that are more concentrated in the centre, with classical bulges being well represented by a \cite{deVaucouleurs} profile with $n=4$ \citep{Carollo1997AJ....114.2366C}, while discs generally have exponential light profiles and are well-modelled with Sérsic indices of $n=1$ \citep[e.g.][]{Freeman1970ApJ, Kormendy1977ApJ}. 
 Although this is true for typical bulges and discs, galaxies are also unique and complicated, with the present-day Universe containing both blue, star-forming bulge-dominated galaxies \citep[e.g.][]{Schawinski2009MNRAS, Kannappan2009AJ, Ferreras2009MNRAS, Barro2013ApJ} and galaxies with prominent discs but little to no star formation activity \citep[e.g.][]{Bamford2009MNRAS, Masters2010MNRAS, Huertas-Company2016MNRAS}. 
 %Therefore, properly separating bulge from disc components remains a highly nontrivial task.

 The different observed properties of bulges and discs suggest that they evolve through different mechanisms. 
 While the exact nature of how and when bulge and disc components assemble remain outstanding questions, there has been remarkable progress in our theoretical understanding of galaxy component evolution. 
 Our current picture of disc formation suggests that proto-galaxies gain their angular momentum through interactions with the gravitational tidal field of neighbouring galaxies \citep[e.g.][]{Hoyle1949MNRAS, Peebles1969ApJ, Doroshkevich1970Ap, White1984ApJ}. 
 Subsequently, a disc is formed as a result of angular momentum conservation during dissipational gas collapse \citep[][]{Fall1980MNRAS, Mo1998MNRAS}.
 %In this scenario, the central parts of the disc will reach the necessary gas density to form stars before the outer part of the disc, resulting in inside-out growth. 
 Various processes, such as tidal forces and mergers can alter a galaxy's angular momentum and warp, destroy, or thicken the disc component, ultimately complicating the scenario described above \citep[e.g.][]{Steinmetz2002NewA, Zavala2012MNRAS}. 
%  On the other hand, ‘classical’ de Vaucouleurs-like bulges, are thought to have formed early by hierarchical mergers (Eggen et al. 1962), and acquired their disks later
 %Indeed, recent studies have observed objects that pose challenges to this picture such as galaxies with counter-rotating discs in which the gas and stars have misaligned spins \citep[e.g.][]{Lu2021MNRAS.503}. 

 The formation of bulges is arguably even more complex as there are numerous ways through which galaxies can form and grow their bulge components. The classical pathway to grow spherical components is through hierarchical mergers of gas-rich galaxies \citep[e.g.][]{Toomre&Toomre1972ApJ, Cole2000MNRAS, Hopkins2009ApJ, Zavala2012MNRAS, Avila-Reese2014MNRAS}. During a merger, the structure that was previously present in the galaxies is destroyed, causing a loss of angular momentum and a burst of star formation. Once this burst of star formation ceases, it leaves behind a bulge-dominated, or even elliptical, galaxy. 
 %At high redshift where pristine gas is more readily abundant, such quenched galaxies in low-density environments are able to re-build their discs by accreting more gas, a process referred to as rejuvenation \citep[e.g.][]{Thomas2010MNRAS}. 
 However, recent developments show that major mergers are rare, and therefore unlikely to be the primary formation mechanism through which bulges and ellipticals form \citep[e.g.][]{Naab2007ApJ, Oesch2010, Newman2012ApJ}. Instead, bulges are now believed to form and grow in-situ from early-on star formation and subsequent secular evolution \citep[e.g.][]{Athanassoula2005MNRAS.358.1477A,Okamoto2013MNRAS} or via violent disc instabilities that drive gas rapidly toward the centre of the galaxy by viscous and dynamical friction \citep[e.g.][]{Noguchi1999ApJ, Dekel2009ApJ, Ceverino2010MNRAS, Krumholz2010ApJ,Genzel2011ApJ, Bournaud2014ApJ,Bournaud2016ASSL}. The bulge components can also be influenced by repeated minor merger events \citep[e.g.][]{Hopkins2010ApJ, Lucia2011MNRAS}. This picture is consistent with \cite{Wellons2016MNRAS}, who find that $\sim50\%$ of compact galaxies at $z=2$ exist as the cores of massive galaxies at  $z=0$. This is also in agreement with recent cosmological magneto-hydrodynamical simulations, which show that, on average, in-situ stars and stars that have migrated to the centre from the disc respectively make up $\sim73\%$ and $\sim23\%$ of the stellar mass in the central 500 parsecs of galaxies at $z=0$ \citep{Boecker2023MNRAS}.

In order to test these theories and simulations, bulge+disc decompositions are necessary so that the properties of discs and bulges can be measured individually. However, at $z\geq1$ (i.e.~over most of cosmic history), such decompositions require exceptionally deep imaging with high spatial resolution. Fortunately, over the last couple of decades, several large surveys, such as HFF and CANDELS, have obtained such imaging and great effort has gone into developing software that can decompose galaxy light profiles into components. For statistical studies with large galaxy samples, galaxy structure needs to be modelled with algorithms that are flexible enough to fit the wide variety of morphological properties observed in galaxies. There are now a plethora of software suites that are well suited for such samples. Among these are \textsc{gim2d} \citep{Simard1998,Simard2002}, \textsc{budda} \citep{2004BUDDA, Gadotti2008MNRAS}, \textsc{pymorph} \citep{Pymorph2010MNRAS}, \textsc{galfit} \citep{Peng2002AJ, Peng2010Galfit}, \textsc{imfit} \citep{Imfit2015ApJ}, \textsc{imcascade} \citep{imcascade}, \textsc{ProFuse} \citep{Robotham2022}, and \textsc{morphofit} \citep{morphofit}. In this work, we choose to model all galaxies with \textsc{GalfitM} \citep{MegaMorph, MegaMorph2}, a code that is based on \textsc{Galfit} \citep{Peng2010Galfit} but allows for objects to be modelled in multiple bands simultaneously, effectively increasing the signal-to-noise ratio ($S/N$) of the data. This feature is crucial as we aim to study low mass systems at high redshift, which appear faint. Additionally, because bulges and discs typically host different stellar populations, they usually have distinct colours. Due to the multi-band nature of \textsc{GalfitM}, such colour differences enable a more reliable separation of the galaxy components and hence lead to more reliable bulge and disc parameters, critical for our goal to examine the properties of bulges and discs separately.
%We aim to use this colour information to obtain more reliable bulge and disc fit parameters, which is only possible by using multi-wavelength data and modelling it with a multi-band approach.

For roughly two decades now, astronomers have been using these software suites, among others, to measure the properties of statistically significant bulge and disc samples to constrain how they have changed over cosmic time \citep[e.g.][]{Simard2002, Allen2006MNRAS, Benson2007MNRAS}.  With modern advances in instrumentation and modelling techniques, fainter as well as higher redshift galaxies have also been studied \citep[e.g.][]{Simard2011,LacknerGunn2012MNRAS, Lang2014ApJ, Bruce2014MNRAS, Meert2015MNRAS, Kennedy2015MNRAS.454..806K}. These previous works have revealed a number of crucial details about the role of the bulge and disc components. For instance, \cite{Dimauro2019} studied the properties of massive bulges and discs in the CANDELS fields with M$_* >$ 2 $\times$ 10$^{10}$ M$_\odot$, finding evidence that massive bulge components can form while largely leaving the disc structure intact. Despite this remarkable progress, bulge+disc decompositions of low mass galaxies at non-local redshifts are largely missing from the literature. In this paper, we extend the redshift regime in which low mass systems can be decomposed %bridge this gap 
by probing the properties of bulges and discs of galaxies with M$_* \geq$ 10$^{7}$M$_\odot$ out to $z=1.5$.

%Bulge and disc properties of low mass galaxies can also be studied by performing bulge+disc decompositions on simulated galaxies  \citep[e.g.][]{Scannapieco2010MNRAS, Bottrell2017MNRAS.467.1033B, Bottrell2017MNRAS.467.2879B}. Unfortunately, some tension still remains between the most powerful hydrodynamic simulations and observations as is noted by \cite{Bottrell2017MNRAS.467.1033B, Bottrell2017MNRAS.467.2879B} who find a deficit of high-mass bulge-dominated galaxies compared to observations.

This paper is organised as follows. The data used for this study are discussed in Section \ref{sec:data}, and in Section \ref{sec:BD_models}, we describe the technical aspects of the bulge+disc decomposition. In Section \ref{sec:methods}, we discuss how we identify reliable bulge and disc component fits and our methods for estimating the stellar mass of the galaxy components. In Section \ref{sec:results}, we present the stellar mass--size relations of bulges and discs individually and discuss the implications of these findings. The positions of individual bulge and disc components on the rest-frame UVJ diagram are presented in Section \ref{sec:comp_UVJ}. Finally, a summary of the paper is in Section \ref{sec:summary}. Appendices include a description of the bulge+disc catalogues that we release along with this paper (Appendix \ref{appendix_catalogue}), tests of various parameters as metrics for identifying whether a given galaxy is better modelled as a one- or two-component system (Appendix \ref{appendix_AIC_BIC}), a discussion of the likelihood that the components are reversed in the modelling (Appendix \ref{appendix:flipped_comp}), and comparisons between different methods for estimating the stellar mass of galaxy components (Appendix \ref{appendix_mass}). Throughout this paper, we use AB magnitudes \citep{Oke1983}, cosmological density parameters $\Omega_{\mathrm{m}}$ = 0.3 and $\Omega_\Lambda$ = 0.7, and a Hubble constant of H$_{0} = 70$ km s$^{-1}$ Mpc$^{-1}$. We assume a \cite{Chabrier2003} initial mass function for all estimates of stellar mass. 

\begin{figure*}
    \centering
    \includegraphics[width=.95\textwidth]{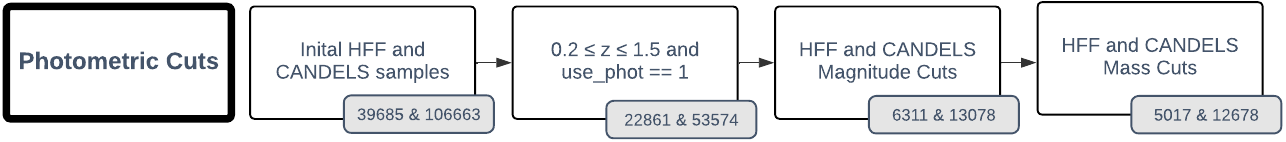}
    \vspace{-0.1cm}
    \caption{Flowchart indicating the photometric cuts that we apply. For each cut, the number of galaxies remaining from the HFF (\textit{left}) and CANDELS (\textit{right}) samples is indicated in the shaded box to the bottom right. See \S\ref{sec:phot_cuts} for further discussion of each quality cut. 
    % make the phot cuts a table and take out of this figure! Or make other Figure -> then all selections in figures
    %In \S \ref{sec:distinction}, we explain the parameters that we investigate for each subgroup in order to separate the total sample into discs and bulges.
    }
    \label{fig:selection_phot_only}
\end{figure*}
% ##################################################################################################
\section{Data} \label{sec:data}
% ##################################################################################################

\defcitealias{Kalina2021}{N21}

\cite{Kalina2021}, hereafter \citetalias{Kalina2021}, present catalogues of structural properties for HFF galaxies modelled with single Sérsic light profiles using the MegaMorph tools \citep[][]{MegaMorph, Vika2013MNRAS, MegaMorph2}. 
They combine these data with CANDELS objects which are modelled in a consistent way (Häußler et al. \textit{in prep}). 
In this work, we extend the \citetalias{Kalina2021} sample and results by decomposing these same galaxies into bulge and disc components.

 We use the same bands as \citetalias{Kalina2021}, which are listed in Table \ref{tab:Bands_used} for all HFF and CANDELS fields, and the tools developed as part of the MegaMorph project -- namely, \textsc{GalfitM} and \textsc{Galapagos-2} -- to obtain two separate bulge+disc decomposition models with different assumptions. In the first set of fits, we constrain the Sérsic index of the disc to $n=1$ and the Sérsic index of the bulge to $n=4$. In the second set of models, we again fix the Sérsic index of the disc to $n=1$, but we allow the Sérsic index of the bulge to be a free parameter because the central structures of galaxies are not always well-modelled by \citeauthor{deVaucouleurs} profiles (e.g.~in the presence of pseudo-bulges or bars). Combined with the results presented in \citetalias{Kalina2021}, we have the following models for every galaxy.
\begin{enumerate}
    \item A single Sérsic profile fit,
    \item a \citeauthor{deVaucouleurs} bulge $+$ exponential disc fit, and
    \item a bulge with a free Sérsic index $+$ exponential disc fit.
\end{enumerate}
Hereafter, we refer to (i) as the single Sérsic profile fit and one-component model interchangeably. We call (ii) the $n=4$ bulge+disc model and (iii) the $n=$ \textit{free} bulge+disc model. We describe the technical aspects of these models in full detail in the following section.

%Our sample, originally presented in \citetalias{Kalina2021}, is a combination of galaxies from the HFF \citep{Lotz2017} and CANDELS \citep{Grogin2011ApJS, Koekemoer2011ApJS} programmes.
In this section, we present our initial sample selection, which consists of a series of cuts based on the photometric properties of the galaxies. These are discussed in more detail below and summarised in Figure \ref{fig:selection_phot_only}, where the number of objects that remain in the sample is indicated in the shaded box to the bottom right of each cut. The smaller of the two numbers corresponds to the HFF sample, while the larger value is the number of galaxies that remain in the CANDELS sample. We note that in \S\ref{sec:modelparamcuts}, we will apply additional selection criteria based on the \textsc{GalfitM}-derived model parameters.

%which gives us three subsamples: a sample of reliable disc components, a sample of reliable bulge components, and a sample of reliable single Sérsic profile fits. These are not mutually exclusive such that a single galaxy can belong to any combination of these three subsamples. To obtain these samples, we first apply a series of cuts based on the photometric properties of the galaxies, followed by cuts on the \textsc{GalfitM}-derived model parameters. These are summarised in Figure \ref{fig:selection_chart} and discussed in more detail below. 

\subsection{Photometric Cuts}\label{sec:phot_cuts}
% We apply a series of quality cuts on the parameters from the photometric HFF and CANDELS catalogues. These photometric cuts are listed on the left-hand side of Figure \ref{fig:selection_chart}, with the number of objects that remain in the sample from the HFF and CANDELS data indicated in the shaded box to the bottom right of each cut. The smaller of the two numbers corresponds to the HFF sample, while the larger value is the number of galaxies that remain in the CANDELS sample. 

We begin with the same initial sample described in \citetalias{Kalina2021}, which includes 39685 HFF objects and 106663 CANDELS objects. In this initial sample, objects from the HFF-DeepSpace images, which have pixel scales of 0.06$''$ per pixel, are matched to the HFF-DeepSpace photometric catalogues \citep{Shipley2018}. Likewise, objects identified with SExtractor \citep{Bertin1996} from the 0.03$''$ per pixel CANDELS images are matched to the 3D-HST catalogues \citep{Skelton2014ApJS}, limiting the spatial separation to be at most 0.2 arcseconds. Additionally, objects from the HFF clusters are required to have redshifts that are consistent with the spectroscopic redshift of the cluster that they belong to, or lower. This is done in order to remove objects which may be affected by lensing effects that make reliable magnitude and flux measurements difficult. The derivation of these redshift limits is discussed in \citetalias{Kalina2021} (see their Fig.~1), but we have reproduced them for each cluster in Table \ref{tab:HFFcluster_properties}, for convenience. 

\begingroup
\setlength{\tabcolsep}{6pt} % Default value: 6pt
\renewcommand{\arraystretch}{1} % Default value: 1
\begin{table}
\caption{List of Advanced Camera for Surveys (ACS) and Wide Field Camera 3 (WFC3) bands used for each field.\vspace{-0.2cm}}
\centering
\begin{tabular}{l l }
\hline
%\multicolumn{1}{l}{ \ } & \multicolumn{5}{c}{ HFF Cluster Fields} \\ 
Field(s) &  Bands Used \\
\hline 
All HFF fields & F435W, F606W, F814W, F105W, F125W,  \\
\ & F140W, F160W  \\
GOODS-N \& GOODS-S & F435W, F606W, F775W, F814W, F850LP, \\
&  F105W, F125W, F140W, F160W \\
COSMOS \& EGS & F606W, F814W, F125W, F140W, F160W\\
UDS & F606W, F814W, F125W, F160W\\

\hline
\end{tabular}
\label{tab:Bands_used}
\end{table}
\endgroup

% \begingroup
% \setlength{\tabcolsep}{6pt} % Default value: 6pt
% \renewcommand{\arraystretch}{1} % Default value: 1
% \begin{table}
% \caption{For each HFF cluster field, we have reproduced the redshift limit above which we exclude background cluster objects in order to avoid biasing any measurements as a result of gravitational lensing. These values are derived in \citetalias{Kalina2021} using the normalised median absolute deviations to identify cluster members.}
% \centering
% \begin{tabular}{l c}
% \hline
% %\multicolumn{1}{l}{ \ } & \multicolumn{5}{c}{ HFF Cluster Fields} \\ 
% HFF Cluster & redshift limit \\
% \hline
% Abell1063 & 0.522\\
% Abell2744 & 0.477\\
% Abell370 & 0.522 \\
% MACS0416 & 0.550\\
% MACS0717 &  0.632\\
% MACS1149 &  0.668\\
% \hline 
% \end{tabular}
% \label{tab:HFFcluster_properties}
% \end{table}
% \endgroup

\begingroup
\setlength{\tabcolsep}{2.6pt} % Default value: 6pt
\renewcommand{\arraystretch}{1.1} % Default value: 1
\begin{table}
\caption{For each HFF cluster field, we reproduce the redshift limits above which we exclude background cluster objects to avoid biasing any measurements as a result of gravitational lensing. These values are derived in \citetalias{Kalina2021} using the normalised median absolute deviations to identify cluster members.\vspace{-0.2cm}}
\centering
\begin{tabular}{c | c | c | c | c | c | c}
\hline
%\multicolumn{1}{l}{ \ } & \multicolumn{5}{c}{ HFF Cluster Fields} \\ 
Abell1063 & Abell2744 & Abell370 & MACS0416 & MACS0717 & MACS1149\\
$z$$\leq$ 0.522 & $z$$\leq$  0.477 &  $z$$\leq$ 0.522 & $z$$\leq$ 0.550 &  $z$$\leq$ 0.632 &$z$$\leq$ 0.668\\
\hline 
\end{tabular}
\label{tab:HFFcluster_properties}
\end{table}
\endgroup

In our sample selection, we first limit the CANDELS and non-cluster HFF samples to galaxies that have redshifts within the range $0.2\leq z \leq 1.5$ (as the objects in the HFF cluster fields are already limited to the redshift limits shown in Table~\ref{tab:HFFcluster_properties}), and a \texttt{use$\_$phot} flag equal to one. This flag removes objects which may have unreliable photometric measurements because they are stars, close to bright stars in the image, or have $S/N$$<$$3$ from the photometry aperture in the F160W band. We note here that the redshift range that we use in this work is different from the one used in \citetalias{Kalina2021}. This choice is motivated by results from \cite{Driver2013}, who infer that at $z$$\sim$$1.7$ galaxies switch from primarily evolving via processes that help form and grow spheroids, to growth via gas infall and minor mergers, which help to form and grow the disc component of galaxies \citep[see also][]{Cook2009MNRAS, Oser2010ApJ}. This is commonly referred to as the two-phase galaxy evolution, where at high redshift, galaxies are believed to evolve through hot processes, which include monolithic collapse, major mergers, and active galactic nuclei (AGN) activity. At $z\lesssim1.7$, cold phase processes such as accretion, minor mergers, secular processes, gas stripping, and strangulation are believed to play a more dominant role in galaxy evolution. This would suggest that the morphology of high redshift galaxy bulges is different from classical \citeauthor{deVaucouleurs} bulges seen at low redshift. 

While recent results based on \textit{James Webb Space Telescope} (JWST) data indicate that there are a significant number of disc-like and/or spheroidal galaxies at high redshift \citep[e.g.][]{Ferreira2022ApJ, Ferreira2023ApJ, Kartaltepe2023ApJ, Robertson2023ApJ}, from \textit{Hubble Space Telescope} (HST) observations, even as a whole, galaxies at $z\gtrsim2$ do not appear to follow the Hubble sequence mostly due to large star-forming clumps that are on the kiloparsec scale (e.g.~\citealt[][]{Buitrago2013MNRAS, Guo2018ApJ}). Then, what does a typical bulge component at high redshift look like? It would likely be blue as it is forming new stars and clumpy from star migration. This is different from low redshift bulges and makes properly classifying bulges across cosmic time highly challenging. Since one might expect both bulges and discs to appear blue in colour at high redshift from HST imaging, using the multi-wavelength information from \textsc{GalfitM} is not as beneficial as it is for lower redshift systems. Therefore, in order to separate discs and bulges using the same criteria across redshift, we choose to limit the redshift range to $z\leq1.5$ in this work.

Because we are decomposing galaxies into components, it is crucial that each object is bright enough for reliable bulge+disc decomposition. Therefore, while in \citetalias{Kalina2021}, galaxies are required to be one magnitude brighter than the 90$\%$ completeness limit, in this work, we require that galaxies are two magnitudes brighter than this limit. This is roughly equivalent to applying a $S/N$$\gtrsim$$14$ cut, but we choose to construct magnitude limited samples instead of applying a cut in $S/N$ as this choice is less biased against low surface brightness (LSB) galaxies. \citetalias{Kalina2021} show that LSB galaxies can be well recovered when a conservative magnitude limit is used, suggesting that any biases against these types of galaxies should be negligible for our study.

Lastly, we apply the same stellar mass cut as \citetalias{Kalina2021} to remove any objects that have stellar masses below $10^7$M$_\odot$ or have stellar mass uncertainties larger than 2dex. In the HFF-DeepSpace and 3D-HST photometric catalogues, stellar mass estimates are derived using \textsc{FAST} \citep{Kriek2009}. We correct these \textsc{FAST}-derived stellar masses for the difference in the F160W magnitude from \textsc{GalfitM} and the photometric catalogues so that the stellar mass is consistent with the profiles used to measure galaxy sizes, following \cite{vdWel2014} and \citetalias{Kalina2021}. Once all of the photometric selection cuts are applied, we have a sample of 5017 HFF and 12678 CANDELS galaxies, as indicated in Figure~\ref{fig:selection_phot_only}.

% ################################################################################

\section{Bulge+Disc Decomposition} \label{sec:BD_models}
% ################################################################################

To decompose these 5017 HFF and 12678 CANDELS galaxies into their main components, we use the MegaMorph tools. However, properly separating galaxies into bulge and disc components has been a long-standing problem since the automated software suites that are available for bulge+disc decompositions suffer from a number of drawbacks. \cite{Lange2016} discuss five common issues that arise when using Levenberg-Marquardt minimisation algorithms, such as \textsc{Galfit}. As \textsc{GalfitM} is largely based on \textsc{Galfit}, many of the same drawbacks are present in our modelling. Therefore, in \S\ref{model_challenges}, we discuss each potential issue that \cite{Lange2016} identify and we present our solutions. These solutions motivate some of the quality cuts that we apply based on the bulge+disc model parameters, which are discussed in \S\ref{sec:modelparamcuts}.

\subsection{Modelling Challenges}\label{model_challenges}
\subsubsection{The best-fitting model converges on local minima}\label{localminima}% as opposed to the global minimum.} 
\textsc{Galfit} and \textsc{GalfitM} are cleverly designed to avoid solutions that have converged on local minima in the $\chi^2$ topology by perturbing the parameters by a "random" amount once the solution has converged. The specific details, such as the step size and direction taken in the parameter space during this perturbation, will depend on the code being used (i.e.~\textsc{Galfit} or \textsc{GalfitM}, which behave slightly differently). Despite this countermeasure, \textsc{Galfit} can still potentially suffer from this issue. \cite{Lange2016} solve this problem by varying the initial input parameters and repeating the fitting process. They then identify a common convergence point, which avoids this issue to some level. Unfortunately, such an approach requires all galaxies to be fit multiple times (e.g.~\citealt{Lange2016} fit all galaxies 40 times), which makes this unfeasible, or at least challenging, for large samples of galaxies. To make matters worse, including additional components to the fit will complicate and possibly introduce further/new degeneracies in the $\chi^2$ topology (see \citealt{Meert2013MNRAS} for further discussion).

Thanks to \textsc{GalfitM}'s multi-wavelength fitting capabilities, this issue is largely mitigated for our fits, as the code already searches for a minimum satisfying the fit in several "independent" images in different filters. \cite{MegaMorph} have demonstrated that fitting galaxy light profiles with multi-wavelength data increases the stability and accuracy of the fit. This is further shown 
for bulge+disc decompositions 
in \cite{MegaMorph2}. The stability and accuracy of the fits are especially increased if the profile parameters are held constant with wavelength, i.e.~the profile fit to each image is the same, as in the case of our bulge+disc decompositions, effectively carrying out several fits to the same object. A caveat here is that this assumes that the intrinsic profiles of the components remain the same with wavelength without any radial gradients in metallicity, stellar populations, or dust content. Although galaxy components can have radial colour gradients, these are small compared to the colour gradients observed in galaxies \citep[e.g.][]{Suess2020ApJ, Suess2021ApJ, Suess2022ApJ...937L..33S, MegaMorph2, Miller2023ApJ...945..155M, Nedkova2024}, which makes them difficult to measure accurately.  
Thus, following \cite{MegaMorph2}, we constrain the sizes of the components to be constant with wavelength throughout the remainder of this work.
%based on the findings of \cite{MegaMorph2} who argue that although differences in colour can exist within galaxy components, these are small compared to the colour differences between the bulge and the disc components, which makes them difficult to measure reliably. 

%Although they use different data, the conclusions are transferable to the data used in this work.
    
In our modelling, we find good agreement between the effective radii of bulges and discs derived from the $n=$ \textit{free} and $n=4$ bulge+disc models in cases where the components are deemed reliable in both models. 
%Specifically, we do not compare component sizes for galaxies where the components are too faint to obtain a reliable comparison. 
Full details regarding the criteria that the bulge, disc, and single Sérsic profile models must satisfy in order to be deemed reliable will be discussed in \S\ref{sec:modelparamcuts}. 
In addition, the magnitudes of the one-component fits and the combined magnitudes of the bulge and disc fits, for both the $n=4$ and $n=$ \textit{free} bulge+disc models, show excellent agreement. As we often find consistent solutions in all three independently-derived best-fitting models, we argue that it is unlikely that they are commonly trapped at local minima.

%  \begin{figure*}
%     \centering
%     \includegraphics[width=0.85\textwidth, angle =90]{1670[1886]_abell2744par_example.pdf}
%     \caption{Fitting results in all seven bands for an example galaxy (id$_{\mathrm{HFF}} = 1886$) from the Abell2744 parallel field. From left to right, the columns show the image, the single profile model followed by its residual. The next four columns show the $n=$\textit{free} model, in which the Sérsic index of the bulge is fit with $n=0.82$. These columns show the bulge+disc model, the residual, the disc model, and the bulge model from left to right. The final four columns show the $n=4$ model, ordered in the same way. The red regions indicate bands in which we do not trust the model due to the nature of the fitting algorithm and the high redshift of this object (see text for details). All images, models, and residuals are oriented such that up is North and left is East.}
%     \label{fig:example_low_n_bulge}
% \end{figure*}

\subsubsection{The best-fitting model returns non-physical solutions.}\label{sec:sect312}  We note that because we are unable to resolve substructures such as pseudo-bulges, bars, and rings due to the high redshift regime studied in this work, all structures that reside in the centres of galaxies are termed "bulges" and that these do not in all cases represent classical \citeauthor{deVaucouleurs} bulges. With this in mind, we have carefully inspected all three models, itemised in \S\ref{sec:data}, for over 1000 galaxies to identify the characteristics of non-physical galaxy models. We find that when the Sérsic index of the bulge is allowed to vary over $n\in[0.2, 12]$ in the $n=$ \textit{free} models, the bulge component is fit with a Sérsic profile with $n<1$ for $\sim40\%$ of galaxies and $n<4$ for $\sim70\%$ of galaxies. In other words, for $\sim40\%$ of galaxies, the Sérsic index of the bulge is smaller than that of the disc component, which is fixed to $n=1$. \cite{Fischer2019MNRAS} find a somewhat similar bulge Sérsic index distribution using better resolved low-redshift data from the Mapping Nearby Galaxies at Apache Point Observatory \citep[MaNGA; ][]{Bundy2015ApJ} Data Release 15, with the majority of their objects being fit with $n_\mathrm{bulge} \sim 1$. When the bulge component is fit with a profile that has a low Sérsic index, it becomes increasingly more difficult to distinguish it from the disc. 

Indeed, one of the reasons why the Sérsic index of bulge components is often fit as a free parameter in the literature \citep[e.g.][]{Allen2006MNRAS} is because galaxies' central substructures are often complex. One example is pseudo-bulges, which are disc-like structures that are well modelled with low Sérsic indices (see \citealt{PseudoBulgeReview2004ASSL} for a review).
While distinguishing such structures from classical bulges in local galaxies yields insight into how these galaxies built up their stellar populations, at high redshift, it is almost impossible to reliably identify and fit pseudo-bulges, especially not at the image resolution present in the data used in this work. Given our redshift regime, we find that instead of providing better fits (e.g. ones where we can distinguish different central substructures), the extra degrees of freedom in the $n=$ \textit{free} models result in fits where the disc and bulge component are fitting similar profiles and splitting the flux of the whole galaxy in a physically meaningless way. A similar effect has also been presented in \cite{MegaMorph2}, where they show that the fainter a (simulated) galaxy component is, the more the fit parameters resemble the other component, i.e.~it becomes increasingly harder to distinguish two similar profiles from each other.

 \begin{figure*}
    \centering
    \includegraphics[width=0.675\textwidth, angle=90]{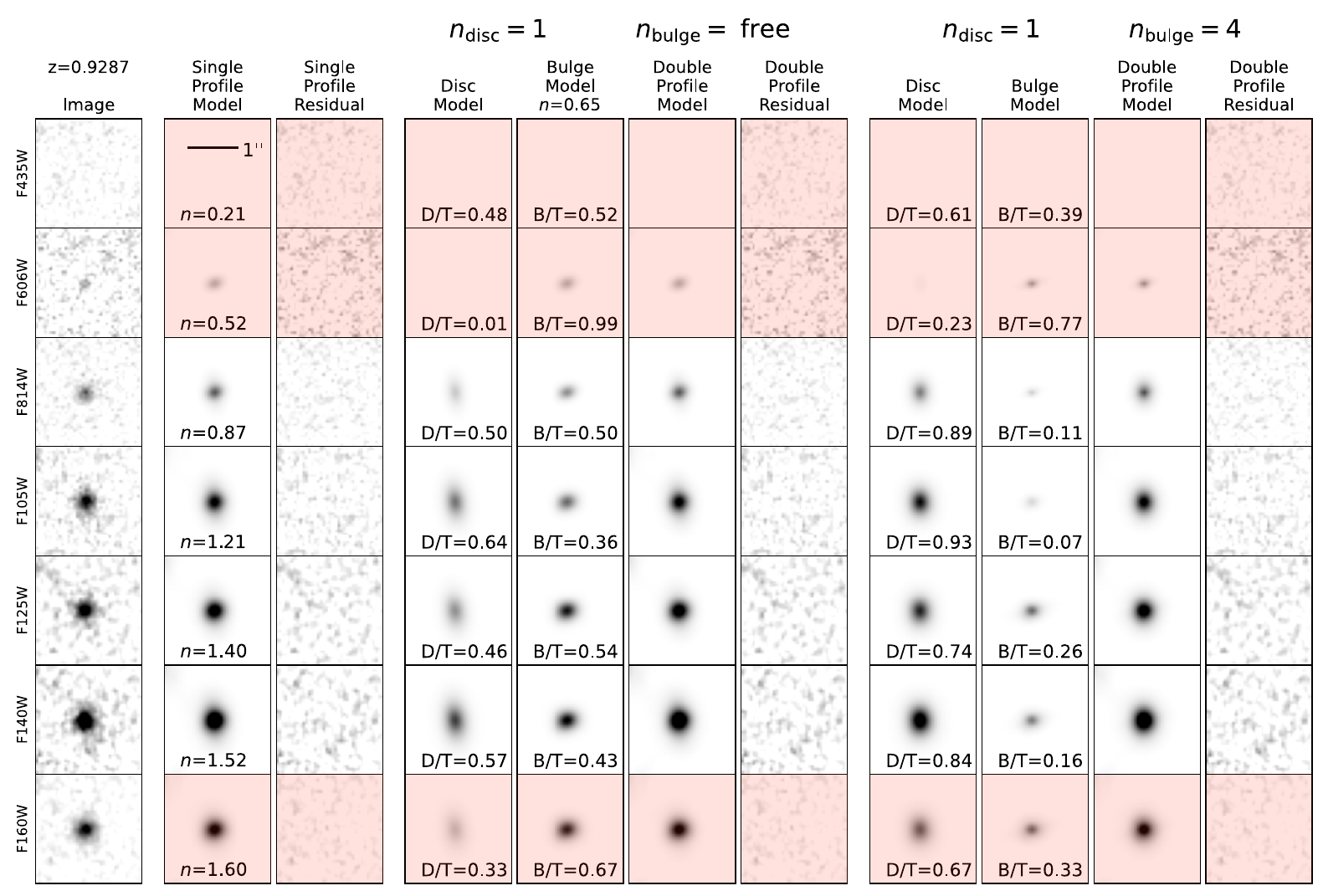}
    \vspace{-0.5cm}
    \caption{Fitting results in all seven bands for an example galaxy (id$_{\mathrm{HFF}} = 1886$) from the Abell2744 parallel field. From left to right, the columns show the image, the single profile model, followed by the residual. The next four columns show the $n=$\textit{free} model, in which the Sérsic index of the bulge is fit with $n=0.65$. From left to right, these columns show the disc model, bulge model, combined bulge+disc model, and residual. The final four columns show the $n=4$ model, ordered in the same way. The highlighted regions indicate bands in which we do not trust the models due to the nature of the fitting algorithm and the high redshift of this object (i.e. dropouts, see text for details). All images, models, and residuals are oriented such that up is North and left is East.
    % comment that this obj was chosen to show several issues. Send some more examples in the ref. reply. With one example we manage to discuss several issues. Rather than showing several different obj
    }\label{fig:example_low_n_bulge}
\end{figure*}

Given these results, we choose to use the \citeauthor{deVaucouleurs} bulge $+$ exponential disc fits (i.e.~the $n=$ 4 models) for the remainder of this paper. We note here that \cite{LacknerGunn2012MNRAS} also choose not to fit the Sérsic index of the bulge as a free parameter since the half-light radius of the bulge is correlated to the bulge Sérsic index \citep[][]{Trujillo2001MNRAS}. In Figure~\ref{fig:example_low_n_bulge}, we show all three independent models for one example galaxy to further justify this choice. While for most objects that we have visually inspected, the $n=$ \textit{free} and $n=4$ models are in fair agreement, this object was specifically chosen because the two models yield very different results. %We discuss and compare the models for this galaxy in more detail below. 

    The first column of Figure~\ref{fig:example_low_n_bulge} shows the image of the example galaxy in each of the seven bands that we use for the HFF fields. The second column shows the single Sérsic profile fit from \citetalias{Kalina2021} and the third shows the difference between the image and the model, i.e., the residual. The following four columns display the fitting results for the  $n=$ \textit{free} bulge+disc model and the last four columns show the fitting results for the $n=4$ bulge+disc model. For these, the first two columns show the disc and bulge models individually, followed by the double profile models and the residuals. In all fits, the disc models are fixed to $n=1$, and the bulge models are fit with $n=0.65$ in the $n=$ \textit{free} model and fixed to $n=4$ in the other. The red regions indicate bands in which the model is less reliable. These include the bluest and reddest bands because \textsc{GalfitM} fits all bands simultaneously with wavelength dependent functions -- specifically, Chebyshev polynomials --  which are less constrained at the edges (i.e.~Runge's phenomenon).Indeed, we often find that the edge bands have the most extreme bulge-to-total ratios (B/T), although this is not case for the example galaxy in Figure~\ref{fig:example_low_n_bulge}. In particular, objects tend to have low B/T in the bluest bands, whereas the largest B/T are usually measured in the reddest bands, likely due to changes in sensitivity across wavelength from young, blue disc stars toward older, red bulge stars.

    Given the redshift of the example galaxy in Figure~\ref{fig:example_low_n_bulge} ($z=0.9287$), it is unlikely to be modelled well in the bluer bands. In fact, in F435W and F606W, the object is barely visible, if at all. Due to this effect, we choose to only use bands whose central wavelengths are redder than 3159\AA \ in the rest frame. As we will use the rest-frame U-band in subsequent analyses,
    %, particularly to obtain rest-frame U--V colours to explore the position of individual components on the UVJ diagram, 
    the 3159\AA \ threshold is obtained by taking 3590{\AA} $-$ 1/2 $\times$ 862{\AA} = 3159{\AA}, where $3590${\AA} is effective wavelength and $862${\AA} is the full-width-half-maximum (FWHM) of the U-band, respectively. This ensures that the quality of the fit is not assessed at wavelengths that fall blueward of the U-band in the rest frame, at which high redshift objects are likely to be poorly modelled, while still allowing us to use the U-band e.g. to derive the positions of galaxies on the rest-frame UVJ diagram.

    In Figure \ref{fig:example_low_n_bulge}, we also report the Sérsic index of the single profile model in each band, where the Sérsic index is allowed to vary with wavelength, as well as the disc-to-total (D/T) and bulge-to-total (B/T) ratios. These ratios show that for the $n=$ \textit{free} bulge + disc model, the flux is being split roughly equally between the disc and the bulge (i.e. the D/T and B/T are both $\sim$ 0.5), while in the $n=4$ model, this galaxy is disc-dominated as the bulge accounts for at most 26\% of the total flux in the bands where we trust the modelling.
    From visual inspection across a range of contrast ratios and scales using SAOImage DS9 \citep{DS92000ascl.soft03002S}, the galaxy shown in Figure \ref{fig:example_low_n_bulge} appears to be a compact disc-dominated object without a significant bulge component. Hence, the $n=4$ bulge + disc fit should be preferred for this galaxy because the $n=$ \textit{free} fit suggests that this object is a bulge + disc system as both components are sufficiently bright and satisfy all of our selection criteria to be classified as reliable components, which we will discuss in full detail in \S\ref{sec:modelparamcuts} and \S\ref{sec:distinction}. 
     %As can be seen from  Figure \ref{fig:example_low_n_bulge}, when this object is fit with two components, one of them is modelled with an unrealistically large radius. Specifically, in the $n=$ \textit{free} bulge+disc model, the disc component is unreliably fit while in the $n=4$ model, the bulge model is unreliable.
     %Furthermore, the unreliable component is fit with such a large radius that its magnitude is very bright despite appearing to be the fainter component in the images, effectively making the B/T and D/T useless.
     %This occurs because $\textsc{GalfitM}$ is minimising the residual of the neighbouring object at the image edge, which causes one component to be fit with an elongated profile and a non-physical size. 
     %Since the bulge is modelled with more degrees of freedom than the disc in the $n=$\textit{free} model, the disc is the unreliable component while the bulge is fit with a low Sérsic index.
     Visual inspection of $\sim1000$ objects for which all models were compared, indicates that fitting the bulge with a free Sérsic profile rarely results in a better model, and in cases where the model is impacted by a neighbouring object, the disc component is usually unreliably fit, even if the galaxy is a disc-like object, which introduces a bias.
     %Given these results, we choose to use the \citeauthor{deVaucouleurs} bulge $+$ exponential disc fits for the remainder of this paper. We note here that \cite{LacknerGunn2012MNRAS} also choose not to fit the Sérsic index of the bulge as a free parameter since the half-light radius of the bulge is correlated to the bulge Sérsic index \citep[][]{Trujillo2001MNRAS}. %, and fitting classical bulges facilitates comparisons with elliptical galaxies.

     Although we will not use the $n=$ \textit{free} bulge + disc fits in our analyses, we show them in Figure~\ref{fig:example_low_n_bulge} for completeness, and in order to justify our choice of using the $n=4$ bulge + disc fits throughout this paper. As several other authors have used free Sérsic indices in their fits, and as such a choice impacts the results, this choice must be carefully made and justified. %Using the $n=4$ bulge + disc fits, we further apply a series of quality cuts to remove any models that have returned non-physical or unreliable solutions, which are described in detail in \S\ref{sec:modelparamcuts}. 

\subsubsection{Reversal of the bulge and disc components.}\label{sec:flippedcomps313} In the fitting, it is possible that the bulge component of the fit is modelling the disc while the disc component of the fit is modelling the bulge. How frequently this occurs strongly depends on the profile types that are being fit to each galaxy. For instance, \cite{Lange2016} fit two Sérsic profiles to their galaxy sample, without constraining the Sérsic index of the disc nor the bulge. Without assuming a light profile shape, both components are free to fit any part of the galaxy light profile. Therefore, the probability of the bulge and disc components being reversed is high. To resolve this problem, \cite{Lange2016} assume that the smaller of the two components is the bulge. While this choice is physically motivated and probably true in the vast majority of cases, it implies physics and can rule out especially interesting objects, i.e. those where the bulge is larger than the disc.

We note that component reversal is not limited to photometric data and is also found in \cite{Fischer2019MNRAS} and \cite{BUDDI_MaNGA2022MNRAS} who use MaNGA integral field spectroscopy (IFU) data. Both of these studies find that galaxies with flipped components tend to result from models where the central component has a Sérsic index $<1$.
In our modelling, we assume an exponential profile for the disc and a \citeauthor{deVaucouleurs} profile for the bulge. Hence, the components are less likely to be reversed, but we further investigate this potential issue in Appendix \ref{appendix:flipped_comp}.

\subsubsection{The uncertainties from \textsc{Galfit} and \textsc{GalfitM} do not reflect the full uncertainty in the final fits.}\label{sec:sect314} Follow-up analyses of the simulations carried out in \cite{haeussler2007ApJS} and \cite{MegaMorph} reveal that the size uncertainty estimates from \textsc{Galfit} are underestimated by a factor $\geq$10 (see also \citealt{vdwel_bell_2012ApJS}), while the uncertainties from \textsc{GalfitM} are underestimated by a factor of $2-2.5$ (H\"au{\ss}ler, private communication, but based on \citealt{MegaMorph2}). While the data used in these follow-up analyses are different from those used here, they can serve as an estimate of this issue.
%and be taken into account at least in a rough manner. 
Thus, as in \citetalias{Kalina2021}, we increase our size uncertainties by a factor of 3 in order to be conservative. We do not consider the uncertainties of the other parameters derived from \textsc{GalfitM} so we do not focus on deriving realistic uncertainties for other parameters in this work.
    
\subsubsection{Identifying if the one- or two-component model better describes a given galaxy is notoriously challenging.}\label{sec:sect315} The obvious metric for identifying the best-fitting model for each galaxy is the $\chi^2$ value returned by the fitting algorithm. However, since the bulge+disc models have more degrees of freedom, the residuals of these models are always cleaner than for the single Sérsic profile. Hence, we investigate the possibility of using reduced $\chi^2$ values (i.e.~$\chi^2_\nu$) in Appendix~\ref{appendix_AIC_BIC}, but we find that neighbouring objects commonly impact these values in the modelling, especially in dense fields, such as the HFF cluster fields.  

\cite{Lange2016} suggest that the Akaike Information Criterion (AIC; \citealt[]{AICAkaike1974}) and Bayesian Information Criterion (BIC; \citealt[]{BICSchwarz1978}) may be used to identify which model better represents the data for multi-wavelength fitting approaches. Indeed, these have been previously tested as methods for this purpose \citep[e.g.][]{Head2014MNRAS,Argyle2018MNRAS}. These works have shown that these criteria alone do not provide a robust method for distinguishing pure bulges, pure discs, and two-component systems; but, when used in combination with other parameters, such as the B/T ratio, the BIC and AIC can provide useful information \citep[see also][]{Bellstedt2024MNRAS}. %Although they find that for their one-band modelling, the AIC and BIC do not provide results that align with their visual classification, we revisit these criteria as possible methods for identifying the best-fitting model in our work. 
ollowing this idea, we have tested using the AIC and BIC in combination with other parameters derived in the modelling. Unfortunately, we find that these criteria yield results that do not align with visual impression. Although we do not use these criteria in this work, for completeness, we discuss some of the details and the reasons why they were not employed in Appendix~\ref{appendix_AIC_BIC}. We describe our approach for distinguishing one-component objects from two-component systems in full detail in \S \ref{sec:distinction}.

%We previously discussed the potential of using the AIC and BIC for distinguishing and separating bulge + disc systems from one-component objects. Indeed, these criteria have been previously tested as methods for reliably identifying whether a given galaxy is best modelled by a single component model or a two component model \citep[e.g.][]{Head2014MNRAS,Argyle2018MNRAS}. These previous works have shown that these criteria alone do not provide a robust method for distinguishing pure bulges, pure discs, and two-component systems; but, when used in combination with other parameters, such as the B/T ratio, the BIC and AIC can provide useful information. Following this idea, we have tested using the AIC and BIC in combination with other parameters derived in the modelling. Unfortunately, we find that these criteria yield results that do not align with visual impression. Although we do not use these criteria in this work, for completeness, we discuss some of the details and the reasons why they were not employed in Appendix~\ref{appendix_AIC_BIC}. 

%  \begin{figure*}
%     \centering
%     \includegraphics[width=1\textwidth]{param_BD_example.png}
%     \caption{Example of how B + D are allowed to vary with wavelength compared to the galaxy as a whole.}
%     \label{fig:wavelength_dependece}
% \end{figure*}

\subsection{Model Parameter Cuts}\label{sec:modelparamcuts}

% \begin{figure*}
%     \centering
%     \includegraphics[width=1\textwidth]{Sample Selection-7.pdf}
%     \caption{\textbf{(a)} Flowchart of our model parameter cuts. For each, the number of galaxies remaining from the HFF (left) and CANDELS (right) samples is indicated in the shaded box as in Figure~\ref{fig:selection_phot_only}. See text for further discussion of each quality cut. \textbf{(b)} Venn diagram showing the number of galaxies that have good single Sérsic fits, good disc fits, good bulge fits, and any combination of the three categories. A `good' fit is one that satisfies all of the quality cuts discussed in \S \ref{sec:data}. 
%     % make the phot cuts a table and take out of this figure! Or make other Figure -> then all selections in figures
%     %In \S \ref{sec:distinction}, we explain the parameters that we investigate for each subgroup in order to separate the total sample into discs and bulges.
%     }
%     \label{fig:selection_chart}
% \end{figure*}

\begin{figure}
    \centering
    \includegraphics[width=0.49\textwidth]{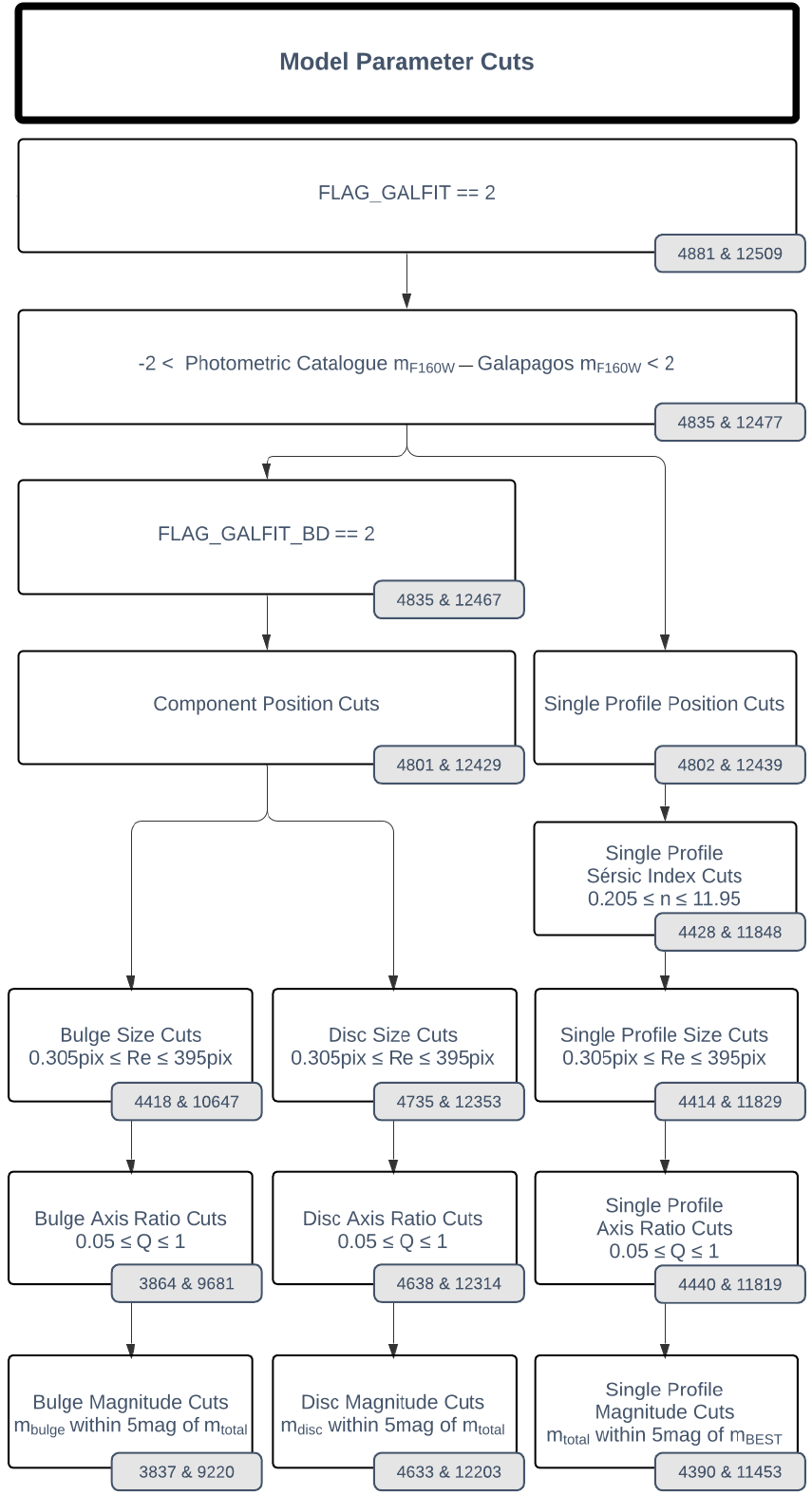}
    \caption{Flowchart of our model parameter cuts. For each, the number of galaxies remaining from the HFF (left) and CANDELS (right) samples is indicated in the shaded box as in Figure~\ref{fig:selection_phot_only}. See \S\ref{sec:modelparamcuts} for further discussion.
    }
\label{fig:selection_chart}
\end{figure}

We now apply a series of quality cuts based on the model parameters that are fit by \textsc{GalfitM}. These quality cuts are shown in Figure \ref{fig:selection_chart} and are used to identify galaxies that have reliable bulge, disc, and/or single component fits. We first remove any objects for which the single Sérsic modelling has not returned a result or was not started. There are a number of reasons why the modelling may not be started, including if the target object is not observed in a sufficient number bands. The minimum number of required bands is a user-specifiable parameter in \textsc{Galapagos-2} that we have set to the same requirements as \citetalias{Kalina2021}, i.e.~the target object must have sufficient data in at least three bands for the HFF and at least two bands for CANDELS. The fit may further crash if \textsc{GalfitM} cannot converge on a solution. The fitting status of objects is indicated by \texttt{flag\_galfit}, which is initially set to 0. This flag set to -1 if the fit is not started and 1 otherwise. If \textsc{GalfitM} completes the fit and converges on a solution, \texttt{flag\_galfit} is set to 2 \citep{MegaMorph}. This flag hence serves as a simple measure of the fitting status.  %For the HFF, we only model galaxies in the F160W footprint 

\begin{figure*}
    \centering
    \includegraphics[width=1\textwidth]{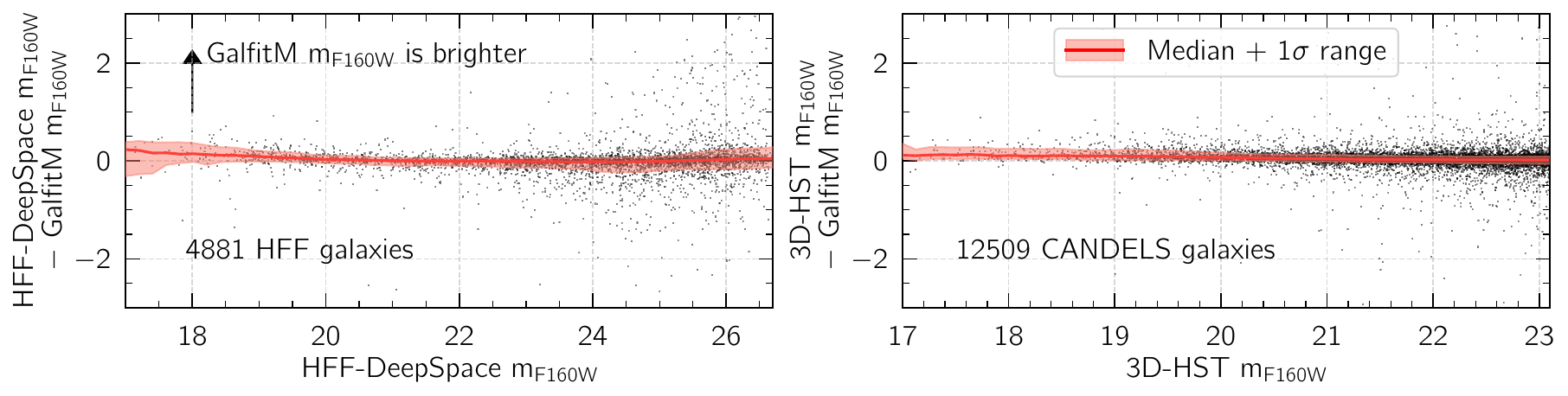}
    \vspace{-0.5cm}
    \caption{\textit{Left:} Difference between the F160W magnitudes from the HFF-DeepSpace photometric catalogues and those derived with \textsc{GalfitM}. \textit{Right:} We again show the F160W magnitude difference, but between the 3D-HST photometric catalogues and \textsc{GalfitM}. In \S \ref{sec:data}, we remove any objects with F160W magnitude differences larger than two magnitudes, which is a negligible fraction of the total sample. 
    %(46 out of 4881 galaxies and 32 out of 12509 galaxies for HFF and CANDELS, respectively) as is shown in this figure.  
    }
    \label{fig:H_mag_comparison}
\end{figure*}
\begin{figure}
    \centering
    \includegraphics[width=0.48\textwidth]{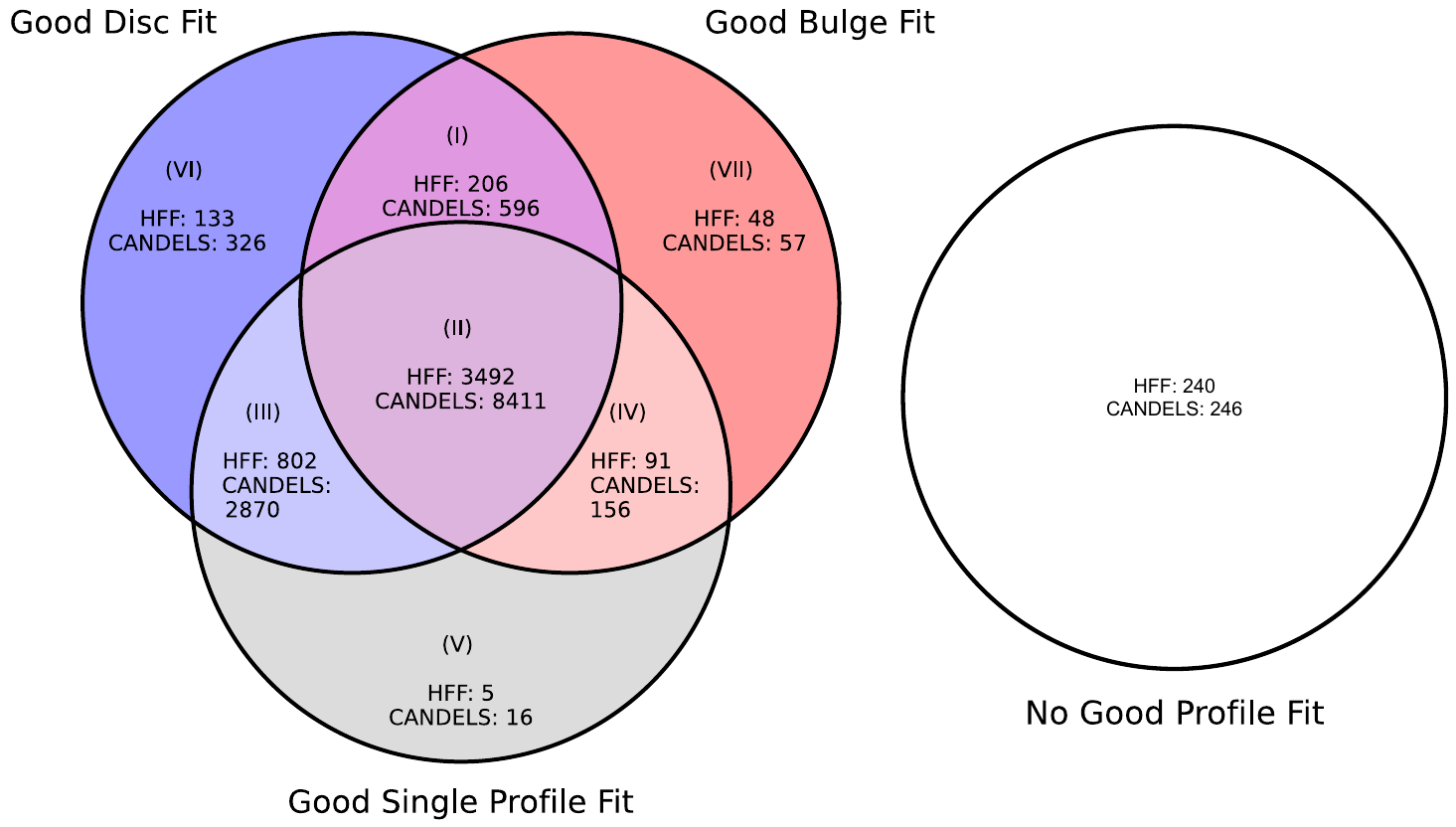}
    \vspace{-0.5cm}\caption{\textit{Left: }Venn diagram of the number of galaxies with good single Sérsic fits, good disc fits, good bulge fits, and any combination of these three categories. A `good' fit is one that satisfies all quality cuts discussed in \S \ref{sec:phot_cuts} and \ref{sec:modelparamcuts}. \textit{Right: }Number of galaxies that do not have a good fit in any profile.
    }
\label{fig:Venn_diagram}
\end{figure}

For the 4881 HFF and 12509 CANDELS galaxies that have \textsc{GalfitM} fits, we remove any galaxies for which the difference between the \textsc{GalfitM}-derived F160W magnitudes from the single Sérsic profile modelling and the F160W magnitudes from the photometric catalogues differ by more than two magnitudes. While we expect the F160W magnitudes from both to be generally consistent, we note here that the magnitudes are derived using fundamentally different techniques. Specifically, in the HFF-DeepSpace and 3D-HST catalogues magnitudes are measured using aperture photometry, whereby the extent of each galaxy is defined and the flux enclosed within that area is summed. We note that for both catalogues, a small AUTO-to-total factor is also applied to correct the \textsc{SExtractor} AUTO flux for the amount of light that falls outside the AUTO aperture for a point source (see \citealt{Skelton2014ApJS} and \citealt{Shipley2018} for more details). On the other hand, \textsc{GalfitM} models galaxy profiles and integrates them to infinity to obtain the total magnitude in each band. As such, a big discrepancy can be used as a measure of `fit failure', as it is likely that the profile fits tried to fit something unphysical in these cases, e.g. by including a neighbouring galaxy.

In Figure \ref{fig:H_mag_comparison}, we compare the H-band magnitudes from the HFF-DeepSpace and 3D-HST catalogues to the single Sérsic profile H-band magnitudes measured with \textsc{GalfitM}. The red line, which indicates the running median difference, is consistent with zero, showing that the independently-derived H-band magnitudes are in good agreement. There are only 46 HFF and 32 CANDELS galaxies that have H-band magnitudes which differ by more than two magnitudes and we have looked at the fits for all of these objects. In general, we find that objects that have fainter magnitudes in the \textsc{Galapagos-2} catalogues than reported in the photometric catalogues are `next' to bright objects, such that the aperture photometry is impacted by these bright neighbours. On the other hand, objects that have brighter \textsc{Galapagos-2} magnitudes seem to have generally unreliable fits (e.g. because the model is also trying to fit a neighbour). 
 %The majority of these get removed anyway by the model parameter cuts that we apply later on. 
 This latter category also contains a few objects that are two galaxies being modelled as one, further highlighting the need to remove these objects from the final sample.

% The remaining quality cuts are applied in an effort to identify which objects have reliable disc models, bulge models, and/or single Sérsic profile models. We discuss these separately and begin with the single Sérsic profile models, for which the majority of cuts are motivated by \citetalias{Kalina2021}. We first check that the central X and Y positions of each fit are within 0.3 arcseconds of the centre of their corresponding postage stamp that is created by \textsc{Galapagos-2}. In \textsc{Galapagos} \citep{Barden2012}, and by extension \textsc{Galapagos-2}, postage stamps are usually centred on individual objects (with an exception for some large objects along the image edge), where the extent of the postage stamp is computed using the size of the object from \textsc{SExtractor}.

The remaining quality cuts are applied separately to the disc models, bulge models, and single Sérsic profile models. The cuts on the single Sérsic profile models are motivated by \citetalias{Kalina2021}, but we discuss all models together as the selection criteria are mostly consistent. First, we check that the bulge+disc modelling has been successfully completed by checking that \texttt{flag$\_$galfit\_BD} is equal to 2. This flag works equivalently to \texttt{flag$\_$galfit} in the case of single Sérsic fits, and can be used as a filter to remove object fits with technical issues. We also require that the central positions of each fit are within 0.3 arcseconds of the centre of their corresponding postage stamp that is created by \textsc{Galapagos-2} (see \citealt{Barden2012} for more details about the postage stamp creation). % WRONG: In \textsc{Galapagos} \citep{Barden2012}, and by extension \textsc{Galapagos-2}, all objects are located in the centre of their respective postage stamp, even if they are located at an image edge, as in such cases, the postage stamps are padded with zeros. % that is being modelled for the HFF [CANDELS] sample. 
Although this criterion removes less than 1$\%$ of the sample, it ensures that any models which are not fitting the central target object are removed.

We next require that the Sérsic index of the single profile model satisfies $0.205\leq n \leq11.95$ following \citetalias{Kalina2021}. %because in the modelling, the fit is constrained to have $n\in$ (0.2, 12). Objects that have a Sérsic index that is approaching one of these limits are running into the fitting constraints and are thus removed from the sample.
As the Sérsic indices of the discs and bulges are fixed to 1 and 4, respectively, we do not apply any Sérsic index cuts to the bulge+disc models. We also check that the size of the galaxy and components is larger than 0.305 pixels and smaller than 395 pixels, unless they are, or belong to, bright cluster galaxies (bCGs) as defined in \cite{Shipley2018}, in which case we do not apply an upper size limit. This selection criterion is justified and described in more detail in \citetalias{Kalina2021}.
%As described in \citetalias{Kalina2021}, these objects are fit in a separate \textsc{Galapagos-2} run, manually checked, and consecutively merged into the object catalogues, so such constraints have neither been used in the fit, nor were they necessary. 
We further remove any galaxies and components that are fit with unphysically elongated profiles that have axis ratios lower than 0.05. This is a rather lenient cut as even the intrinsic thickness of discs is generally $\sim0.2$ \citep[e.g.][]{Lambas1992MNRAS, Kado-Fong2020ApJ}. Finally, we check that the magnitude in all bands is within 5 magnitudes from the magnitude derived with SExtractor \citep{Bertin1996}, with an empirically derived offset between bands as is used in the \textsc{Galapagos-2} setup for multi-band data. This is another lenient cut that is motivated by \citetalias{Kalina2021}.

Although there are a significant number of model parameter cuts applied, combined they remove $\lesssim5$\% of the sample. We note here that we also require that the bulge, disc, and single Sérsic profile magnitudes satisfy 0 $<$ m $<$ 40 in all bands. These cuts do not remove any objects, so we have not listed them in Figure \ref{fig:selection_chart}, but we mention them here for completeness.
%, as they are used within \textsc{Galapagos-2}. 
The final number of galaxies with reliable disc, bulge, and single profile fits are shown in Figure~\ref{fig:Venn_diagram} in the Venn diagram on the left. We note that for a large majority of our objects, none of the fits have run into fitting constraints. Finally, objects for which none of the profile fits were reliable are shown on the right of Figure~\ref{fig:Venn_diagram}. Although there are only a few of these, we note that most are removed because the \textsc{GalfitM} fits were not completed (i.e.~\texttt{flag$\_$galfit} $\neq2$).

%Next, we discuss the cuts applied to the bulge and disc fits. First, we check that the bulge+disc modelling has been successfully completed by checking that \texttt{flag$\_$galfit\_BD} is equal to 2. This flag works equivalently to \texttt{flag$\_$galfit} in the case of single Sérsic fits, and can be used as a simple filter to remove object fits with technical issues. Similar to the single profile fit, we check that the X and Y positions of the disc and the bulge are close to the centre of the postage stamp, which is expected to be the centre of the object that is being modelled. As the Sérsic index of the disc is fixed to 1 and the Sérsic index of the bulge is fixed to 4, we do not apply any Sérsic index cuts to the bulge+disc models. The effective radii and the axis ratios of the components are however not fixed, and therefore we constrain both of these in the same way as for the single profile fits. Finally, we check that the magnitudes of the bulges and discs are within five magnitudes of the magnitude corresponding to half of the total flux in all bands (i.e. the flux of the single Sérsic profile fit is split into two, and then the bulge and disc are checked to be within five magnitudes of that resulting magnitude).

% \begin{figure}
%     \centering
%     \includegraphics[width=0.47\textwidth]{failed_bulge_reason_hist.pdf}
%     \caption{Reasons why the bulge has failed while single \& disc fits do not. }
%     \label{fig:n_dist_SS}
% \end{figure}

%%%%%%%%%%%%%%%%%%%%%%%%%%%%%%%%%%%%%%%%%%
%%%%%%%%%%%%%%%%%%%%%%%%%%%%%%%%%%%%%%%%%%
%%%%%%%%%%%%%%%%%%%%%%%%%%%%%%%%%%%%%%%%%%
\section{Methods} \label{sec:methods}

As discussed in \S\ref{sec:sect315}, identifying whether a given galaxy is better modelled as a one-component object or a bulge+disc system is highly nontrivial. In this section, we present our methods for distinguishing one- from two-component galaxies in \S \ref{sec:distinction}. In \S\ref{sec:mass}, we present our methods for measuring the stellar masses of individual components.

%%%%%%%%%%%%%%%%%%%%%%%%%%%%%%%%%%%%%%%%%%
%%%%%%%%%%%%%%%%%%%%%%%%%%%%%%%%%%%%%%%%%%
%%%%%%%%%%%%%%%%%%%%%%%%%%%%%%%%%%%%%%%%%%

%%%%%%%%%%%%%%%%%%%%%%%%%%%%%%%%%%%%%%%%%%
%%%%%%%%%%%%%%%%%%%%%%%%%%%%%%%%%%%%%%%%%%
%%%%%%%%%%%%%%%%%%%%%%%%%%%%%%%%%%%%%%%%%%

%%%%%%%%%%%%%%%%%%%%%%%%%%%%%%%%%%%%%%%%%%
%%%%%%%%%%%%%%%%%%%%%%%%%%%%%%%%%%%%%%%%%%
%%%%%%%%%%%%%%%%%%%%%%%%%%%%%%%%%%%%%%%%%%
\subsection{Distinguishing between one- and two-component systems} \label{sec:distinction}
%%%%%%%%%%%%%%%%%%%%%%%%%%%%%%%%%%%%%%%%%%
%%%%%%%%%%%%%%%%%%%%%%%%%%%%%%%%%%%%%%%%%%
%%%%%%%%%%%%%%%%%%%%%%%%%%%%%%%%%%%%%%%%%%

Our goal in this subsection is to obtain a sample of reliable bulges and a sample of reliable discs. The bulge sample will consist of bulge-dominated elliptical galaxies and the bulge components of bulge+disc systems, while the disc sample will be comprised of disc galaxies and the disc components of multi-component galaxies. We note here that these samples are not complete in that if a particular galaxy makes it into our disc sample but not into the bulge sample, this does not imply that this galaxy has $no$ bulge component. It only means that we are able to reliably model and measure the properties of the disc, but not the bulge.

\begin{figure*}
    \centering
    \includegraphics[width=.97\textwidth]{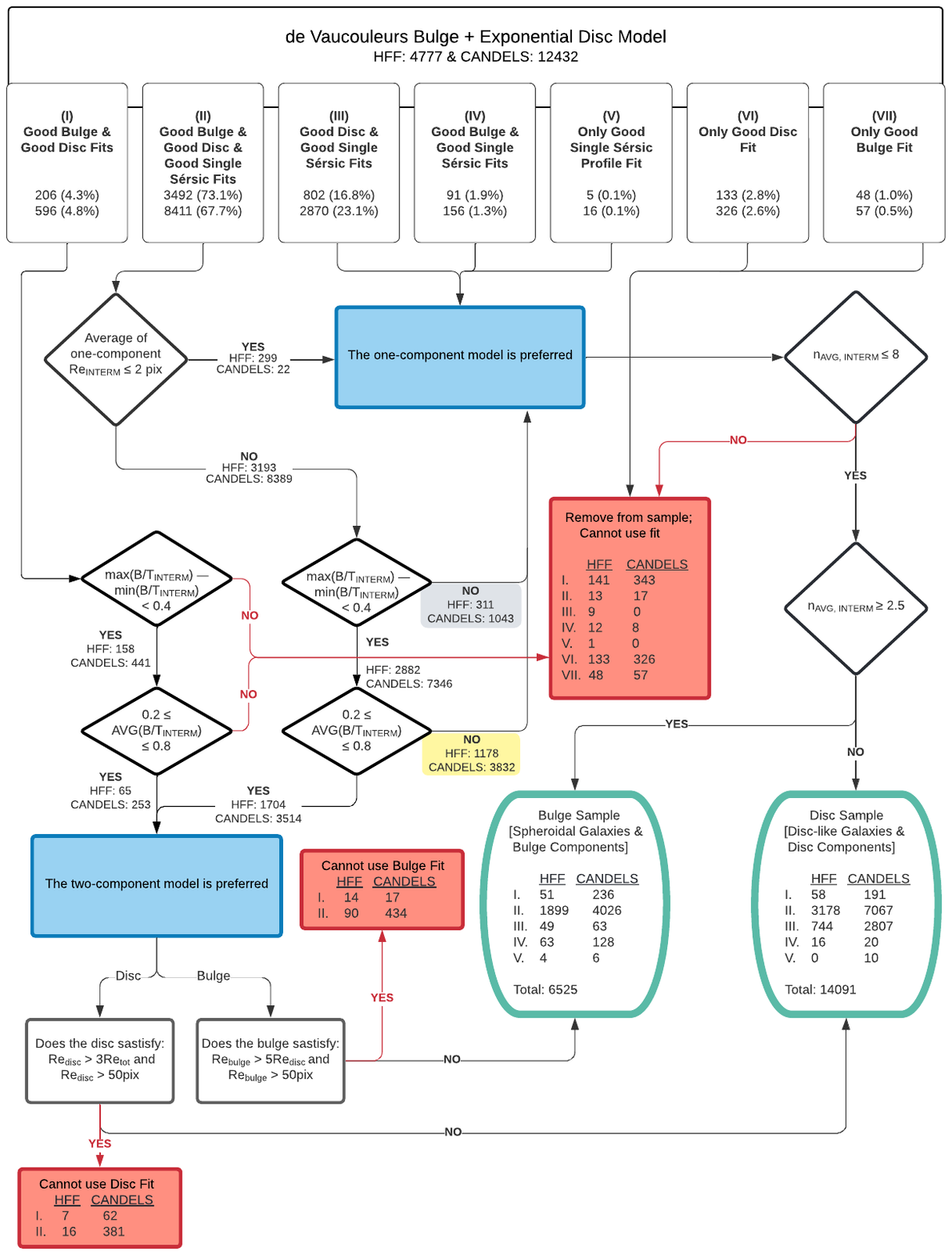}
    \caption{Flowchart indicating the criteria that the modelled galaxies in each of the seven categories shown in Figure~\ref{fig:Venn_diagram} must satisfy in order to make it into the final bulge sample and final disc sample. Each path is discussed in detail in the text.}
    \label{fig:flowchart}
\end{figure*}

To accomplish this task of obtaining reliable bulge and disc samples, we must first identify which galaxies are better modelled as single component objects as opposed to two-component systems. In the literature, there have been four general methods used to do this, although the details naturally vary between studies. One option is to devise a logic filter (i.e.~a physically motivated flagging system) that uses a series of tests to categorise galaxies. This was first employed by \cite{Allen2006MNRAS} and subsequently used by a number works \citep[e.g.][]{LacknerGunn2012MNRAS, Mendel2014ApJS, Meert2015MNRAS, MA2017A&A}. The so-called `F-test', first used in \cite{Simard2011}, is another alternative that compares the $\chi^2$ residuals of different models to establish which model best represents the data. 
%\cite{Lange2016} presented yet another way to assess the appropriateness of different model decompositions. 
Another method is presented in \cite{Lange2016} who run many installations of their modelling, 
%starting each with different initial conditions, and find where the models converge,
as previously discussed, and visually classify their best-fitting galaxy models as one- or multi-component. 
The Galaxy Zoo citizen science project also provides visually determined morphologies for $>10^{5}$ galaxies \citep{Lintott2008MNRAS}. More recently, deep-learning techniques have proven to be quite successful in selecting the best-fitting model \citep[e.g.][]{Huertas2015ApJS,Dimauro2018MNRAS, Ghosh2020ApJ}.
As can be seen, there are a number of different approaches, each with its own advantages. In this work, we have chosen to develop our own logic filter following \cite{Allen2006MNRAS}. Since we have divided our sample into seven subgroups, shown in Figure \ref{fig:Venn_diagram}, we use these as starting points. The decision tree that we have used to classify galaxies is shown in Figure \ref{fig:flowchart} and in the following subsections, we provide a detailed discussion of the set of criteria that each group of galaxies must meet in order to be considered in the final bulge and disc samples.

% \begin{figure*}
%     \centering
%     \includegraphics[width=1\textwidth]{flow_chart_v7.pdf}
    
%     \caption{Flowchart indicating the criteria that each model category must satisfy in order to make it into the final bulge sample and final disc sample. Each path is discussed in detail in the text. }
%     \label{fig:flowchart}
% \end{figure*}

\subsubsection{Group (I)}
%Objects with Good Bulge and Good Disc Fits}
\label{sec:GroupI}
We begin by discussing objects for which the single Sérsic profile fit has run into a fitting constraint, but the bulge and disc component fits have not. This group is shown in the left-most box at the top of Figure \ref{fig:flowchart} and is called group (I). This group consists of 206 HFF galaxies and 596 galaxies from the CANDELS fields. As the single Sérsic profile fit is unreliable for this set of galaxies, the one-component model is never preferred and we base all of our cuts on the bulge and disc model properties. 

We first check that the difference between the maximum and minimum B/T in the intermediate bands is smaller than 0.4, or in other words, the B/T does not vary by more than 40\% across bands. While the B/T is expected to vary somewhat with wavelength \citep[see e.g.][]{Vulcani2014MNRAS}, large variations indicate that the fit is not well constrained, and this selection helps to avoid such objects. If this condition on the B/T is satisfied, then the two-component model is preferred; otherwise, since there is no reliable single component model to revert to, the model is thrown out of the sample entirely. We make an important note here that the B/T is only considered in the intermediate bands, defined in \S \ref{sec:sect312}.
% . We have discussed why we deem some bands less reliable than others in \S \ref{sec:BD_models} but briefly reiterate this point here. We exclude the red-most and blue-most bands because of the way \textsc{GalfitM} fits Chebyshev polynomials to the parameters measured in each band. In the fitting, the bluest and reddest bands are the least constrained because they are endpoints, an effect referred to as Runge's phenomenon. Indeed, we often find that the edge bands have the most extreme B/T values. Additionally, high redshift galaxies are often ill-constrained in bluer bands. For example, in Figure \ref{fig:example_low_n_bulge}, the galaxy, which is at redshift $z=0.9287$, is barely visible in F435W and F606W. Hence any measurements in those bands are unreliable. We therefore decide to only use those bands that are at longer wavelengths than 3159\AA \ in the rest frame. We use 3159\AA \ because it is the effective wavelength of the U-band minus half of the filter's FWHM, as discussed in \S \ref{sec:data}. For the example galaxy in Figure \ref{fig:example_low_n_bulge}, this means using only the F814, F105W, F125W, and F140W bands to calculate the average B/T and checking for the extreme values.

Next, we check that the average bulge-to-total luminosity ratio in the intermediate bands satisfies B/T $\in (0.2, 0.8)$. Numerous works \citep[e.g.][]{Allen2006MNRAS, Meert2015MNRAS, MegaMorph2, Ghosh2023ApJ} have independently shown that if the B/T is larger than 0.8, then the decomposition is likely unreliable because the object is a pure-bulge or pure-spheroid, implying that no true disc component exists. This idea is also used by \cite{Trujillo2004MNRAS}, who replace objects with B/T $>$ 0.6 with single Sérsic profiles. Similarly, a low B/T (i.e.~$\leq 0.2$) indicates that the object is a disc-dominated galaxy. This limit is also used by \cite{Meert2015MNRAS}, who suggest that single profile fits be used for objects modelled with B/T $\leq 0.2$ in their work as the bulge+disc decomposition for such objects is unreliable. Additionally, \cite{Dimauro2018MNRAS} and \cite{MegaMorph2} use simulated galaxies across a range of redshifts to show that bulge magnitude uncertainties increase significantly in systems with B/T $< 0.2$. Moreover, \cite{Davari2016ApJ} have shown that due to the small angular sizes of bulges at $z\sim2$, their properties can only be reliably recovered for galaxies with B/T $\geq 0.2$. Based on these findings, we argue that we cannot properly measure component properties when one component dominates over the other by more than 80\%.
Hence, if the average B/T is outside these bounds for a given object, we treat it as a one-component system.

In the local Universe, however, $\sim70\%$ of spiral galaxies are expected to have B/T$\leq 0.2$ \citep[e.g.][]{Weinzirl2009ApJ}. Given that galaxies with such low B/T are common, we re-derive our stellar mass--size relations for individual components, that will be presented in \S\ref{sec:results}, imposing B/T $\in (0.1, 0.8)$ instead of B/T $\in (0.2, 0.8)$. We find that the stellar mas--size relations of the disc and bulge components remain largely unchanged. Therefore, while we choose to treat galaxies with B/T $\leq 0.2$ as one-component systems, we emphasise that this choice does not strongly impact our results.

%In general, we find that in cases where B/T $\leq 0.2$, the single Sérsic profile model parameters are very similar to those of the disc model, and similar to the parameters from the bulge model when B/T $\geq 0.8$. But, we choose not to consider the bulge models when B/T $\geq 0.8$ nor the disc models when B/T $\leq 0.2$ in place of the single Sérsic profile models as for this group of galaxies, the single profile fits have not satisfied all of our selection criteria discussed in \S \ref{sec:data}. This choice is motivated by the robustness of the model parameter cuts in removing unreliable models. As we do not have a reliable single Sérsic profile model to revert to for group (I) galaxies, objects with  B/T $\leq 0.2$ or  B/T $\geq 0.8$ are removed from the sample. 

From Figure \ref{fig:flowchart}, it can be seen that for group (I), 141 HFF and 343 CANDELS galaxies are removed from the sample. For the remaining galaxies, the two-component model is preferred and so we now independently check whether the bulge or disc models have returned physical parameters. For the disc component, we check if the effective radius of the disc is larger than 50 pixels. For galaxies in group (II), where we have a reliable single Sérsic profile fit, we also check if the effective radius of the disc is more than $3\times$ larger than the average effective radius of the single component model. 
%If this condition is not met, then the disc component fit is included in the final disc sample. 
If either of these conditions are met, then the disc component fit is likely nonphysical and is removed. This ensures that unrealistically large or elongated discs (i.e. bad fits) do not enter our
`good' sample.
%Although we do not use the $n=$\textit{free} models, an example of such a case can be seen in the middle panels of Figure \ref{fig:example_low_n_bulge}, where the disc component is fit with a much larger effective radius than the single Sérsic profile model. In fact, the disc fit would have been removed for this object based on the behaviour of the B/T, highlighting the efficacy of these checks. 
For the bulge component, we remove the bulge from the sample if the effective radius of the bulge is five times larger than the effective radius of the disc component, in an effort to remove any bulge components which are fit with nonphysical sizes. Similar to the discs, we also check whether the size of the bulge is larger than 50 pixels. This cut does not remove any bulges, as all of these are already flagged by previous cuts. Otherwise, the bugle component fit is included in the final bulge sample. In total, from group (I), we have 51 and 236 objects in the final bulge sample and 58 and 191 objects in the final disc sample from HFF and CANDELS, respectively.

\subsubsection{Group (II)}%Objects with Good Single Profile, Good Bulge, and Good Disc Fits}
\label{sec:all_good}
We next discuss the set of galaxies for which none of the fits have run into fitting constraints. We call this set of objects group (II) and they make up $\sim70\%$ of the total sample. As shown in Figure \ref{fig:flowchart}, we first check whether the effective radius from the single Sérsic profile fit is on average smaller than two pixels in the intermediate bands. 
%As in \S \ref{sec:GroupI}, the intermediate bands exclude the reddest and bluest band and any bands that have rest-frame wavelengths that fall bluewards of 3159\AA. 
Visual impression suggests that objects with effective radii $\leq2$pixels are simply too small to be properly decomposed into bulges and discs. Hence, the one-component model is preferred for the 299 HFF and 22 CANDELS objects that satisfy this condition. We note here that more HFF galaxies are removed mostly because the HFF images have larger pixel scales. If the one-component model is preferred, we check that the average Sérsic index of the single profile fit from the intermediate bands is less than eight, motivated by \citetalias{Kalina2021}. Objects with $n_{\mathrm{AVG,\ INTERM}}>8$ are removed from the sample, while objects with $n_{\mathrm{AVG,\ INTERM}}\leq8$ are separated into the final `bulge' and `disc' samples depending on the value of the average Sérsic index of the single profile fit, again considered only in the intermediate bands. For galaxies with $n \geq 2.5$, the single profile fit is included in the final bulge sample, similar to e.g.~\citet{Trujillo2007MNRAS} and \cite{ Roy2018MNRAS}. If the average Sérsic index is less than 2.5 in the intermediate bands, then the single profile fit is included in the final disc sample.

% Now, we need to distinguish and separate bulge+disc systems from one-component objects. We previously discussed the potential of using the AIC and BIC for this purpose. Indeed, these criteria have been tested as methods for reliably identifying whether a given galaxy is best modelled by a single component model or a two component model \citep[e.g.][]{Head2014MNRAS,Argyle2018MNRAS}. Unfortunately, our results have shown that these criteria alone do not provide a robust method for distinguishing pure bulges, pure discs, and two-component systems. However, when used in combination with other parameters, such as the B/T ratio, the BIC and AIC can provide useful information. Following this idea, we obtain the AIC and BIC using equations \ref{AICeq} and \ref{BICeq}, respectively. 

\begin{figure}
    \centering
    \includegraphics[width=0.48\textwidth]
    {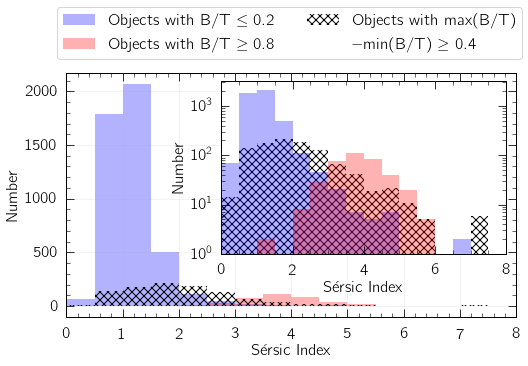}
    \vspace{-0.6cm}
    \caption{Sérsic index distribution of the single profile fits for objects that have B/T $\leq 0.2$ (blue) or B/T $\geq$ 0.8 (red). This sample is highlighted in yellow in Figure \ref{fig:flowchart}. The hatched area shows the single profile Sérsic index distribution of objects for which the B/T varies by more than 0.4 across bands. This sample is highlighted in grey in Figure \ref{fig:flowchart}. The inset shows the same distributions but on a log-scale y-axis to highlight that the distribution for the B/T $\leq$ 0.2 sample mostly contains objects with low Sérsic indices while the B/T $\geq$ 0.8 distribution  primarily contains objects with high Sérsic indices. However, this is not true for all objects, implying that a separation based on Sérsic index will not be perfect. }
    \label{fig:SersicDistributions}
\end{figure}

We largely base our selection on the behaviour of the B/T, similar to group (I) galaxies discussed in \S \ref{sec:GroupI}. As group (II) galaxies have reliable single Sérsic profile fits, instead of removing galaxies which do not satisfy the two B/T criteria, we revert to the one-component model. The number of HFF and CANDELS galaxies that remain after each decision point are indicated in Figure \ref{fig:flowchart}, but as the criteria that the galaxies must satisfy are the same as for group (I) galaxies, we do not discuss them again here. Instead, we investigate the properties of galaxies which do not satisfy the B/T criteria that we impose. We first look into the set of objects for which the B/T ratio varies by more than 0.4 across bands, which consists of a total of 1354 galaxies and is highlighted in grey in Figure \ref{fig:flowchart}. We find that the distribution of the average Sérsic index of the single profile fit peaks at $n\sim1.6$ but has a wide spread, such that some galaxies from this sample are classified as discs and others as bulges. This distribution is shown as a hatched region in Figure \ref{fig:SersicDistributions}. We have visually inspected $\sim$500 randomly chosen galaxies from this sample, finding that they are a mix of galaxy types. This is already suggested by the histogram as it encompasses parts of the blue distribution, which we find are mostly disc-like galaxies and the red distribution, which we find is made up of primarily bulge-dominated systems. 

We also investigate whether the galaxies with extreme B/T (i.e., B/T $\leq 0.2$ or B/T$\geq$0.8) end up in the final disc sample or the final bulge sample, based on their single profile Sérsic index. In Figure \ref{fig:SersicDistributions}, the galaxies with B/T $\leq 0.2$ are shown in blue. For these, the distribution strongly peaks around $n\sim1.1$, showing that the large majority of these galaxies are classified as discs, in agreement with our visual impression. On the other hand, the galaxies with B/T $\geq 0.8$ are shown in red and for these, the average single profile Sérsic index in the intermediate bands peaks around $n=3.7$, showing that the majority of galaxies with high B/T are classified as bulge systems, again in agreement with our visual impression. Indeed, we generally find that in cases where B/T $\leq 0.2$, the single Sérsic profile model parameters are similar to those of the disc model, and similar to the parameters from the bulge model when B/T $\geq 0.8$. Finally, we note that the blue and red distributions are not representative of the full galaxy sample. They are specifically shown for galaxies with extreme B/T values and we find that disc-dominated galaxies are classified as discs while the bulge-dominated galaxies are predominantly classified as bulges, lending confidence to our modelling and classification.

Lastly, we stress that while the one-component model is preferred for most objects from group (II), we do not imply that the majority of these galaxies are true single-component systems. In fact, we expect most galaxies at the redshifts that we are studying to be bulge + disc objects. Hence, as in the beginning of \S \ref{sec:distinction}, we again caution that our samples are not complete. For example, if a galaxy makes it into our final disc sample but not into the final bulge sample, this does not mean that this galaxy has no bulge component, but only that we are unable to reliably model and measure the properties of the bulge.

\subsubsection{Groups (III), (IV), and (V)}%Objects with Good Disc and Good Single Profile Fits, Good Bulge and  Good Single Profile Fits, and Only Good Single Profile Fits\\} 
\label{sec:groupIII}

We discuss group (III), group (IV), and group (V) objects in this section together as the one-component model is preferred for all of these galaxies. We begin with group (III), which consists of objects for which only the bulge component has run into fitting constraints. 
%The object shown in Figure~\ref{fig:example_low_n_bulge} is an example of a group (III) galaxy. Because both the radius and the axis ratio of the bulge are non-physical, we only believe the disc component and the single profile fit. 
For these, we revert back to the one-component model and classify the single Sérsic profile fit as a bulge or a disc based on the Sérsic index. The main reason behind this decision is that although the disc component of the bulge+disc model has not run into any fitting constraints, the two components are modelled together, such that the disc component model will depend on the bulge component model and vice versa. Hence, the one-component model will be preferred for galaxies in group (IV) as well. For group (V) galaxies, we naturally prefer the one-component model as it is the only one that has not run into fitting constraints.
%, and use the Sérsic index to decide whether to classify each object as disk or bulge dominated, as is often done in the literature. 
For galaxies that fall into one of these three groups, we follow the same steps to divide the single Sérsic fits into bulges and discs as for all objects for which the one-component model is preferred.
%Specifically, we first check whether the Sérsic index of the single profile fit satisfies $n_\mathrm{AVG,~INTERM}\leq8$ in order to remove any objects that are likely point sources or unreliably fit. Galaxies that do not satisfy this condition are removed from the sample as the fit cannot be used. In order to separate disc-like galaxies from ellipticals, we use the Sérsic index of the single component model. Objects with $n_\mathrm{AVG,~INTERM}\geq2.5$ are classified as bulges while the rest are included in the disc sample.

\subsubsection{Groups (VI) and (VII)} \label{sec:groupVI}
We remove any galaxies that only have a good disc fit or
good bulge fit because we revert back to the one-component model for galaxies for which one of the components has run into a fitting constraint as discussed in \S \ref{sec:groupIII}. As there is no reliable single profile fit for us to revert to for group (VI) and (VII) galaxies, they are removed. Altogether, these galaxies account for only $\sim3$\% of the total sample.

%%%%%%%%%%%%%%%%%%%%%%%%%%%%%%%%%%%%%%%%%%
%%%%%%%%%%%%%%%%%%%%%%%%%%%%%%%%%%%%%%%%%%
%%%%%%%%%%%%%%%%%%%%%%%%%%%%%%%%%%%%%%%%%%
\subsection{Determining the Stellar Masses of Bulges and Discs} \label{sec:mass}
%%%%%%%%%%%%%%%%%%%%%%%%%%%%%%%%%%%%%%%%%%
%%%%%%%%%%%%%%%%%%%%%%%%%%%%%%%%%%%%%%%%%%
%%%%%%%%%%%%%%%%%%%%%%%%%%%%%%%%%%%%%%%%%%
\begin{figure*}
    \centering
    \includegraphics[width=1\textwidth]{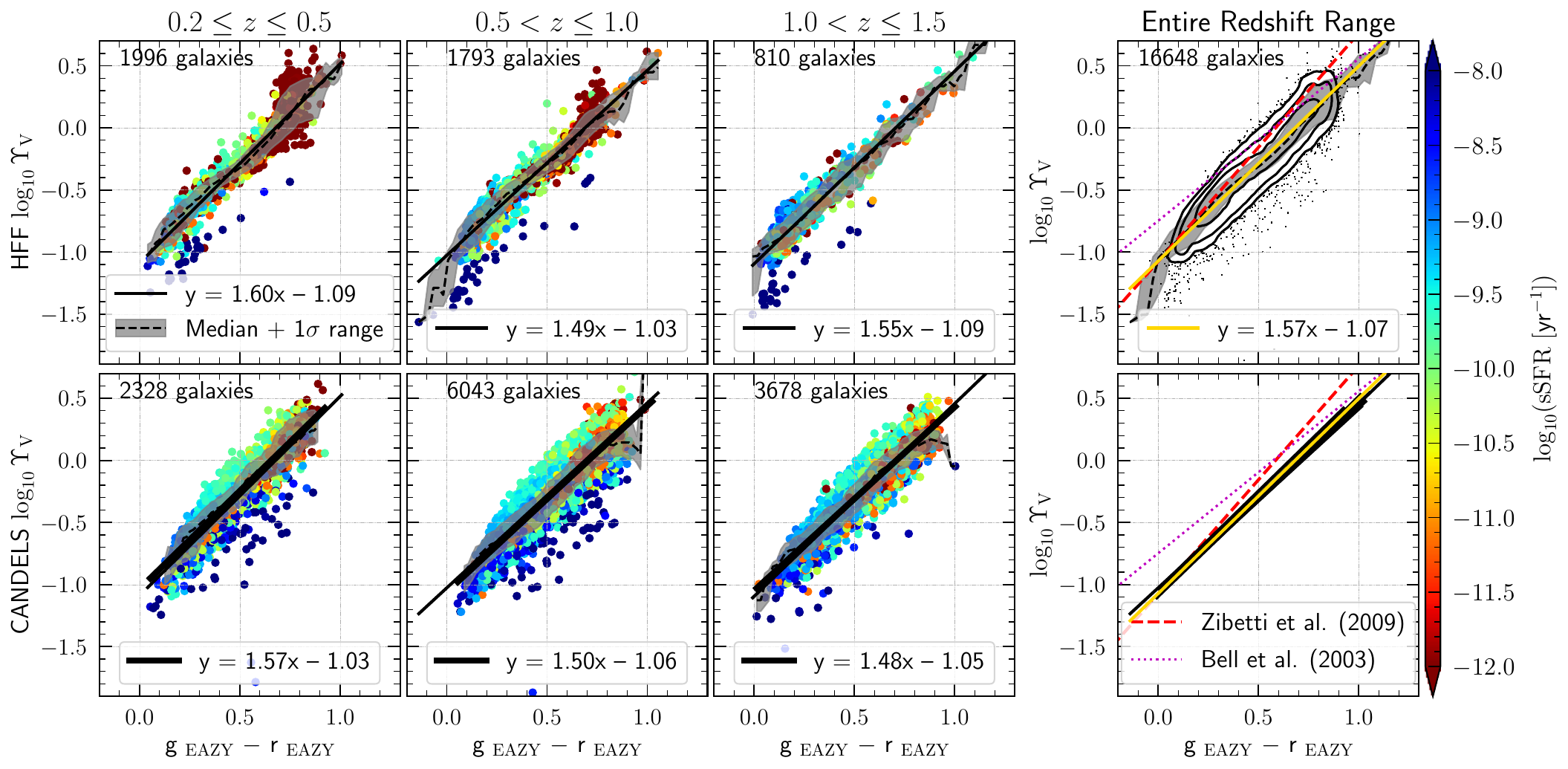}
    \vspace{-0.4cm}
    \caption{V-band mass-to-light ratio as a function of rest-frame g-r colour as derived from \textsc{eazy} \protect\citep[][]{eazy_brammer} for HFF (top three panels) and CANDELS (bottom three panels) galaxies across three redshift bins, where objects are colour-coded by their sSFR. The right-most panels show the entire sample, with the yellow line showing the best-fitting $\log_{10}(\Upsilon_\star)$ -- colour relation. The contours in the right-most top panel show the number density. The relations derived from all redshift bins are overplotted in the bottom right-most panel, showing that the yellow best-fitting line to all data is representative of the entire redshift range.}
    \label{fig:g_r}
\end{figure*}

Global galaxy stellar masses are obtained from the HFF-DeepSpace and 3D-HST catalogues and are corrected for the difference between the F160W magnitude from the photometric catalogues and those derived from fitting Sérsic profiles with \textsc{GalfitM}, as briefly explained in \S\ref{sec:phot_cuts}. The stellar masses reported in the photometric catalogues are derived by fitting spectral energy distributions (SEDs) to multi-band photometry using FAST, assuming a \cite{Chabrier2003} initial mass function (IMF), solar metallicity, exponentially declining star formation histories with a minimum e-folding time of $\log_{10}$($\tau$/yr) = 7, and the \cite{Calzetti2000} dust attenuation law, in both the HFF-DeepSpace and 3D-HST catalogues. There are a few minor differences -- namely, the HFF-DeepSpace catalogues assume a minimum age of 10 Myr and $0<\mathrm{A}_{\mathrm{V}}<6$mag, while the 3D-HST catalogues use a minimum age of 40 Myr and $0<\mathrm{A}_{\mathrm{V}}<4$mag. As a result of these differences, we treat the CANDELS and HFF samples separately for the remainder of this section, although these differences do not significantly impact the recovered galaxy properties.

\begin{figure}
    \centering
    \includegraphics[width=0.48\textwidth]{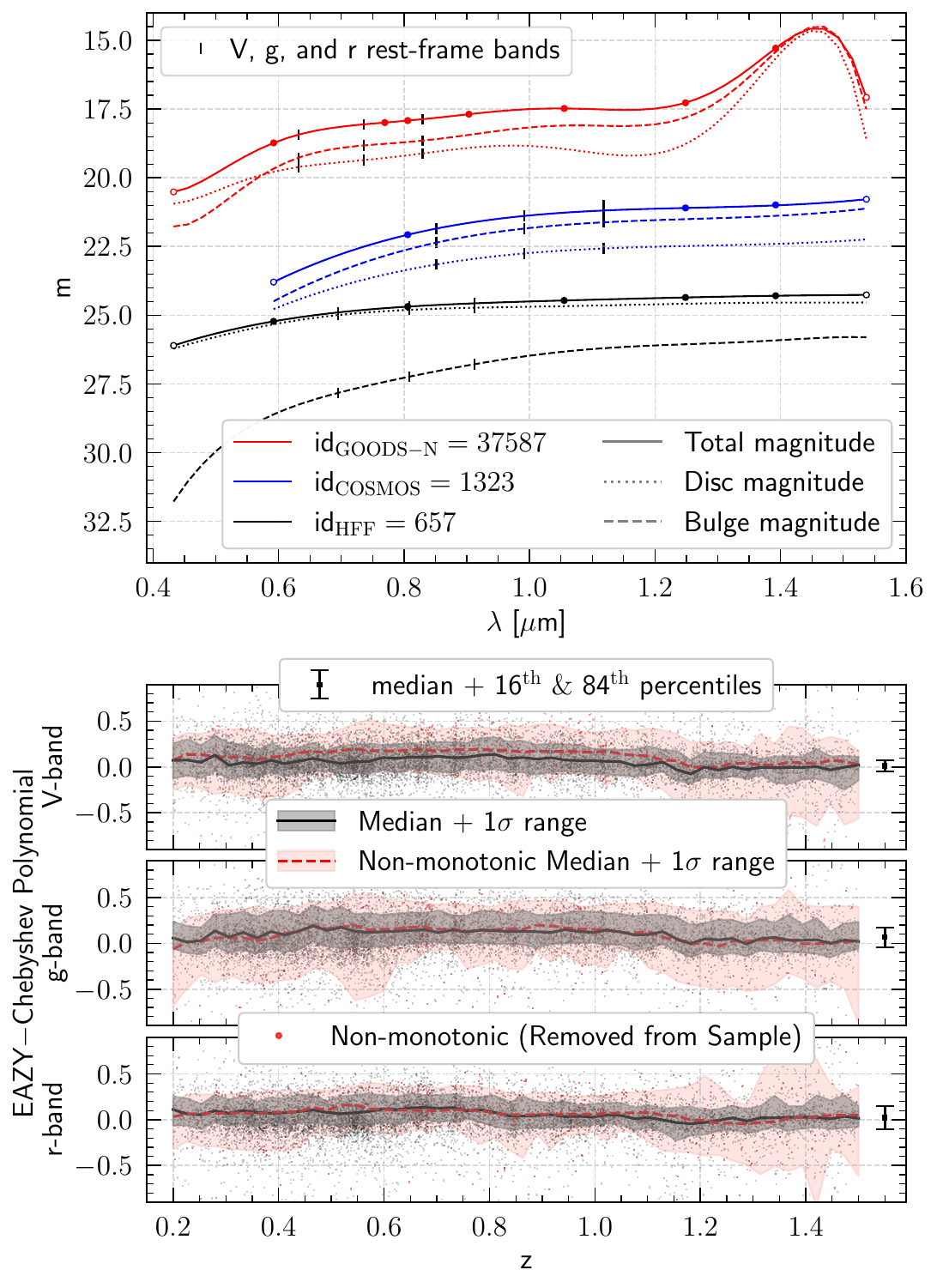}
    \vspace{-0.5cm}
    \caption{\textit{Top}: Total, bulge, and disc magnitudes as a function of wavelength for three example galaxies from the GOODS-N, COSMOS, and HFF fields, shown in red, blue and black, respectively. All observed bands used in the modelling are indicated as circles, with the ones which we deem less reliable (see \S \ref{sec:sect312}) shown as empty circles. The black vertical lines indicate the rest-frame V-, g-, and r-band magnitudes. \textit{Bottom}: Difference between the total magnitude from the \textsc{GalfitM} Chebyshev polynomials and \textsc{eazy} for each rest-frame band that we use to estimate the stellar mass of the bulge and disc components. These are generally in good agreement as shown by the median and $1\sigma$ range. Objects removed by our requirement that galaxies' magnitude polynomials change monotonically with wavelength are shown in red.} 
    \label{fig:wavelength_dependence}
\end{figure}

While many well-studied methods exist for deriving the global stellar masses of galaxies from photometric and spectroscopic data, deriving stellar masses for galaxy components is significantly more challenging, not least of which because the photometry derived will be less accurate. Obtaining stellar masses relies on accurate SED fitting, which itself strongly relies on many assumptions such as the stellar population synthesis models, IMF, dust extinction, and more (see \citealt{Pacifici2023ApJ} for an analysis of different SED fitting codes). One possible method for obtaining the stellar masses of bulge and disc components is to determine the flux emitted by the bulge and disc components from the B/T and D/T light ratios in each band, and fit the bulge and disc photometry with FAST, as in e.g.~\cite{Dimauro2019}. More recent techniques with codes such as \textsc{ProFuse} \citep{Robotham2022} that simultaneously perform SED fitting to galaxy components in conjunction with structural decomposition of their light profiles, allow component stellar masses to be measured directly as an optimised parameter rather than being derived post model fitting \citep[see also][]{Bellstedt2024MNRAS}. Unfortunately, this technique is not feasible at the higher redshifts explored in this work due to limitations in resolution.
%present stellar mass -- size relations of massive bulge and disc components using this method. 

In this paper, we derive stellar masses for the components of HFF and CANDELS galaxies using the empirical relation between optical colour and the stellar mass-to-light ratio ($\Upsilon_\star$). This relation, which was first proposed by \cite{Tinsley1972A&A}, has been studied extensively \citep[e.g.][]{Bell2001,Bell2003,Driver2006MNRAS,Zibetti2009, Taylor2011,Into2013,vandeSande2015,Zhang2017, Miller2023ApJ...945..155M, Arjen2024ApJ} and is a popular technique for obtaining stellar masses for galaxy components. For instance, \cite{Cook2019MNRAS} use the \cite{Zibetti2009} $\log_{10}(\Upsilon_\star)$ -- (g$-$i) relation, while \cite{Lange2016}, \cite{ Costantin2020ApJ}, and \cite{Barsanti2021ApJ} either use the \cite{Taylor2011} relation directly, or some modified version of it \citep[see][for a discussion of the underlying assumptions and sources of error for different $\log_{10}(\Upsilon_\star)$ -- colour relations]{Hon2022MNRAS}. While we opt to use this method to obtain the stellar masses of bulge and disc components, we find that the stellar masses obtained by splitting galaxies' total flux and fitting the bulge and disc photometry with \textsc{FAST} are in excellent agreement. We show detailed comparisons between the two approaches in Appendix \ref{appendix_mass}.

Although the $\log_{10}(\Upsilon_\star)$ -- colour relation has been well studied, the impacts of the assumptions made on the star formation history (SFH), stellar age, metallicity, and dust introduce significant scatter around this relation \citep[][]{Bell2001, Gallazzi2009ApJS, Zhang2017, Hon2022MNRAS}. Additionally, the choice of IMF and dust attenuation laws strongly impact the $\log_{10}(\Upsilon_\star)$ -- colour relation. As a result, the derived empirical relations vary across studies. For instance, because \cite{Zibetti2009} assume a bursty SFH, their relation is much steeper than the \cite{Zhang2017} relation, who focused on dwarf galaxies. We therefore derive our own best-fit parameters to this relation for a \cite{BruzualCharlot2003} stellar population synthesis model library with a \cite{Chabrier2003} IMF, keeping all of the same parameters that were used to derive physical properties in the HFF-DeepSpace and 3D-HST photometric catalogues.

We have chosen to derive the $\log_{10}(\Upsilon_\star)$ -- (g$-$r) relation, which is shown in Figure \ref{fig:g_r}. We use the g- and r-bands because we find that the relations are tighter when using these bands compared to other rest-frame filters. We also obtain $\log_{10}(\Upsilon_\star)$ in the rest-frame V-band because \cite{Zhang2017} have shown that when using optical colours to estimate $\log_{10}(\Upsilon_\star)$, the V-band is the least susceptible to systematic dependencies on SFHs, metallicities, and dust extinction. All global rest-frame colours are obtained using \textsc{eazy} \citep{eazy_brammer}. 
%due to the different assumptions that were used to derive the physical properties of galaxies, discussed at the beginning of the section. 

In Figure~\ref{fig:g_r}, we split our sample into three redshift bins and by survey (i.e.~HFF and CANDELS).
We fit the $\log_{10}(\Upsilon_\star)$ -- (g$-$r) relation in each survey and redshift bin individually, shown as black lines (thick for CANDELS; thin for HFF in their respective panels). The running median and 1$\sigma$ range are shown as dashed lines and shaded regions, respectively and the best fit HFF relation is overlaid the CANDELS relation at the same redshift for easier comparison. The best-fitting relations for each panel are reported in the legend. We find that these best-fitting relations are consistent across HFF and CANDELS, and largely independent of redshift. Given these results, we fit one $\log_{10}(\Upsilon_\star)$ -- (g$-$r) relation to all of our galaxies, which is shown in yellow in the right-most panels of Figure \ref{fig:g_r}, and compared to others from the literature.

Once we have our empirical $\log_{10}(\Upsilon_\star)$ -- colour relation, we use the magnitude Chebyshev polynomial of the discs and bulges to obtain the rest-frame V-, g-, and r-band magnitudes. %, as this is the only information available for the individual components (i.e.~\textsc{EAZY} cannot be used here). 
Using these values, we then derive the stellar mass of the individual components from our best-fitting empirical $\log_{10}(\Upsilon_\star)$ -- (g--r) relation and the estimated rest-frame V-band magnitude. Generally, it is not advisable to obtain rest-frame magnitudes from the Chebyshev polynomial from \textsc{GalfitM} (see e.g.~\citealt[]{MegaMorph}; \citetalias{Kalina2021}) because the magnitude polynomial is allowed to vary freely as a function of wavelength. This is done in order to recover the complex wavelength dependence of galaxy SEDs as best as possible, but unfortunately, it can also lead to Runge's phenomenon.
%, in which the function can oscillate wildly between the fixed points where data is available. 
%As such, we need one more safety net in our procedure to check that the polynomial varies smoothly and does not oscillate as a function of wavelength. 
Thus, we additionally check that the polynomial varies smoothly and does not oscillate as a function of wavelength.
In practice, we find that this is true for the majority of magnitude polynomials. We show for these polynomials for three randomly chosen example galaxies in Figure \ref{fig:wavelength_dependence}. The solid lines represent the magnitude of the whole galaxy, and the dashed and dotted lines show the magnitudes of the bulge and disc components, respectively. The circles show the magnitudes in the observed bands, with the empty circles indicating bands in which we do not trust the B/T and D/T because they are endpoints or fall blueward of 3159{\AA} in the rest frame, as discussed in \S\ref{sec:data}. Although the object from the GOODS-N field, shown in red, appears to suffer from Runge's phenomenon to some extent at the reddest bands, this is a relatively low redshift galaxy such that the rest-frame V-, g-, and r-bands, shown as black vertical lines, are at wavelengths where the polynomial is well constrained. Motivated by the lack of oscillations in the magnitude Chebyshev polynomials, we devise a method for identifying objects for which we can reliably obtain such intermediate magnitude values from the Chebyshev polynomial in order to obtain g$-$r colours for the galaxy components. For each desired rest-frame wavelength, we check if the magnitude polynomial within the observed bands that enclose it is monotonically decreasing or increasing with wavelength, as is the case for the three example galaxies shown in Figure \ref{fig:wavelength_dependence}. Objects for which the magnitude Chebyshev polynomial oscillates between data points that do not enclose rest-frame bands remain in the sample since rest-frame values are not obtained from the portion of the polynomial that suffers from Runge's phenomenon.

While this check primarily excludes faint galaxies, which typically have larger magnitude uncertainties and more variation in their magnitude Chebyshev polynomials, it ensures that we only use rest-frame magnitudes for galaxies that have well-constrained polynomials. This requirement removes 952 galaxies ($\sim6\%$ of the total sample). To further assess the reliability of the rest-frame values that we obtain from the polynomials, we compare the rest-frame V-, g-, and r-band magnitudes from the Chebyshev polynomials for the whole galaxy and those derived by \textsc{eazy} \citep{eazy_brammer} in the bottom three panels of Figure \ref{fig:wavelength_dependence}, finding very good agreement between the two for all three bands across our entire redshift range. The medians and 1$\sigma$ variations of the V-, g-, and r-band magnitude differences for the objects that are rejected by our monotonic requirement are shown as a red dashed lines and red shaded regions, respectively. We find that the 1$\sigma$ variation for these is much larger than it is for the monotonic sample, providing added justification for excluding these objects.

% ##################################################################################
\section{Stellar Mass--Size Relations}
\label{sec:results}
% ##################################################################################

In the previous sections, we used several model parameters to reliably select samples of bulges and discs and derived their stellar masses. In this section, we present the stellar mass--size relations for these components. We use two methods to obtain stellar mass--size relations -- one for galaxies where the one-component model is preferred and another for galaxies for which the two-component model is preferred. For the one-component galaxies, we take the corrected stellar mass and the 5000\AA \ effective radius from the single Sérsic profile size Chebyshev polynomial. This approach is equivalent to the one used by \citetalias{Kalina2021}.
%and the resulting stellar mass--size relations, which are shown in panels (d) -- (f) of Figures \ref{fig:Mass_size_disk} and \ref{fig:Mass_size_bulge}, are consistent with the ones presented in \citetalias{Kalina2021}, as expected.
%as seen from the middle panels of Figures \ref{fig:Mass_size_disk} and \ref{fig:Mass_size_bulge}. 
For the two-component galaxies, we obtain the stellar mass of each component using the methods described in \S \ref{sec:mass}. The effective radii of the bulges and discs are obtained from the bulge and disc size Chebyshev polynomials, respectively. As discussed in \S\ref{localminima}, sizes are not allowed to change as a function of wavelength for the components, i.e.~they are the same at all wavelengths. Due to these model assumptions, the components' sizes can be measured at any wavelength, but we choose to measure them at 5000\AA \ for consistency.
%it is irrelevant at which wavelength we choose to measure the components' sizes. 
%We constrain the sizes of the components to be constant with wavelength based on the findings of \cite{MegaMorph2}, who argue that although differences in colour can exist within galaxy components, these are small compared to the colour differences between the bulge and the disc components, which makes them difficult to measure reliably. 
The resulting stellar mass--size relations are discussed in more detail in \S\ref{sec:discussion_discs} and \ref{sec:discussion_bulges} for our disc and bulge samples, respectively.

Throughout this section, we colour-code all galaxies and components by their sSFRs. For galaxies, as a whole, we obtain sSFRs from the 3D-HST and HFF-DeepSpace photometric catalogues. For the components, we derive bulge and disc photometry (see also Appendix~\ref{appendix_mass}) and obtain physical parameters by fitting this photometry with \textsc{FAST}. In short, we convert the component magnitude measurements from \textsc{GalfitM} in each filter listed in Table~\ref{tab:Bands_used} to fluxes. This assumes that the magnitudes in each band are measured well, but as bulge+disc decomposition is highly challenging, this is not always true. As the measurement errors increase in cases where the magnitudes are poorly constrained, this uncertainty can be, to some extent, taken into account.
We derive uncertainties for our flux measurements by adding the uncertainty from the \textsc{GalfitM} modelling and the uncertainty from the 3D-HST and HFF-DeepSpace photometric catalogues in quadrature.

Another complication is that the 3D-HST and HFF-DeepSpace catalogues contain flux measurements for more filters than we have used in our modelling, which means that the SEDs that were used to derive the physical properties of the galaxies presented in \cite{Shipley2018} and \cite{Skelton2014ApJS} are better constrained thanks to these additional data. To better constrain the SEDs of our discs and bulges, we devise a method to include filters redder than the H-band, which are crucial for obtaining reliable properties, such as stellar mass. We obtain B/T and D/T measurements redward of the H-band by fitting a line to the available B/T and D/T measurements as a function of wavelength, taking into consideration that the bulge and disc cannot account for more than 100\% or less than 0\% of the total flux (i.e. we impose 1 and 0 as limits). We then take the total flux from the photometric catalogues and divide it by the B/T [D/T] from the fitted line to obtain the flux of the bulge [disc] at each filter.

% ####################################
% ####################################
\subsection{Stellar Mass--Size Relation for Discs and Disc-Like Galaxies} \label{sec:discussion_discs}
% ####################################
% ####################################

\begin{figure*}
    \centering
    \includegraphics[width=1\textwidth]{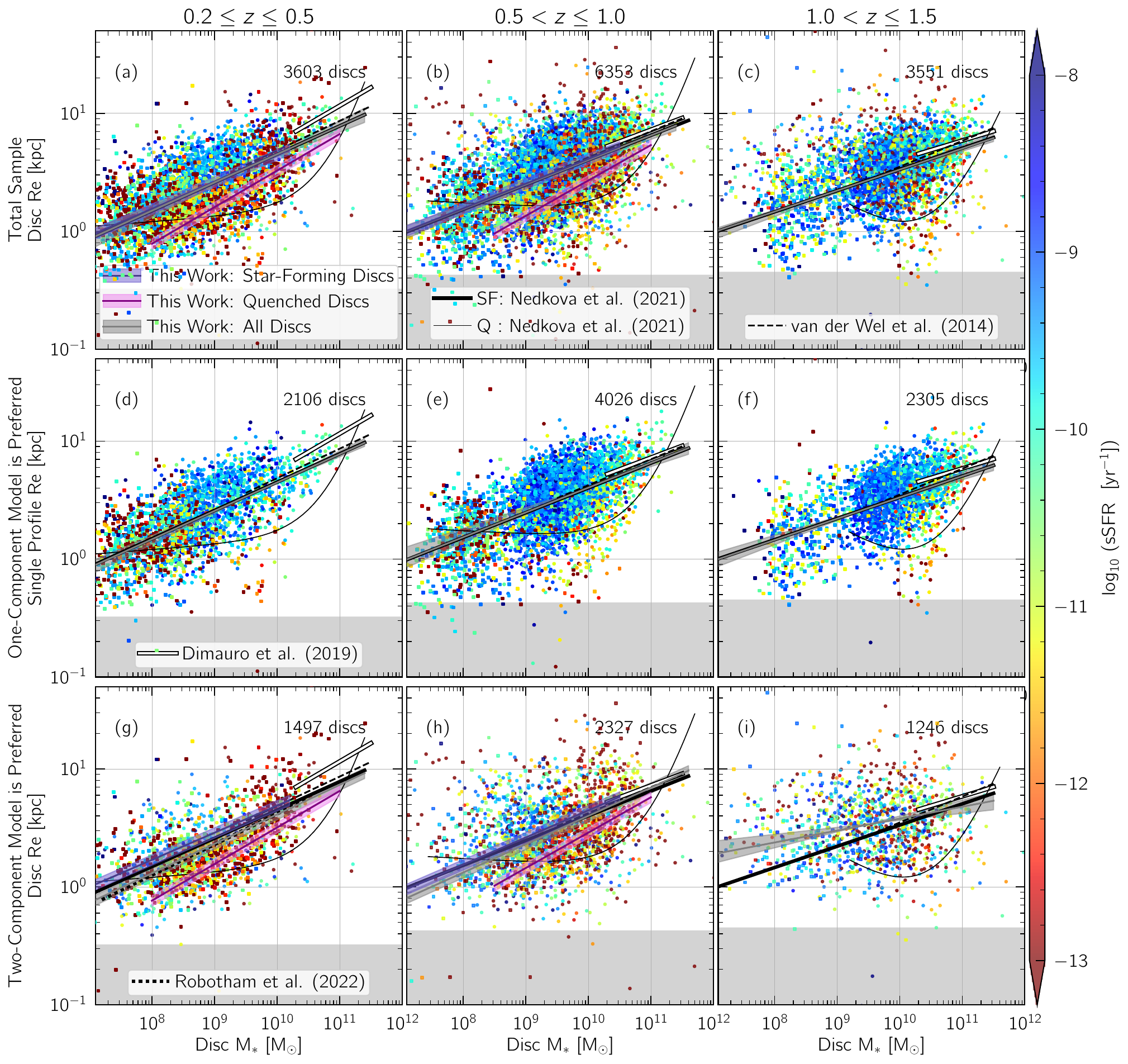}
    \vspace{-0.6cm}\caption{Disc component stellar mass--size relations across three redshift bins, shown in each column. We present stellar mass--size relations for the total disc sample (top row), the disc sample for which the one-component model is preferred (middle row) and the disc components of bulge+disc systems (bottom row). In panels (a), (b), (g), and (h), discs are divided into star-forming and quiescent, with the relations for the star-forming and quenched discs shown in blue and purple, respectively. Objects from the HFF fields are shown as squares and CANDELS galaxies are shown as points. The stellar mass--size relations for star-forming galaxies from \protect\citet[][dashed lines]{vdWel2014} and \protect\citetalias{Kalina2021} (thick black lines) are shown to highlight the agreement between the mass--size relation of discs and star-forming galaxies. Our results are also in qualitative agreement with \protect\cite{Dimauro2019} and \protect\cite{Robotham2022}, indicated by the white and dotted black lines, respectively. All objects and components are colour-coded by their sSFR (see text for details). The grey areas in each panel indicate sizes smaller than half of the FWHM of the F160W PSF at the maximum redshift of each panel, that are potentially difficult to measure (see also \protect\citetalias{Kalina2021}). The number of objects is indicated in the top right of each panel. 
    %trend in green with SFR
    }
    \label{fig:Mass_size_disk}
\end{figure*}

We present stellar mass--size relations for discs and disc-like galaxies in Figure \ref{fig:Mass_size_disk}, where panels (a) -- (c) show the entire disc sample, panels (d) -- (f) show the relations of galaxies for which the one-component model is preferred, and panels (g) -- (i) at the bottom of the figures show the results for objects for which the two-component model is preferred. All galaxies and components are colour-coded by their sSFRs, which are obtained as described above.%. We obtain these for galaxies, as whole, directly from the 3D-HST and HFF-DeepSpace photometric catalogues, while the sSFRs of individual components are obtained by running \textsc{FAST} described in \S\ref{sec:results}.

%we generate catalogues of bulge and disc flux by splitting the total flux from the photometric catalogues by the B/T and D/T luminosity ratios in each band, where the B/T and D/T are obtained from the modelling. We then use these catalogues as inputs to \textsc{FAST} \citep{Kriek2009} to derive the sSFR. Full details about the construction of the photometric
% catalogues for bulges and discs are provided in Appendix~\ref{appendix_mass}.

% 

% \begin{table*}
% \caption{Best-fitting parameters for disc-like galaxies and disc components according to the single power law shown in Equation \ref{eq:singlepl} for each redshift bin. We report the best-fitting parameters for the entire disc sample, for the subsample of discs for which the one-component model is preferred, and for the subsample for which the two-component model is preferred.  }
% \centering
% \begin{tabular}{c c c c c c c }
% \hline
% \multicolumn{1}{l}{ \ } & 
% \multicolumn{2}{c}{Total Sample} &
% \multicolumn{2}{c}{One-Component Model} &
% \multicolumn{2}{c}{Two-Component Model}\\
% % \multicolumn{1}{l}{ \ } & 
% % \multicolumn{1}{c}{\underline{(M$_{*}$ $\geq$ 10$^{9.5}$ M$_{\odot}$) }}
% % \multicolumn{4}{l}{\underline{(M$_{*}$ $\geq$ 10$^{7}$ M$_{\odot}$) }}\\ 

% $z$& $\log_{10}$($A$) & $B$ &  $\log_{10}$($A$) & $B$    &$\log_{10}$($A$) & $B$  \\
% \hline 

% 0.2$\leq{z}\leq$0.5 & 
% $0.82\pm{0.02}$&$0.24{\pm0.01}$ & $0.82{\pm0.02}$&$0.24{\pm0.01}$ & $0.82{\pm0.02}$&$0.24{\pm0.01}$\\

% 0.5$<$ $z$ $\leq$1.0 &
% $0.77{\pm0.01}$&$0.19{\pm0.02}$ & $0.75{\pm0.01}$&$0.21{\pm0.01}$ & $0.76{\pm0.01}$&$0.21{\pm0.01}$\\

% 1.0$<$ $z$ $\leq$1.5 & 
% $0.64{\pm0.01}$ & $0.17{\pm0.01}$ & $0.65{\pm0.01}$ & $0.18{\pm0.01}$ & $0.65{\pm0.03}$ & $0.10{\pm0.04}$\\

% \hline

% \end{tabular}
% \label{tab:disc_params}
% \end{table*}

% \begingroup
% \setlength{\tabcolsep}{2.6pt} % Default value: 6pt
% \renewcommand{\arraystretch}{1.1} % Default value: 1
\begin{table*}
\caption{Best-fitting parameters for disc-like galaxies and disc components according to the single power law from Equation \ref{eq:singlepl} for each redshift bin. We report these parameters for the total disc sample, for the subsample of discs for which the one-component model is preferred, and for the subsample for which the two-component model is preferred. For total and two-component samples at $0.2\leq z \leq 1.0$, we further divide discs into star-forming and quiescent based on their sSFRs and report the best-fitting relations for each individually.}
\centering
\begin{tabular}{c c c c c c c }
\hline
\multicolumn{1}{l}{ \ } & 
\multicolumn{6}{c}{\textbf{Total Sample}} \\
\multicolumn{1}{l}{ \ } & \multicolumn{2}{c}{All Discs} & \multicolumn{2}{c}{Star-Forming Discs} & \multicolumn{2}{c}{Quenched Discs} \\
% \multicolumn{1}{l}{ \ } & 
% \multicolumn{1}{c}{\underline{(M$_{*}$ $\geq$ 10$^{9.5}$ M$_{\odot}$) }}
% \multicolumn{4}{l}{\underline{(M$_{*}$ $\geq$ 10$^{7}$ M$_{\odot}$) }}\\ 

$z$& $\log_{10}$($A$) & $B$ & $\log_{10}$($A$) & $B$  & $\log_{10}$($A$) & $B$ \\%&  $\log_{10}$($A$) & $B$    &$\log_{10}$($A$) & $B$  \\
\hline 

0.2$\leq{z}\leq$0.5 & 
$0.81\pm{0.02}$&$0.24\pm{0.01}$ &$0.81\pm{0.02}$ & $0.23\pm{0.01}$ & $0.73\pm{0.02}$ & $0.31\pm{0.01}$ \\%& 

0.5$<$ $z$ $\leq$1.0 &
$0.74\pm{0.01}$&$0.21\pm{0.02}$ & $0.76\pm{0.01}$& $0.21\pm{0.02}$ & $0.64\pm{0.01}$ & $0.30\pm{0.01}$\\%& 

1.0$<$ $z$ $\leq$1.5 & 
$0.64\pm{0.01}$ & $0.17\pm{0.01}$ \\% & $0.65\pm{0.01}$ & $0.18\pm{0.01}$ & $0.65\pm{0.03}$ & $0.10\pm{0.04}$\\

\hline

\multicolumn{1}{l}{ \ } & \multicolumn{6}{c}{\textbf{One-Component Model}} \\
%$z$& $\log_{10}$($A$) & $B$ \\%&  $\log_{10}$($A$) & $B$    &$\log_{10}$($A$) & $B$  \\
\hline 
0.2$\leq{z}\leq$0.5 & 
$0.82\pm{0.02}$&$0.24\pm{0.01}$\\%& $0.82\pm{0.02}$&$0.24\pm{0.01}$\\

0.5$<$ $z$ $\leq$1.0 &
$0.75\pm{0.01}$&$0.21\pm{0.01}$ \\%& $0.76\pm{0.01}$&$0.21\pm{0.01}$\\

1.0$<$ $z$ $\leq$1.5 & 
$0.65\pm{0.01}$ & $0.18\pm{0.01}$ \\%& $0.65\pm{0.03}$ & $0.10\pm{0.04}$\\

\hline
\multicolumn{1}{l}{ \ } & \multicolumn{6}{c}{\textbf{Two-Component Model}}\\
%$z$& $\log_{10}$($A$) & $B$ \\%&  $\log_{10}$($A$) & $B$    &$\log_{10}$($A$) & $B$  \\
\hline 
0.2$\leq{z}\leq$0.5 & 
$0.79\pm{0.02}$&$0.25\pm{0.01}$ & $0.85\pm{0.03}$& $0.23\pm{0.01}$ & $0.72\pm{0.02}$ & $0.31\pm{0.01}$\\

0.5$<$ $z$ $\leq$1.0 &
$0.77\pm{0.01}$&$0.24\pm{0.01}$ &  $0.80\pm{0.02}$ & $0.23\pm{0.01}$ & $0.67\pm{0.02}$ &  $0.30\pm{0.01}$\\

1.0$<$ $z$ $\leq$1.5 & 
 $0.65\pm{0.03}$ & $0.10\pm{0.04}$\\
\hline

\end{tabular}
\label{tab:disc_params}
\end{table*}

The best-fitting stellar mass--size relations with their 1$\sigma$ confidence ranges for each panel of Figure\ref{fig:Mass_size_disk} are indicated in grey, and are derived in a way that is consistent with \citetalias{Kalina2021}. Specifically, we fit the disc relations with 
 
 \begin{equation}
    R_e = A \Big( \frac{M_*}{5 \times 10^{10} \ \mathrm{M}_\odot} \Big) ^ {B}
    \label{eq:singlepl}
\end{equation}

\noindent as is commonly adopted in the literature \citep[e.g.][]{vdWel2014, Dimauro2019, Nedkova2024}, where $R_e$ is the effective radius, M$_*$ is the stellar mass and $A$ and $B$ are fit parameters, describing the normalisation and slope of the relation, respectively. 

% To provide easier comparisons with the quenched galaxy stellar mass--size relations presented in \citetalias{Kalina2021}, we fit the bulge stellar mass--size relations with a double power law following
% %As the bulge components exhibit a flattening in the stellar mass -- size relation, we fit those with

% \begin{equation}
% R_e = \gamma \big( M_* \big)^{\alpha} \Big( 1 + \frac{M_*}{10^{\delta}} \Big)^{\beta - \alpha}
% % \log_{10}R_e &= \log_{10}\gamma + \alpha \log_{10} M_* + (\beta - \alpha)  \log_{10}\Big( 1 + 10^{\log_{10} M_* -\delta} \Big)
% \label{eq:doublepl}
% \end{equation} 

% %\noindent following \cite{Shen2003MNRAS}, \cite{Lange2016}, and \citetalias{Kalina2021},
% \noindent where $\gamma$ is the normalisation, or the effective radius at a stellar mass of 1 M$_\odot$. The slopes at the low and high mass ends are described by $\alpha$ and $\beta$ respectively, and 10$^\delta$ represents the transition stellar mass between the low and high mass slope. 

The best-fitting parameters are obtained using a fitting procedure that explores the parameter space with a Markov Chain Monte Carlo (MCMC) algorithm exploiting the \textsc{emcee} package \citep{emcee2013PASP}. We assume uniform priors for all fit parameters and the contribution to the model by each galaxy is inversely weighted by its uncertainty in mass and size such that objects with larger uncertainties contribute less. In brief, the size uncertainties are obtained from \textsc{GalfitM} and increased by a factor of three as discussed in \S \ref{sec:sect314}, while the uncertainty on the stellar mass depends on whether the one- or two-component model is preferred. When the one-component model is preferred, the mass uncertainty is obtained from \textsc{FAST}. When the two-component model is preferred, the stellar mass uncertainty is taken to be the uncertainty on the corrected stellar mass from \textsc{FAST} added in quadrature with 0.1 dex of the component stellar mass in order to account for the spread in the $\log_{10}(\Upsilon_\star)$ -- colour relation, shown in Figure \ref{fig:g_r}. We provide our best-fitting parameters for the disc stellar mass--size relations in Table \ref{tab:disc_params}.

Figure \ref{fig:Mass_size_disk} shows that there is good agreement between the stellar mass--size relations that we derive for all discs in this work, shown in grey, and those of star-forming galaxies derived in \citetalias{Kalina2021}, shown as thick black solid lines. Similar trends are present -- namely, the normalisation decreases with redshift -- suggesting that high redshift discs are smaller in size and more compact than low redshift discs. \cite{Dokkum2015ApJ} find that while individual galaxies likely have complex formation histories including merger events and star burst episodes, as a population galaxies follow specific tracks on the size--mass plane that are determined by the dominant mode of growth of the ensemble. For star-forming galaxies, this dominant mode of growth is star formation. Hence, the interpretation of disc components and disc-like galaxies following stellar mass--size relations that are consistent with those of star-forming galaxies is that both disc-like galaxies and the disc components of galaxies indeed evolve via star formation.
%This appears to hold true for both discs and host galaxies that are quenched and star-forming, as indicated by the colour-coding.
%As previously discussed, the sSFRs are obtained by running \textsc{FAST} \citep{Kriek2009} on the disc SEDs, which we derive by splitting the total flux of the galaxies by the D/T at each wavelength (see \S \ref{sec:results} and Appendix \ref{appendix_mass} for more details). 
%Galaxies for which the one-component model is preferred are colour-coded by the sSFRs that are reported in the HFF-DeepSpace \citep{Shipley2018} and 3D-HST \citep{Skelton2014ApJS} photometric catalogues. 

In the nearby Universe, \cite{Abramson2014ApJ} have shown that galaxies with massive bulges have lower global sSFRs. They argue that star formation primarily occurs in the disc components of galaxies and while bulge components contribute to the total stellar mass of galaxies, they do not significantly contribute to the overall star formation, resulting in a suppressed global sSFR. This directly influences the so-called galaxy ``main sequence'', a tight correlation between galaxy star
formation rates and stellar masses \citep[e.g.][]{Whitaker2012ApJ...754L..29W, Popesso2023MNRAS.519.1526P}. Specifically, there is a flattening of the main sequence at the massive end where galaxies are more likely to be bulge-dominated.
At higher redshifts,
\cite{Whitaker2015ApJ} have found that the scatter of the main sequence is related to galaxy structure, such that the slope of the main sequence is roughly unity for disc-like galaxies, while galaxies with more dominant bulges exhibit a shallower slope on the SFR-stellar mass plane. Combined, these results suggest that pure disc galaxies and disc components lie on the star-forming main sequence. Our findings from Figure~\ref{fig:Mass_size_disk} are in general agreement with this picture as we also find disc-like galaxies and disc components are evolving primarily via star formation. %\cite{Dokkum2015ApJ} have shown are evolving primarily via star formation. 

% While individual galaxies likely have complex formation histories including merger events and star burst episodes, of compaction, merg- ers, and star bursts, the population of massive galaxies should follow a particular track in the size-mass plane that is deter- mined by the dominant mode of growth when the evolution of many galaxies is averaged. Tracks derived from observa- tions and simulations are shown in Fig. 22. The blue and red tracks show the evolution of galaxies matched by their cumu- lative number density, for (relatively) low mass galaxies (van Dokkum et al. 2013, blue) and high mass galaxies (Patel et al. 2013, red). The solid parts of the curves are for 1.5 < z < 3 and the dotted parts for 0 < z < 1.5. Low mass galaxies evolve along a single track with a slope of ∼ 0.3. High mass galaxies evolve along a similar track from z ∼ 3 to z ∼ 1.5 but then turn “upward”, around the time when star formation ceases and the growth becomes dominated by dry mergers (see Sect. 9.1).

An interesting feature of the stellar mass--size relation of disc galaxies for which we prefer the one-component model, shown in panels (d), (e), and (f) of Figure \ref{fig:Mass_size_disk}, is that at $z\lesssim 1$, lower mass disc galaxies seem to have lower sSFRs. Typically, low-mass galaxies tend to have higher sSFRs \citep[e.g.][]{Feulner2005ApJ, Whitaker2014ApJ, Belfiore2018MNRAS}, or sSFRs that are consistent with those of high-mass galaxies \citep[e.g.][]{Ramraj2017MNRAS, Cedres2021ApJ}, making this observed trend somewhat unexpected. We note that these low-mass objects are some of the faintest in our sample, which in turn implies that their SEDs are less reliable. This possibly explains some of the scatter we measure in the sSFRs. However, we would expect the low-mass discs in other panels to be at least equally impacted by poorly constrained SEDs, yet we observe more low-mass, low-sSFR disc galaxies in panel (d) than (e) and (f). This suggests that this trend in the stellar mass--size relation of disc galaxies is not driven by uncertainties in our sSFR measurements. We also find that $\sim74\%$ of the objects in panels (d) and (e) of Figure \ref{fig:Mass_size_disk} with stellar masses $\leq10^{9}$ M$_\odot$ and sSFR $\leq10^{-11}$ yr$^{-1}$ are from the HFF cluster fields, while over our full sSFR range, $\sim43\%$ of low-mass discs are from the HFF clusters. These results are in broad agreement with \cite{Tan2022ApJ}, who show that disc-like quiescent galaxies dominate the HFF quiescent population at $10^{8.5}$M$_\odot \lesssim$ M$_* \lesssim 10^{9.5}$M$_\odot$. In dense environments, galaxies are affected by environmental processes such as ram pressure stripping that can remove the gas necessary for any further star formation from a galaxy while leaving its disc morphology intact \citep[e.g.][]{Weinmann2006MNRAS}. Low-mass galaxies are more likely to be stripped due to their shallower gravitational potential wells, which would result in low-mass galaxies with disc-like morphologies but low sSFRs, consistent with our findings in Figure \ref{fig:Mass_size_disk}.

Another feature of our stellar mass--size relations can be most clearly seen in the bottom panels of Figure \ref{fig:Mass_size_disk}, where we plot the disc sample for which the two-component is preferred. %(i.e.~the stellar mass -- size relation is derived using the mass and size of the disc components from \textsc{GalfitM}).
At $z\leq1$, we find that disc components with low sSFRs tend to lie below discs with higher sSFRs on the size--mass plane. This suggests that discs with low sSFRs tend to be smaller than discs with higher sSFRs, at least when the effective radius is measured at 5000\AA. This may constitute evidence for compaction \citep[e.g.][]{Zolotov2015MNRAS} where disc components of galaxies that are quenching or have recently quenched move `downwards' onto the stellar mass -- size relation of quenched galaxies. We note here that in panel (g), this trend holds over a large stellar mass range (i.e.~$10^8$M$_\odot\lesssim$M$_* \lesssim10^{11}$M$_\odot$). In other words, this is not a trend that we observe only for low mass systems which are typically faint and therefore have less reliable bulge+disc decompositions. Indeed, we find that the median uncertainty on the $\log_{10}$(sSFR) measurements across all redshift bins is roughly $\sim0.4$ dex for both galaxies with higher sSFRs ($\log_{10}$(sSFR)$\geq$-11 yr$^{-1}$) and lower sSFRs ($\log_{10}$(sSFR)$<$-11 yr$^{-1}$).

We first assess whether these quenched discs are truly discs or if they are elliptical galaxies that are misclassified as discs. We find that there are 568 passive discs in panel (g) of Figure \ref{fig:Mass_size_disk}, which are identified by selecting galaxies with $\log_{10}$(sSFR) $\leq -11$ yr$^{-1}$ in this panel. Of these, 424 are from the HFF fields, which suggests that there may be an environmental effect that is responsible for shaping these discs. Indeed, previous works, such as \cite{Chan2021ApJ}, have shown that there is an excess of quenched discs in cluster environments. This effect also explains some of the redshift dependence that we observe -- namely, that we find a significant sample of these quenched discs at $0.2\leq z \leq1.0$ but not at higher redshifts -- because all HFF clusters have spectroscopic redshifts $z<1.0$.

Given the difference between star-forming and quenched discs on the size--mass plane, we divide our disc samples in panels (a), (b), (g), and (h) of Figure~\ref{fig:Mass_size_disk} using a $\log_{10}$(sSFR)$=-11$ yr$^{-1}$ threshold following e.g.~\cite{Ilbert2010ApJ, Ilbert2013A&A} and \cite{Sarron2021MNRAS}. 
We do not separate the sample in this way in other panels as there are not enough quenched discs to obtain reliable stellar mass--size relations. 
In Figure~\ref{fig:Mass_size_disk}, the relations for the star-forming and quenched discs are shown in blue and purple, respectively. 
In general, we find that the relations of star-forming discs agree well with those of the full disc sample (grey lines) and star-forming galaxies from \citet[][dashed lines]{vdWel2014} and \citetalias{Kalina2021} (thick black lines). This implies that star-forming disc components are evolving in mass and size via star formation \citep{Dokkum2015ApJ}. Interestingly, quenched discs appear to lie on different stellar mass--size relations. Specifically, their relations are steeper and have lower normalisations (see Table~\ref{tab:disc_params}). This is consistent with our findings that discs with low sSFRs tend to be smaller and suggests that quenched discs are evolving via mechanisms that are different from those through which star-forming discs evolve. These mechanisms could include minor mergers which cause significant changes in galaxies' sizes but leave their stellar mass mostly unchanged \citep[e.g.][]{Buitrago2008ApJ, Bezanson2009ApJ, Naab2009ApJ, Bluck2012ApJ}. Hence, processes such as minor mergers could explain the steeper slope of the stellar mass--size relation of quenched discs, but future work is needed to confirm this.

Next, we compare our stellar mass--size relations to those from previous works. The bottom panels of Figure \ref{fig:Mass_size_disk} show the disc sample where the two-component model is preferred. At $z\leq1$, we find good agreement with the stellar mass--size relations of star-forming galaxies from \citetalias{Kalina2021} (thick black line), suggesting that discs and star-forming galaxies are likely -- and perhaps somewhat unsurprisingly -- evolving due to the same mechanisms, or least through mechanisms that impact size and mass in similar ways. However, at $1.0\leq z\leq1.5$ (panel i), we find a shallower relation for disc components than at low redshift. This is likely due to the smaller sample size in this panel coupled with the challenges of measuring component properties at high redshift. At higher redshifts, galaxies are expected to be clumpier and more irregular which would also lead to larger measured scatter. Due to cosmology, the resolution at $z=1.5$ is also reduced by almost a factor of 3 compared to the resolution at $z=0.2$. Hence, with increasing redshift, we are not only sampling less evolved galaxies, but also losing sensitivity and resolution, all of which make it harder to separate the bulge and disc components in our highest redshift bin. 
When combined with data from the one-component discs, this effect is mitigated as seen in panel (c) of Figure \ref{fig:Mass_size_disk}. In comparing our disc stellar mass--size relations with those from \cite{Dimauro2019} for massive discs with  M$_* >$ 2 $\times$ 10$^{10}$ M$_\odot$ from the CANDELS fields (white lines), we find good agreement at the high mass end, especially at $z\geq0.5$. In our lowest redshift bin, we find a lower normalisation than \cite{Dimauro2019}, as can be seen in panels (a), (d), and (g). This could be a result of the different redshifts used in this work and in \cite{Dimauro2019} as we only include galaxies with $z\geq0.2$; however, we note that our disc stellar mass--size relations and the relation derived by \cite{Robotham2022} at $z\sim0$ agree within 1$\sigma$. Therefore, these differences are more likely due to different sample selections. We note that \citetalias{Kalina2021} and \cite{Dimauro2019} also find different stellar mass--size relations for star-forming galaxies due to differences in the sample selection (see Appendix B of \citetalias{Kalina2021} for a discussion).

\begin{figure}
    \centering
    \includegraphics[width=0.485\textwidth]{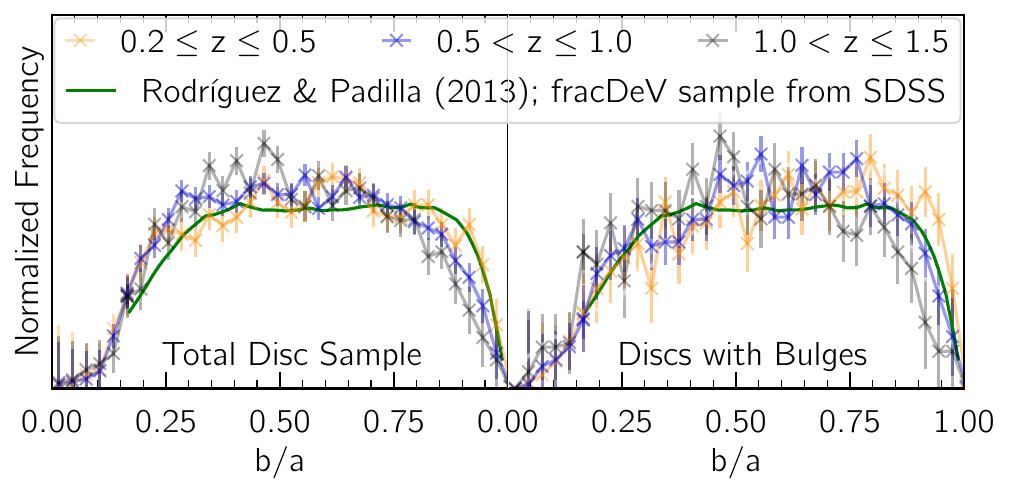}
    \vspace{-0.5cm}
    \caption{Normalised number distribution of the axis ratios in our three redshift bins for the sample of all disc components and disc-like galaxies on the left and objects with both a disc and a significant bulge component (i.e.~B/T $\geq$0.5) on the right. Uncertainties are obtained by bootstrapping the samples and results from \protect\cite{Rodrigues2013MNRAS} for SDSS galaxies at $z\sim0$ are reproduced in green.}
    \label{fig:q_dist}
\end{figure}

Finally, we note that this sample of disc components and one-component disc-like galaxies, as an ensemble, should have low axis ratios when viewed edge-on and be approximately round when viewed face-on. Hence, histograms of their axis ratios should be nearly flat, with a cut-off at the intrinsic thickness, around $b/a\lesssim0.2$ \citep[e.g.][]{vdwel2014ApJ...792L...6V}. Figure~\ref{fig:q_dist} demonstrates this, providing added evidence that this sample does indeed consist of disc structures. This is both a technical test, showing that the bulge+disc decompositions are inclination dependent, and a physical test indicating that objects with $n\sim1$ are shaped like flattened, rotating discs. In green, we show results from \cite{Rodrigues2013MNRAS}, where disc-like galaxies are selected using the fracDeV parameter, originally defined in \cite{Abazajian2005AJ}. This parameter indicates whether a galaxy's light profiles is closer to an exponential or a \citeauthor{deVaucouleurs} profile. For both disc samples, shown in the left and right panels of Figure~\ref{fig:q_dist}, we find excellent agreement with these results. We note that number distribution for the sample of disc galaxies with significant bulge components, shown in the right panel, has larger uncertainties because it is a smaller sample but it also better aligns with the results from \cite{Rodrigues2013MNRAS}. The striking agreement between the axis ratio distributions from \cite{Rodrigues2013MNRAS} and those for disc galaxies with bulge components from our work indicates that most $z\sim0$ galaxies likely have significant bulge components (see also \citealt{Huertas-Company2016MNRAS}). Lastly, we note that these distributions are also consistent with \cite{vdwel2014ApJ...792L...6V} and \cite{Pandya2023arXiv231015232P}, when their results are combined over the stellar mass ranges used in this work.

% The right panel of Figure~\ref{fig:q_dist} displays the same normalised number distribution but for the bulge components and one-component elliptical galaxies. As for the disc sample, this distribution is qualitatively consistent with expectations for spheroidal objects \citep[e.g.][]{Lambas1992MNRAS, Vincent2005ApJ,Padilla2008MNRAS, Rodrigues2013MNRAS}. However, we find a slightly higher fraction of bulges and elliptical galaxies with low axis ratios than \cite{Rodrigues2013MNRAS}. This is likely due to central structures such as bars, which are elongated, coupled with our lower resolution compared to nearby galaxies from the Sloan Digital Sky Survey (SDSS; \citealt{York2000AJ}). We discuss the bulge component and elliptical galaxy sample and its properties in greater detail in the following section. 

\subsection{Stellar Mass--Size Relation for Bulges and Elliptical Galaxies} \label{sec:discussion_bulges}

\begin{figure*}
    \centering
    \includegraphics[width=1\textwidth]{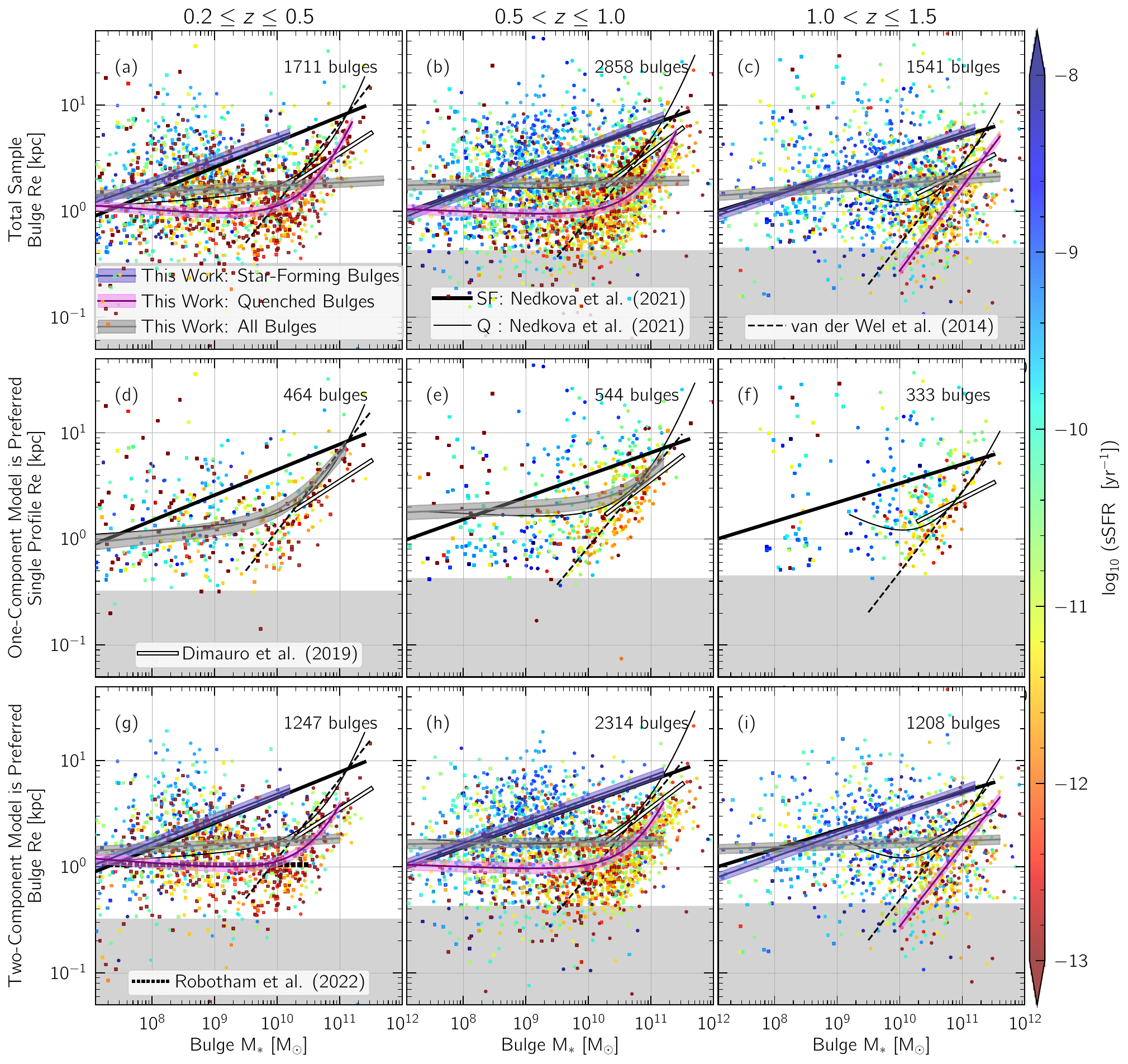}
    \vspace{-0.6cm}\caption{Same as Figure \ref{fig:Mass_size_disk}, but for bulge components. We provide fits for star-forming bulges in blue, quenched bulges in purple, and all bulges in grey. The stellar mass--size relations for quiescent and star-forming galaxies from \protect\citetalias{Kalina2021} are shown as a double and single power law, respectively, as well as the  stellar mass--size relations for quiescent from \protect\cite{vdWel2014}, in each redshift bin to highlight the level of agreement between the relations of bulges and quiescent galaxies. At the high mass end, the stellar mass--size relations of bulges generally agree with those derived by \protect\cite{Dimauro2019}, indicated by the white lines and at the low mass end, our results are consistent with \protect\citet[][black dotted line]{Robotham2022}.}
    \label{fig:Mass_size_bulge}
\end{figure*}

In Figure~\ref{fig:Mass_size_bulge}, we present the stellar mass--size relation of bulge components and bulge-dominated, or elliptical, galaxies. We first note that there are more galaxies for which we prefer the two-component model, shown in panels (g) -- (i) than galaxies for which we prefer the one-component model, shown in panels (d) -- (f). This difference becomes more significant with increasing redshift, with the one-component galaxies making up $\sim27\%$ of the total sample at $0.2\leq z \leq 0.5$ and $\sim20\%$ at $z > 0.5$. This result is consistent with elliptical galaxies forming at later times, which could explain the fewer one-component bulges at $1.0\leq z \leq 1.5$. It could also be an effect of high redshift objects not being well described by Sérsic profiles, which makes it easier to reproduce those galaxies with two profiles, rather than one. %In either case, this result is in qualitative agreement with \cite{Huertas-Company2016MNRAS}, who measure stellar mass functions for disc-like, spheroidal, and bulge+disc galaxies. They find that the number of galaxies that host bulge components has increased from $z\sim3$ to the present. 

\begin{table*}
\caption{Estimated best-fitting parameters for bulge-dominated galaxies and bulge components. One-component bulges and quenched bulges at $z\leq1.0$ are fit with the double power law shown in Equation~\ref{eq:doublepl}. All other fits are derived using Equation~\ref{eq:singlepl} but we do not derive one-component best-fitting relations for the $1.0\leq z \leq 1.5$ bin as our relations are dominated by scatter. The best-fit parameters for the quenched bulges in the highest redshift bin are reported in brackets as we fit these with a single power law.\vspace{-0.2cm}}
\centering
\begin{tabular}{c c c c c c c c c c  }
\hline
\multicolumn{1}{l}{ \ } & 
\multicolumn{8}{c}{\textbf{Total Sample}} \\ 

% \multicolumn{1}{l}{ \ } & 
% \multicolumn{1}{c}{\underline{(M$_{*}$ $\geq$ 10$^{9.5}$ M$_{\odot}$) }}
% \multicolumn{4}{l}{\underline{(M$_{*}$ $\geq$ 10$^{7}$ M$_{\odot}$) }}\\ 

\multicolumn{1}{l}{ \ }  & \multicolumn{2}{c}{All Bulges } & \multicolumn{2}{c}{Star-Forming Bulges } & 
\multicolumn{4}{c}{Quenched Bulges }\\
$z$& $\log_{10}$($A$) & $B$& $\log_{10}$($A$) & $B$ &  $\alpha$ [$\log_{10}$($A$)] & $\beta$ [$B$]& $\log_{10}(\gamma)$  & $\delta$ \\
\hline 

0.2$\leq{z}\leq$0.5 & 
$0.27\pm{0.03}$ & $0.03\pm{0.02}$ & $0.84\pm{0.02}$ & $0.21\pm{0.01}$ & 
$-0.02\pm{0.02}$&$0.75\pm{0.07}$&$0.39\pm{0.07}$ & $10.71\pm{0.09}$  \\

0.5$<$ $z$ $\leq$1.0 &
$0.28\pm{0.02}$&$0.01\pm{0.01}$&$0.78\pm{0.02}$ & $0.23\pm{0.01}$ & $-0.02\pm{0.02}$ & $2.04\pm{0.08}$ &$0.19\pm{0.08}$& $10.86\pm{0.10}$  \\

1.0$<$ $z$ $\leq$1.5 &
$0.24\pm{0.01}$&$0.01\pm{0.01}$&$0.79\pm{0.02}$ & $0.21\pm{0.01}$ & $[-0.01\pm{0.03}]$ & $[0.79\pm{0.03}$] \\

% 1.0$<$ $z$ $\leq$1.5 & 
% $0.64\pm{0.01}$ & $0.17\pm{0.01}$\\

\hline

\multicolumn{1}{l}{ \ } & 
\multicolumn{8}{c}{\textbf{One-Component Model}} \\ 
\multicolumn{1}{l}{ \ } & \multicolumn{4}{c}{All Bulges}& \multicolumn{2}{c}{\ }\\

$z$&   $\alpha$ & $\beta$ & $\log_{10}(\gamma)$  & $\delta$ & &   \\
\hline 
0.2$\leq{z}\leq$0.5 &  
$0.06\pm{0.02}$&$0.82\pm{0.08}$&$-0.46\pm{0.03}$ & $10.32\pm{0.11}$  \\

0.5$<$ $z$ $\leq$1.0 &  
$0.02\pm{0.01}$&$1.29\pm{0.13}$&$0.08\pm{0.03}$ & $11.16\pm{0.12}$  \\

% 1.0$<$ $z$ $\leq$1.5 & 
% $0.65\pm{0.03}$ & $0.10\pm{0.04}$\\

\hline

\multicolumn{1}{l}{ \ } & 
\multicolumn{8}{c}{\textbf{Two-Component Model}} \\ 
\multicolumn{1}{l}{ \ }  & \multicolumn{2}{c}{All Bulges } & \multicolumn{2}{c}{Star-Forming Bulges } & 
\multicolumn{4}{c}{Quenched Bulges }\\

$z$& $\log_{10}$($A$) & $B$& $\log_{10}$($A$) & $B$&   $\alpha$ [$\log_{10}$($A$)] & $\beta$ [$B$] & $\log_{10}(\gamma)$  & $\delta$ \\
\hline 
0.2$\leq{z}\leq$0.5 & 
$0.21\pm{0.01}$&$0.01\pm{0.01}$&$0.85\pm{0.01}$ & $0.23\pm{0.01}$ & 
$-0.04\pm{0.02}$ & $-1.65\pm{0.10}$ & $0.39\pm{0.06}$ & $11.22\pm{0.12}$\\

0.5$<$ $z$ $\leq$1.0 &
$0.24\pm{0.01}$&$0.01\pm{0.01}$&$0.79\pm{0.02}$ & $0.21\pm{0.01}$ & 
$-0.02\pm{0.02}$ & $1.94\pm{0.09}$ & $0.17\pm{0.05}$ & $11.12\pm{0.12}$\\

1.0$<$ $z$ $\leq$1.5 &
$0.24\pm{0.02}$&$0.02\pm{0.01}$&$0.66\pm{0.01}$ & $0.21\pm{0.01}$ & $[-0.03\pm{0.02}]$ & $[0.76\pm{0.02}$] \\
\hline
\end{tabular}
\label{tab:bulge_params}
\end{table*}

Across our entire redshift range, we find a significant number of bulge components, which could suggest that bulges -- at least occasionally -- form early on in galaxies that also have a disc component.
More detailed analyses, for instance, dynamical decompositions in high resolution IFU data, are needed to better constrain when bulges form. While this might be an interesting project, it is well beyond the scope of this work.
We also note here that \citetalias{Kalina2021} report significantly more quenched galaxies at these redshifts despite using the same initial sample, but we remind the reader that our samples are not complete such that if a particular galaxy makes it into our disc sample but not into the bulge sample, this does not mean that this galaxy has $no$ bulge component. We also only include galaxies that are two magnitudes brighter than the 90$\%$ completeness limit whereas \citetalias{Kalina2021} include fainter objects and we apply additional cuts to the sample based on the bulge and disc model parameters, as discussed in \S \ref{sec:phot_cuts} and \ref{sec:modelparamcuts}. Combined, these effects naturally result in the smaller sample used in this work.

To provide easier comparisons with the quenched galaxy stellar mass--size relations presented in \citetalias{Kalina2021}, we fit the one-component bulge stellar mass--size relations with a double power law following
% %As the bulge components exhibit a flattening in the stellar mass -- size relation, we fit those with

\begin{equation}
R_e = \gamma \big( M_* \big)^{\alpha} \Big( 1 + \frac{M_*}{10^{\delta}} \Big)^{\beta - \alpha}
% \log_{10}R_e &= \log_{10}\gamma + \alpha \log_{10} M_* + (\beta - \alpha)  \log_{10}\Big( 1 + 10^{\log_{10} M_* -\delta} \Big)
\label{eq:doublepl}
\end{equation}

%\noindent following \cite{Shen2003MNRAS}, \cite{Lange2016}, and \citetalias{Kalina2021},
\noindent where $\gamma$ is the normalisation, or the effective radius at a stellar mass of 1 M$_\odot$. The slopes at the low and high mass ends are described by $\alpha$ and $\beta$ respectively, and 10$^\delta$ represents the transition stellar mass between the low and high mass slope. As for Equation~\ref{eq:singlepl}, we assume uniform priors and use an identical fitting procedure.

The best-fitting relations for bulges for which the one-component model is preferred are shown in panels (d) and (e) of Figure~\ref{fig:Mass_size_bulge}. We find that the double power-law fit that we derive agrees reasonably well with the stellar mass--size relations of quenched galaxies. We note however, that $\sim39\%$ of the one-component bulges have some ongoing star formation based on their sSFRs. Thus, improved number statistics are needed to better constrain the stellar mass--size relations of one-component bulges. Additional data would also allow one-component bulges to be separated into star-forming and quenched, yielding insight into the effects of star-formation on the sizes of one-component bulges. Lastly, we find the modelling does not converge in panel (f) due to the large scatter and limited data.

In all other panels of Figure~\ref{fig:Mass_size_bulge}, we find that a double power-law fit is not warranted for the full sample of bulges. We therefore fit these with the single power law from Equation~\ref{eq:singlepl}, finding that the best-fitting relation is mostly flat (i.e.~mass-independent). These fits are shown in grey and are consistent with \cite{Lima2024MNRAS}, who similarly find that bulge size is not correlated with global galaxy stellar mass. However, the bulges that are quenched occupy a region of the size--mass plane that is distinct from the locus of the star-forming bulges. We thus further divide our bulge sample into star-forming and quenched, using the $\log_{10}$sSFR$= -11$ threshold that we also use for the disc sample. At $z\leq 1.0$, we find that the double power law better describes the quenched bulges (shown in purple) while a single power law function is preferred for the star-forming bulges (shown in blue). The best-fitting parameters for each fit are presented in Table~\ref{tab:bulge_params}.

In all panels of Figure~\ref{fig:Mass_size_bulge}, we reproduce the star-forming and quiescent stellar mass-size relations from \citetalias{Kalina2021} as thick solid black lines and thin black curves, respectively. The quiescent stellar mass-size relations from \citet[][dashed line]{vdWel2014} and the relations for bulge components from \citet[][white line]{Dimauro2018MNRAS} and \citet[][horizontal dotted line]{Robotham2022} are also shown. In comparing the relations that we derive for quiescent bulges (in purple) to the relations of quiescent galaxies from \citetalias{Kalina2021} and \cite{vdWel2014}, we find that, on average, quiescent bulge components are smaller than quiescent galaxies of the same mass. This is also found for local classical bulges in \cite{Gadotti2009MNRAS}. Moreover, at the low mass end, quiescent bulge components with M$_* \lesssim 10^{10}$ M$_\odot$ lie on a nearly flat relation on the size-mass plane. From MaNGA data, both \cite{Johnston2013MNRAS.428.1296J} and \cite{Jegatheesan2024} find that less massive bulges formed over extended timescales, while more massive bulges mostly built up their mass both rapidly and early-on. These differences in formation mechanisms may explain why these massive and less massive quiescent bulges follow distinct relations on the mass--size plane.

While we find that both star-forming and quiescent discs lie on similar stellar mass--size relations (see \S\ref{sec:discussion_discs}), star-forming bulges appear to lie in a distinct region of the mass--size plane compared to quiescent ones. In particular, we find good agreement between the relations we derive for our star-forming bulge sample (shown in blue) and that of star-forming galaxies from \citetalias{Kalina2021}. This suggests that the location of galaxy bulges on the mass--size plane is strongly dependent on the primary mechanism through which galaxies and their components are evolving.

More recently, \cite{M_A2021MNRAS} decomposed a sample of 129 local galaxies with IFU data from the CALIFA survey \citep{CALIFA} into bulges and discs. They used the SFHs of their objects to study the stellar mass--size relation of their $z\sim0$ sample at $z=1$, 1.5, and 2 to place constraints on how the main galaxy components assemble. They find that most of their bulges formed early-on but have not significantly evolved in mass with cosmic time. Their findings are qualitatively consistent with \cite{Dimauro2019}, who also find that bulges lie on a shallower stellar mass--size relation compared to quiescent galaxies. On the other hand, \cite{Robotham2022} find a nearly flat stellar mass--size relation for bulge components from GAMA (Galaxy And Mass Assembly; \citealt[]{Driver2011MNRAS}) data at $z\sim0$, similar to our results for all bulges but with lower normalisation. Our results for the bulges are in qualitative agreement with all of these findings. This may seem contradictory as these claims are not compatible, but can be explained by the mass ranges examined by those works, which barely overlap, but \textit{both} of which we cover in this work. First, at the massive end ($>10^{10}$M$_\odot$), we find that both star-forming and quiescent bulges lie on mass-dependent stellar mass--size relations, in broad agreement with \cite{Dimauro2019} and \cite{M_A2021MNRAS}. While we find steeper relations for massive quenched bulges than \cite{Dimauro2019}, our relations for star-forming bulges are shallower. As 
\cite{Dimauro2019} do not separate their bulges into star-forming and quiescent, they find shallower slopes for the bulge mass--size relation compared to quiescent galaxies and infer that this indicates that bulges undergo milder size evolution than quiescent galaxies. This milder evolution is likely an effect of fitting star-forming and quiescent bulges together, as we have shown that these two categories of bulges have distinct stellar mass--size relations.
%From Figure \ref{fig:Mass_size_bulge}, it can be seen that the slopes at the high mass end derived in this work compared to the \cite{Dimauro2019} best-fitting lines shown as white lines, are in good agreement, although the normalisation is different. 
Second, for bulges with stellar masses below $\sim10^{10}$M$_\odot$, our relations are nearly flat and qualitatively consistent with \cite{Robotham2022}. %, although the bulges themselves are often found to be non-star-forming.

%Additionally, \cite{Huertas-Company2016MNRAS} studied the morphological stellar mass functions of disc-like galaxies, spheroidal galaxies, and bulge+disc galaxies, finding that the abundance of bulge-less galaxies decreases with redshift such that the vast majority of local galaxies have bulge components.

% From MaNGA data, \cite{Johnston2013MNRAS.428.1296J} find that less massive bulges formed over extended timescales, while more massive bulges mostly built up the their mass rapidly and early-on. These differences in formation mechanisms may explain why these galaxies follow distinct relations on the mass--size plane.

%Finally, we note that in general, both the disc and bulge stellar mass--size relations have more scatter than those obtained in \citetalias{Kalina2021} for star-forming and quiescent galaxies. This scatter is possibly the result of two effects. First, galaxy components are more challenging to model accurately compared to whole galaxies and therefore, their physical parameters will naturally show larger scatter on the size--mass plane. However, this effect could also be real, indicating that there is more variety within disc populations and bulge populations than there is among galaxies, as a whole.  

\section{UVJ Diagrams of Galaxy Components } \label{sec:comp_UVJ} 

\begin{figure*}
    \centering
    \includegraphics[width=\textwidth]{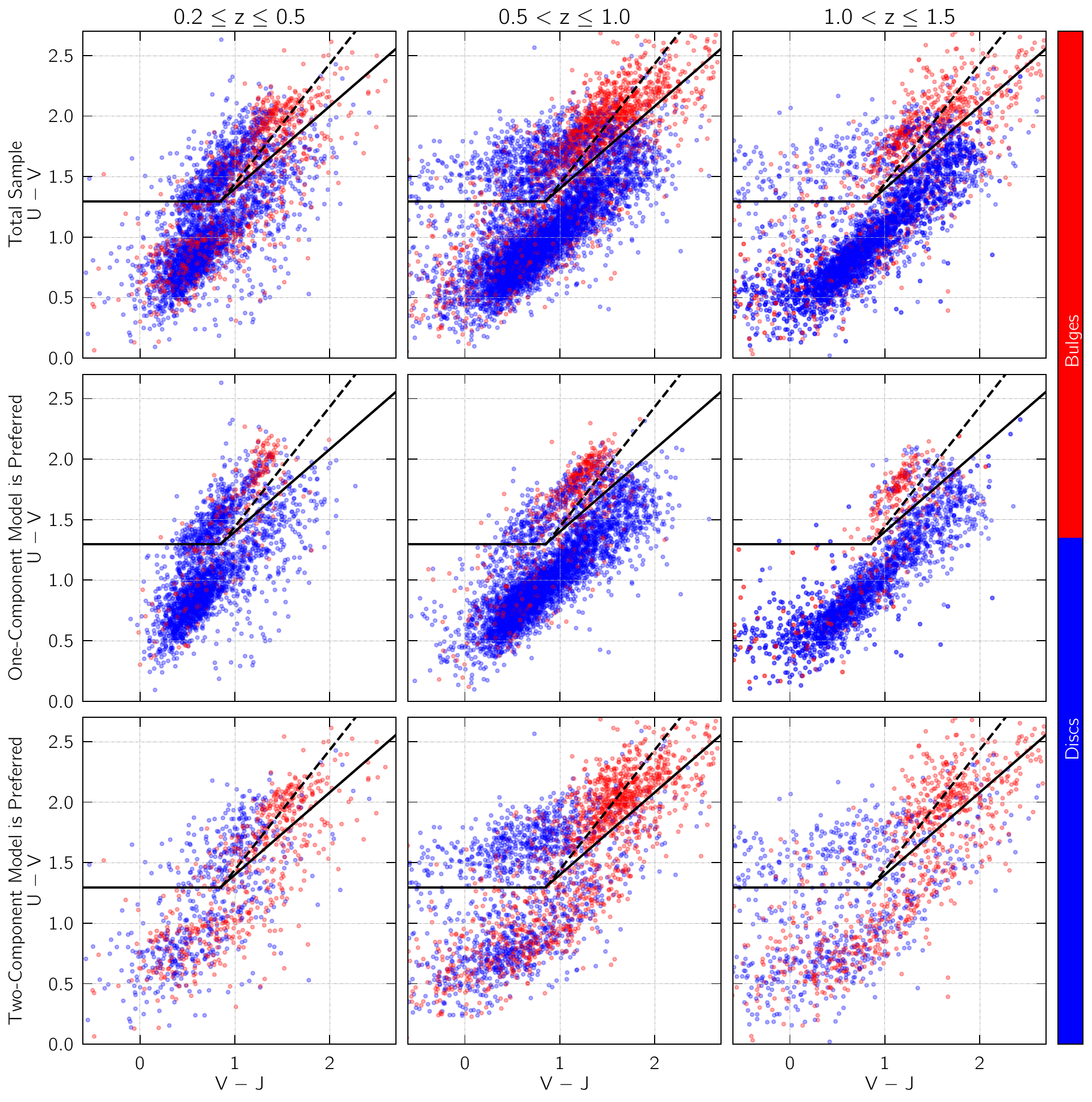}
    \vspace{-0.6cm}\caption{Rest-frame U$-$V and V$-$J colours for the bulge and disc components, shown in red and blue, respectively. The top three panels show the entire sample, the middle three panels show galaxies for which the one-component model is preferred, and the bottom three show galaxies for which the two-component model is preferred. The dashed lines which separate the star-forming and quiescent galaxies are matched to \protect\citetalias{Kalina2021}, while the solid lines are those derived in this work to separate star-forming and quiescent components. }
    \label{fig:comp_UVJ}
\end{figure*}

\begin{figure*}
    \centering
    \includegraphics[width=0.98\textwidth]{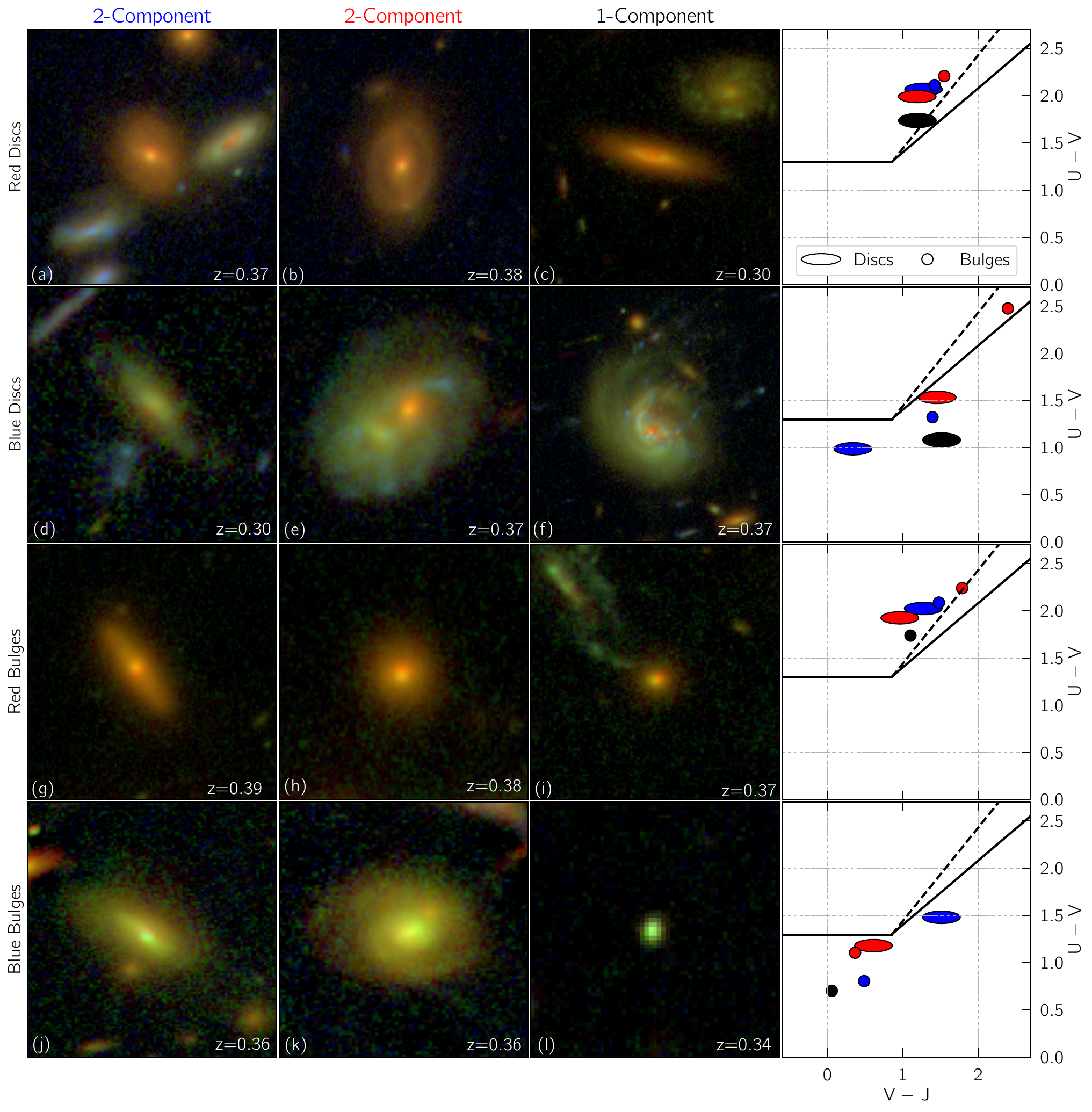}
    \vspace{-0.3cm}\caption{Visually selected example galaxies at $z\sim0.3$. The first two columns show objects for which the two-component model is preferred, and the third column shows galaxies for which the one-component model is preferred as discussed in \S \ref{sec:distinction}. The right-most column shows the UVJ diagram of the components where the components of galaxies from panels (a), (d), (g), and (j) are shown in blue, the components of galaxies from panels (b), (e), (h), and (k) are shown in red, and the one-component galaxies are shown in black.  We show examples of red and blue discs and red and blue bulges as labelled. The redshifts of the galaxies are indicated in the bottom right of each panel. All example galaxies shown in this figure are from the Abell370 cluster field and the composite RGB maps are based on 3-band HST imaging observations with a pixel scale of 0.06$''$ (blue: F435W, green: F606W, red: F814W).}
    \label{fig:RGB_quenched_disk}
\end{figure*}

We derive rest-frame U$-$V and V$-$J colours for the bulge and disc components 
%in order to obtain the positions of these individual components on the UVJ diagram. This allows us 
to establish whether they have rest-frame colours that are consistent with galaxies as a whole, and whether the components are star-forming or quenched. Recently, \cite{Dimauro2022MNRAS} presented rest-frame colours for the bulge and disc components of CANDELS galaxies. They find that, in general, discs and bulges exhibit significant colour differences and that the mean colour of their bulge sample is concentrated in the quenched region of the UVJ diagram. On the other hand, disc component of star forming galaxies are mostly found in the star forming region of the UVJ diagram, while disc components hosted in quenched galaxies mostly occupy the quenched region. %Below, we compare our findings with these results 

We show the UVJ diagram across our three redshift bins in Figure \ref{fig:comp_UVJ}, where the top rows include the entire sample, the panels in the middle row include galaxies for which the one-component model is preferred, and the bottom row shows galaxies for which the two-component model is preferred, similar to Figures \ref{fig:Mass_size_disk} and \ref{fig:Mass_size_bulge}. Visual impression suggests that the UVJ separations commonly used in the literature to distinguish star-forming and quiescent objects \citep[e.g.][]{Wuyts2007,Williams2009ApJ...691.1879W, Whitaker2011ApJ...735...86W, Martis2016} are not optimal for the bulge and disc components in our sample (see also \citealt{Akhshik2021ApJ, Akhshik2023ApJ} who demonstrate that commonly used bimodal UVJ separations are not well suited to describe the rest-frame colour evolution of a strongly lensed quiescent galaxy experiencing a late rejuvenation episode). In Figure \ref{fig:comp_UVJ}, we show the separating lines used in \citetalias{Kalina2021} as dashed lines. These cleanly divide the one-component sample, as expected; however, for the two-component galaxies, we find $\sim38\%$ of the total bulge sample, shown in the top row of Figure \ref{fig:comp_UVJ} fall in the dusty star-forming region of the UVJ diagram when using the dashed separating lines. These bulges, on average, have low sSFRs, suggesting that they are quenched. Hence, we visually derive new separations, shown in  equation \ref{eq:UVJ}.
\begin{equation}
\begin{split}
    U-V = 1.3 \ \ \mathrm{for} \ \ V-J < 0.75 \\
    U-V = 0.68 \ (V-J) + 0.72 \ \ \mathrm{for} \ \ V-J \geq 0.75
    \label{eq:UVJ}
\end{split} 
\end{equation}
While the separation for galaxies with $V-J$ colours that are bluer than 0.75 remain the same, for galaxies with redder $V-J$ colours, we have made the dividing line shallower as it provides a better separation between quenched and star-forming components. This is especially apparent in the median redshift range, where the traditional line goes straight through the bulge population, while the new (solid black) line follows its `edge' nicely.

% Panels (a) and (b) show two quenched galaxies for which the two-component model is preferred and both the disc and bulge component are reliable. These are specifically chosen to be face-on galaxies so that their structure can be easily seen.  Panel (c) shows one example quiescent galaxies for which the one-component model is preferred and it is classified as a disc based on the Sérsic index of its single Sérsic profile fit.

%%%%%

From Figure \ref{fig:comp_UVJ}, we find that most bulges and bulge-dominated galaxies have redder colours than discs, as expected, and consistent with previous studies \citep[e.g.][]{Cameron2009ApJ...699..105C,Dimauro2022MNRAS}. Red colours can be produced by large amounts of reddening due to dust, evolved stellar populations, or a combination of these effects (see also \citealt{Nedkova2024}). 
Nonetheless, this difference in colour between bulges and discs is likely due to differences in their formation and evolution histories. In Figure \ref{fig:comp_UVJ}, there is a non-negligible population of disc-like galaxies in the quenched region of the UVJ diagram. A few examples of these can be seen in panels (a), (b), and (c) of Figure \ref{fig:RGB_quenched_disk}, which shows example red discs, blue discs, red bulges, and blue bulges from the Abell 370 cluster field. The first two columns show example galaxies for which the two-component model is preferred and the third column shows galaxies that are better modelled as one-component objects, but classified as `discs' or `bulges' according to Figure \ref{fig:flowchart}. The final column shows the positions of the discs on the UVJ diagram as ellipses and the positions of the bulges as circles, where those that are colour-coded blue correspond to the galaxy shown in the left-most column, those that are colour-coded red correspond to the galaxy shown in the second column, and those shown in black correspond to the objects for which the one-component model is preferred.

First, we note that the positions of the components on the UVJ diagram align extremely well with expectations based on the components' colours from the RGB images. These galaxies were visually selected, without any prior information of their location on the UVJ diagram. This result reinforces the need for modified definitions of quiescence for components, that we have shown in Equation \ref{eq:UVJ}, and lends confidence to our bulge and disc catalogues that we generate by splitting the total flux of the galaxies by the B/T and D/T luminosity ratios. Second, we find that the majority of discs that lie in the quenched region of the UVJ diagram are also colour coded as passive in Figure \ref{fig:Mass_size_disk} indicating that they typically have low sSFRs. This is expected as the SED fitting codes from which the sSFRs are derived use the same colour information. We note that these galaxies lie on stellar mass--size relations that are slightly steeper than the one derived for star-forming galaxies in \citetalias{Kalina2021} and the relation derived for disc components from \cite{Robotham2022} at $z\sim0$,  as previously noted in \S\ref{sec:discussion_discs}. % Specifically, in panels (g) and (h) of Figure~\ref{fig:Mass_size_disk}, there appear to be more red points below the \citetalias{Kalina2021} and \cite{Robotham2022} relations, than above them as previously noted in \S\ref{sec:discussion_discs}. 
%The galaxies shown in the top row of Figure~\ref{fig:RGB_quenched_disk} are examples of such galaxies from the Abell 370 cluster field. These objects in panels (a) and (b) are specifically chosen to be face-on so their disc-like structure can be clearly seen in order to highlight that they are not misclassified.

%In this subsection, we further investigate the properties of this sample.
While the vast majority of quenched galaxies have early type morphologies in agreement with expectations \citep[e.g.][]{Strateva2001AJ, Bell2012ApJ...753..167B}, there exist environmental mechanisms that can quench a galaxy without significantly impacting its structure. One such mechanism is ram-pressure stripping \citep{GunnGott1972ApJ...176....1G, Bekki2009MNRAS.399.2221B, Boselli2022}, which occurs in large evolved galaxy clusters, like the cluster fields from the HFF. Ram-pressure stripping strips the halo and disc of cold gas such that it can no longer be used to fuel star formation, causing the galaxy to quench without destroying the stellar disc structure \citep[e.g.][]{Weinmann2006MNRAS}. Mass quenching mechanisms such as heating from AGN can also cease star formation without destroying a galaxy’s disc component \citep[e.g][]{Tabor1993MNRAST, Fabian1994ARA&A..32..277F}. Although the formation processes of quenched discs are complex and likely involve a combination of these mechanisms, further investigation of this interesting subsample could yield insight into the environmental effects on galaxy sizes, which has been a topic of some debate \citep[see e.g.][]{Maltby2015MNRAS, Afanasiev, Strazzullo2023A&A, Figueira2024}.

Blue bulges, shown in the bottom row of Figure~\ref{fig:RGB_quenched_disk}, are another interesting class of objects. Their colours indicate that they likely experienced a recent episode of star formation fuelled by enriched gas \citep[e.g.][]{Johnston2014MNRAS}. Recently, \cite{Jegatheesan2024} showed that low-redshift spiral galaxies with a bulge that is younger and more metal-rich than the disc are not uncommon. In our work, we find that blue bulges are typically hosted in galaxies that also have a blue disc, i.e.~objects such as the one-component blue bulge in panel (l) of Figure~\ref{fig:RGB_quenched_disk} are much rarer than the blue bulges shown in panels (j) and (k). This is illustrated in Figure~\ref{fig:comp_UVJ}, where bulge components in the bottom row make up $\sim$41\% of the objects in the star-forming region of the UVJ diagram while one-component bulges in the middle row constitute only $\sim4\%$ of star-forming galaxies. While we find that bulge components that occupy the star-forming region of the UVJ diagram are fairly common across $0.2\leq z \leq 1.5$, additional observations and work are needed to further investigate the properties of blue bulges. For instance, higher resolution data are needed to reveal if star-forming bulges at these redshifts are structurally different from classical bulges.

%%%%%%%%%%%%%%%%%%%%%%%%%%%%%%%%%%%%%%%%%%%%%%%%%%
%%%%%%%%%%%%%%%%%%%%%%%%%%%%%%%%%%%%%%%%%%%%%%%%%%
%%%%%%%%%%%%%%%%%%%%%%%%%%%%%%%%%%%%%%%%%%%%%%%%%%

\section{Summary} \label{sec:summary}
In this paper, we derive stellar mass--size relations for bulge and disc components over $0.2 \leq z \leq 1.5$, using a sample of $\sim$17,000 galaxies over a larger stellar mass range than previously studied at these redshifts. The depth of the HFF imaging and the multi-wavelength fitting capabilities of the MegaMorph tools, combined with the large area sampled by the CANDELS fields allow these relations to be extended to include lower mass objects, and their components. Our main results are as follows. 

\begin{itemize}
    % \item At high redshift, it becomes increasingly difficult to identify pseudo-bulges, which are disc-like structures, and to distinguish them from disc components. Although we have modelled each galaxy with two separate bulge+disc models (i.e., one where the Sérsic index of the bulge is fit as a free parameter and another where the bulge component is fit with a Sérsic index fixed to $n=4$), we choose to use the models where the bulge Sérsic index is fixed to $n=4$ because they better align with visual classifications, although we forfeit the ability to find pseudo-bulges.
    % \item While the AIC and BIC of the \textsc{GalfitM} models can provide clues about whether a given galaxy is a one- or two-component system, we find that these values do not line up with visual classification, especially for HFF galaxies which live in denser environments compared to CANDELS galaxies. 
    
    % \item The main results of this paper are presented in Figures \ref{fig:Mass_size_disk} and \ref{fig:Mass_size_bulge}, where we separate our disc and bulge samples into galaxies for which the single Sérsic profile is preferred (`disc dominated' and `bulge dominated' galaxies) and galaxies for which the two-component model is preferred (`discs' and `bulges'). We present our results for these two groups separately. 
    
    \item We find that galaxy discs with low sSFR are typically smaller than those with higher sSFR (i.e.~they lie below discs with high sSFRs on the size--mass plane as shown in Fig.~\ref{fig:Mass_size_disk}). 
    %This could suggest an evolutionary track where the disc components of galaxies that are quenching or have recently quenched move `downwards' onto the stellar mass -- size relation of quenched galaxies. 
    This may constitute evidence for compaction \citep[][]{Zolotov2015MNRAS} where disc components of galaxies that are quenching or have recently quenched move `downwards' onto the stellar mass -- size relation of quenched galaxies.

    \item{By separating our disc sample into star-forming and quenched, we show that star-forming discs lie on stellar mass -- size relations that are consistent with those star-forming galaxies from e.g.~\citetalias{Kalina2021}, suggesting that they primarily evolve via star formation. However, the relations for quenched discs are steeper and have lower normalisations than the stellar mass -- size relations of star-forming discs. This implies that quenched discs evolve through different mechanisms, possibly including minor mergers.}
    
%    \item The stellar mass -- size relations of our one-component and two-component disc samples are consistent with those derived by \citetalias{Kalina2021} for star-forming galaxies in all panels except panel (i) (high redshift, two-component galaxies), in which we derive a flatter stellar mass--size relation as a result of the large scatter. The amount of scatter in this panel is likely a result of the difficulty of decomposing galaxies at these high redshifts and larger uncertainties; however, it could also be a result of galaxy discs being clumpier and more irregular at higher redshifts. 
    
    \item We show that the stellar mass -- size relations of galaxy bulges is mostly flat, i.e.~mass-independent. This result extends recent work by \cite{Robotham2022}, who show that bulges are consistent with having no size dependency on component stellar mass, to $z\sim1$.

    \item While the stellar mass--size relations of bulges is flat, star-forming bulges lie on relations that are in broad agreement with those of star-forming galaxies from \citetalias{Kalina2021}. This result suggests that if star-forming, bulges' size growth is mostly driven by star formation.
    
    \item The stellar mass--size relations of quiescent bulges, on the other hand, exhibit a flattening at $\sim10^{10}$M$_\odot$ (see Fig.~\ref{fig:Mass_size_bulge}), similar to quiescent galaxies. One notable difference is that at a given stellar mass, quenched bulges are on average smaller than quiescent galaxies, which could suggest that quiescent galaxies contain a disc component that increases their measured effective radius. These discs can be small and/or faint and hence, not visible or separable in our data. Alternatively, bulge components could simply become more compact due to different physical processes acting on them than those imposed onto most quenched galaxies. At $z>1.0$, we do not have a sufficient sample of low-mass quenched discs to constrain this flattening.
 
 % \item We also find that at stellar masses $\lesssim10^{10}$ M$_\odot$, most bulges are embedded in star-forming galaxies while above this stellar mass, most bulges reside within galaxies with lower sSFR. Finally, from Figure \ref{fig:Mass_size_bulge}, we find that the assumed functional form for the stellar mass--size relation is not suitable at $z\geq1$ given the scatter in our sample.
    
     % \item We find that the stellar mass--size relations for both discs and bulges have more scatter than those for star-forming and quiescent galaxies. This is most likely due to the challenges of obtaining reliable bulge and disc samples and securely measuring their properties, but it may also hint that there is more variety within disc populations and bulge populations. 

     \item Finally, we present rest-frame U$-$V and V$-$J colours for bulge and disc components in Figure~\ref{fig:comp_UVJ}. We find that the dividing lines commonly used in the literature to distinguish star-forming from quiescent galaxies \citep[e.g.][]{Williams2009ApJ...691.1879W, Whitaker2011ApJ...735...86W, Martis2016} are not optimal for galaxy components. We visually derive new separations, shown in Equation \ref{eq:UVJ}, and find that the classifications based on the UVJ diagram agree well with the visual impression of component colour, as demonstrated in Figure \ref{fig:RGB_quenched_disk}.

\end{itemize}

%%%%%%%%%%%%%%%%%%%%%%%%%%%%%%%%%%%%%%%%%%%%%%%%%%
%%%%%%%%%%%%%%%%%%%%%%%%%%%%%%%%%%%%%%%%%%%%%%%%%%
%%%%%%%%%%%%%%%%%%%%%%%%%%%%%%%%%%%%%%%%%%%%%%%%%%
\section{Acknowledgements} 

\noindent{We thank the anonymous referee for exceptionally thorough and constructive report that has helped to improve the quality, clarity, and readability of this paper. KVN gratefully acknowledges support from the John F. Burlingame and Kathryn A. McCarthy Graduate Fellowships in Physics at Tufts University. KVN also thanks ESO, Chile, for a seven-month scientific visit, which was key in making this work possible. KVN and DM acknowledge the very generous support of the National Science Foundation under Grant Number 1513473 and by HST-AR-14302 and HST-AR-14553, provided by NASA through a grant from the Space Telescope Science Institute, which is operated by the Association of Universities for Research in Astronomy, Incorporated, under NASA contract NAS5-26555. EJJ acknowledges support from FONDECYT Iniciaci\'on en investigaci\'on 2020 Project 11200263 and the ANID BASAL project FB210003.}

%%%%%%%%%%%%%%%%%%%%%%%%%%%%%%%%%%%%%%%%%%%%%%%%%%

\section*{Data Availability}

Data directly related to this publication and its figures are available on request from the corresponding author.



\bibliographystyle{mnras}
\bibliography{my}

%\begin{thebibliography}{99}
%\end{thebibliography}

%%%%%%%%%%%%%%%%%%%%%%%%%%%%%%%%%%%%%%%%%%%%%%%%%%

%%%%%%%%%%%%%%%%% APPENDICES %%%%%%%%%%%%%%%%%%%%%
\appendix
\section{HFF Bulge + Disc Catalogues}\label{appendix_catalogue}

\begingroup
\setlength{\tabcolsep}{4pt} % Default value: 6pt
\renewcommand{\arraystretch}{1} % Default value: 1

\begin{table*}
\centering
\caption{Columns from the released HFF bulge+disc catalogues. id$_\mathrm{HFF}$ corresponds to the ID numbers from the HFF-DeepSpace catalogue \citep{Shipley2018} and we create flags `use\_bulge', `use\_disc', `use\_single\_as\_bulge', and `use\_single\_as\_disc' based on our criteria summarised in the flowchart in Figure~\ref{fig:flowchart}. All other parameters are derived with the \textsc{GalfitM} and \textsc{Galapagos-2} codes. Column names that end with `BD' indicate parameters for the bulge+disc fitting, while those ending in `B[D]' indicate that there is such a parameter for the bulge that ends with `B' and for the disc that ends with `D'. All columns for which the description begins with `\textbf{[7]}' are seven-element arrays. Re and Re error estimates are reported in pixels because the quality cuts that we apply are based on the measurements in pixels. Hence, a pixel scale of 0.06$''/$pixel is necessary to convert the measured Re into units of arcseconds. Position angles are defined such that 0 is `up' (in respect to the image; this does not necessarily mean `North'), increasing anticlockwise.}\vspace{-0.2cm}
\begin{tabular}{ l l}
\hline
Column & Description  \\
\hline

id$_{\mathrm{HFF}}$ &  HFF-DeepSpace catalogue ID  \\
RA & \textsc{SExtractor} ALPHA\_J2000  \\
DEC & \textsc{SExtractor} DELTA\_J2000  \\
FLAG\_GALFIT\_BD & Sérsic fit flag for the bulge+disc decomposition: $-1=$ not enough bands with data; not attempted, $0=$ not attempted, \\
& $1=$ fits started, but crashed, $2=$ fits completed  \\
NEIGH\_GALFIT\_BD & Number of neighbouring profiles fit (or fixed) during the modelling \\
CHISQ\_GALFIT\_BD & \textsc{GalfitM} $\chi^2$ value \\
CHISQ\_GALFIT\_BD\_PRIME & \textsc{GalfitM} $\chi^2$ value within the primary ellipse \\
NDOF\_GALFIT\_BD & Degrees of freedom (DoF) allowed during the fit, includes number of pixels. Used to derive $\chi^2$ and $\chi^2/\nu$ \\
NDOF\_GALFIT\_BD\_PRIME  & same as NDOF\_GALFIT\_BD but for the fit within the primary ellipse  \\
CHI2NU\_GALFIT\_BD  &  \textsc{GalfitM} reduced $\chi^2$: $\chi^2/\nu$  \\
CHI2NU\_GALFIT\_BD\_PRIME & same as CHI2NU\_GALFIT\_BD but for the fit within the primary ellipse \\
use\_bulge &modelling parameter flag: $1=$ bulge model is reliable; $0=$ otherwise  \\
use\_disc & modelling parameter flag: $1=$ disc model is reliable; $0=$ otherwise   \\
use\_single\_as\_bulge & modelling parameter flag: $1=$ single profile model is reliable and  spheroidal; $0=$ otherwise  \\
use\_single\_as\_disc & modelling parameter flag: $1=$ single profile model is reliable and disc-like; $0=$ otherwise  \\
X\_GALFIT\_DEG\_B[D] & DoF of x-position; not allowed to vary with wavelength  \\
X\_GALFIT\_BAND\_B[D] & \textbf{[7]} x-position for F125W, F435W, F606W, F814W, F105W, F140W, and F160W, in this order \\
... & \\
MAG\_GALFIT\_DEG\_B[D]& DoF of apparent magnitude; full freedom has been allowed \\
MAG\_GALFIT\_BAND\_B[D] & \textbf{[7]} apparent magnitude measured at each band \\
MAGERR\_GALFIT\_BAND\_B[D] & \textbf{[7]} apparent magnitude uncertainties at each band  \\
MAG\_GALFIT\_CHEB\_B[D] & \textbf{[7]} magnitude Chebyshev polynomial coefficients -- as we allow full freedom, all values are nonzero  \\
MAGERR\_GALFIT\_CHEB\_B[D] & \textbf{[7]} uncertainties on the magnitude Chebyshev polynomial coefficients \\
RE\_GALFIT\_DEG\_B[D] & DoF of effective radius (Re)  \\
RE\_GALFIT\_BAND\_B[D] & \textbf{[7]} Re [in pixels] at each band  \\
REERR\_GALFIT\_BAND\_B[D] & \textbf{[7]} Re uncertainty [in pixels] at each band \\
RE\_GALFIT\_CHEB\_B[D] & \textbf{[7]} Re Chebyshev polynomial coefficients -- as we use a 2nd order Chebyshev polynomial, the first 3 values are nonzero  \\
REERR\_GALFIT\_CHEB\_B[D] & \textbf{[7]} uncertainties on the Re Chebyshev polynomial coefficients \\
N\_GALFIT\_DEG\_B[D] & DoF of Sérsic index ($n$)  \\
N\_GALFIT\_BAND\_B[D] & \textbf{[7]} $n$  at each band\\
NERR\_GALFIT\_BAND\_B[D] &  \textbf{[7]} $n$  uncertainty at each band \\
N\_GALFIT\_CHEB\_B[D] & \textbf{[7]} $n$ Chebyshev polynomial coefficients\\
NERR\_GALFIT\_CHEB\_B[D] &  \textbf{[7]} uncertainties on the $n$ Chebyshev polynomial coefficients \\
Q\_GALFIT\_DEG\_B[D] & DoF of axis ratio; not allowed to vary with wavelength  \\
Q\_GALFIT\_BAND\_B[D] & \textbf{[7]} axis ratio at each band \\
... & \\
% QERR\_GALFIT\_BAND & \textbf{[7]} axis ratio uncertainty in each band & \{0.001, 0.001, 0.001, 0.001, 0.001, 0.001, 0.001\}\\
% Q\_GALFIT\_CHEB & \textbf{[7]} axis ratio Chebyshev polynomial coefficients & \{0.1438, 0.0, 0.0, 0.0, 0.0, 0.0, 0.0\}\\
% QERR\_GALFIT\_CHEB & \textbf{[7]} uncertainties on the axis ratio Chebyshev polynomial coefficients & \{0.0007, 0.0, 0.0, 0.0, 0.0, 0.0, 0.0\}\\
PA\_GALFIT\_DEG\_B[D] & DoF of position angle; not allowed to vary with wavelength  \\
PA\_GALFIT\_BAND\_B[D] & \textbf{[7]} position angle at each band \\
... & \\
% PAERR\_GALFIT\_BAND & \textbf{[7]} position angle uncertainty at each band & \{0.11, 0.11, 0.11, 0.11, 0.11, 0.11, 0.11\} \\
% PA\_GALFIT\_CHEB & \textbf{[7]} position angle Chebyshev polynomial coefficients &\{76.21, 0.0, 0.0, 0.0, 0.0, 0.0, 0.0\} \\
% PAERR\_GALFIT\_CHEB & \textbf{[7]} uncertainties on the position angle Chebyshev polynomial coefficients& \{0.05, 0.0, 0.0, 0.0, 0.0, 0.0, 0.0\}\\

\hline
\end{tabular}

\label{tab:HFFcat}
\end{table*}
\endgroup

Along with this paper, we release a bulge+disc decomposition catalogue for each of the 12 HFF cluster and parallel fields, where the disc component is modelled with a Sérsic index of $n=1$ and the bulge component with $n=4$. Although we also use galaxies from the five CANDELS fields for this study, which were modelled in a consistent way, these data will be released by Häu\ss ler et al. (\textit{in prep}). The catalogues that we release extend the \citetalias{Kalina2021} HFF structural catalogues as we present the bulge and disc properties of the galaxies modelled by \citetalias{Kalina2021}. In Table \ref{tab:HFFcat}, we describe the columns in the released catalogues, where the ID numbers of the objects are matched to the HFF-DeepSpace catalogues \citep{Shipley2018}. Apart from the ID numbers and positions of the objects, we do not reproduce values already presented in \citetalias{Kalina2021} and we therefore suggest that the catalogues that accompany this work be matched to the ones presented in \citetalias{Kalina2021}, by position or, trivially, by their ID numbers.

\section{Metrics for Identifying Best-Fitting Models}\label{appendix_AIC_BIC}

\begin{table*}
    \caption{Number of free parameters for each field for the single Sérsic profile model (top) and for the bulge+disc model (bottom). CANDELS galaxies have been modelled using a different number of bands in each field, and hence are listed independently, while we use the same number of bands for all HFF fields. }\vspace{-0.2cm}
    \centering
    \begin{tabular}{c c c c c c c c }
    \hline
    \multicolumn{1}{l}{ \ } & 
    \multicolumn{1}{l}{ \ } & 
    \multicolumn{6}{c}{Single Sérsic Profile Model} \\
         &  Component  & HFF & GOODS-N  & GOODS-S & COSMOS &  EGS & UDS   \\
        \hline 
        Apparent Magnitude & total & 7 & 9 & 9 & 5 & 5 & 4 \\
        Size & total  & 3 & 3 & 3 & 3 & 3 & 3  \\
        Sérsic Index & total & 3 & 3 & 3 & 3 & 3 & 3 \\
        Axis Ratio & total & 1 & 1 & 1 & 1 & 1 & 1   \\
        Position Angle & total & 1 & 1 & 1 & 1 & 1 & 1  \\
        X \& Y Position & total  & 2 & 2 & 2 & 2 &  2 & 2\\
        \hline
        \multicolumn{2}{c}{ Total number of free parameters:} & 17 & 19 & 19 & 15 & 15 & 14 \\ 
        \hline
        \hline
        \multicolumn{2}{l}{ \ } & 
        \multicolumn{6}{c}{Bulge + Disc Model} \\ 
        & & HFF & GOODS-N  & GOODS-S & COSMOS &  EGS & UDS \\
        \hline
        Apparent Magnitude & bulge & 7 & 9 & 9 & 5 & 5 & 4 \\
        & disc & 7 & 9 & 9 & 5 & 5 & 4\\
        Size & bulge & 1 & 1 & 1 & 1 & 1 & 1  \\
        & disc & 1 & 1 & 1 & 1 & 1 & 1   \\
        Sérsic Index & bulge & 0 & 0 & 0 & 0 & 0 & 0  \\
        & disc  & 0 & 0 & 0 & 0 & 0 & 0   \\
        Axis Ratio  & bulge & 1 & 1 & 1 & 1 & 1 & 1 \\
        & disc & 1 & 1 & 1 & 1 & 1 & 1  \\
        Position Angle & bulge & 1 & 1 & 1 & 1 & 1 & 1  \\
        & disc & 1 & 1 & 1 & 1 & 1 & 1  \\
        X \& Y Position & bulge \& disc & 2 & 2 & 2 & 2 & 2 & 2 \\
        \hline
        \multicolumn{2}{c}{ Total number of free parameters:} & 22 & 26 & 26 & 18 & 18 & 16  
    \end{tabular}
    \label{tab:free_param}
\end{table*}

As discussed throughout the paper, identifying whether the one- or two-component model is the most appropriate choice for a given galaxy is nontrivial. Although the AIC, BIC, and $\chi^2_\nu$ parameters are commonly used metrics to assign preference to models, we find that these parameters are unreliable predictors of the best-fitting model. In this appendix, we provide more details regarding these parameters and the tests carried out to assess whether these parameters can reliably distinguish one- and two-component systems.  We obtain the AIC and BIC using equations \ref{AICeq} and \ref{BICeq}, respectively. 

\begin{equation}
    AIC = \chi^2 + 2k
\label{AICeq}
\end{equation}
\begin{equation}
BIC = \chi^2 + k\ln(N).
\label{BICeq}
\end{equation}
For both criteria, the second term is a penalty function that increases with increasing $k$, the total number of free model parameters, and $N$, the number of data points. Although $\chi^2$ and $N$ are derived directly from \textsc{GalfitM}, $k$ is not. Hence, we obtain $k$ by counting the number of free parameters for each model, as shown in Table \ref{tab:free_param}. 
%For instance, for the one-component fit for the HFF fields, we allow full degrees of freedom (DoF) for the magnitude, three DoF for the effective radius, three DoF for the Sérsic index, one DoF for the axis ratio, one DoF for the position angle, one DoF for the x-position, and one DoF for the y-position. This totals 17 DoF for the single Sérsic profile fit, as is shown in Table \ref{tab:free_param}. While we consistently model the HFF galaxies in seven bands, there are a different number of bands available for the five CANDELS fields, therefore we count $k$ for each field independently, as shown in the table. 
We make the important distinction here that we use the `\texttt{\_prime}' $\chi^2$ values from \textsc{GalfitM} because we wish to obtain the BIC and AIC for the primary object only. In theory, the `\texttt{\_prime}' values remove the contributions from neighbouring objects, allowing the two criteria to be calculated for the object in question only.

 We compare the classifications based on the BIC and AIC with visual classifications of $\sim$1000 randomly selected galaxies, and find that the two classification schemes are often in disagreement. 
%We find that the primary ellipse often extends beyond the target galaxy. 
Because the HFF programme has targeted particularly dense cluster fields, we find that the primary ellipse of the primary object, within which the primary $\chi^2$ and the number of data points are calculated, often extends beyond the target galaxy and usually contains neighbouring objects anyway (i.e. the masking employed by \textsc{Galapagos-2} to derive these `\texttt{\_prime}' values is not stringent enough). This suggests that our BIC and AIC values are unreliable since the `\texttt{\_prime}' values are still affected by neighbouring objects. We also note that in an ideal case, we would be able to take the BIC and AIC of the bulge+disc and single Sérsic profile fits in the bands that are redder than 3159{\AA}, as we have done for the B/T; however, as the BIC and AIC are based on the $\chi^2$ of the model across all bands, this is unfeasible, as the individual bands are not separated in the calculation of $\chi^2$. As a result of these drawbacks, we argue that the BIC and AIC cannot be used to robustly distinguish between one- and two-component systems for our sample.

We additionally test the reduced $\chi^2$ parameter as a way to identify the best-fitting model. In a similar fashion, we use the `\texttt{\_prime}' $\chi^2_\nu$ from the \textsc{GalfitM} modelling. However, as we find that the masking that \textsc{Galapagos-2} uses to obtain `\texttt{\_prime}' values still contain some contamination from neighbouring objects, we again argue that the $\chi^2_\nu$ is not a reliable parameter for this task.

%%%%%%%%%%%%%%%%%%%%%%%%%%%%%%%%%%%%%%%%%%
%%%%%%%%%%%%%%%%%%%%%%%%%%%%%%%%%%%%%%%%%%
%%%%%%%%%%%%%%%%%%%%%%%%%%%%%%%%%%%%%%%%%%
\section{Flipped Components} \label{appendix:flipped_comp}

\begin{figure}
    \centering
    \includegraphics[width=0.48\textwidth]{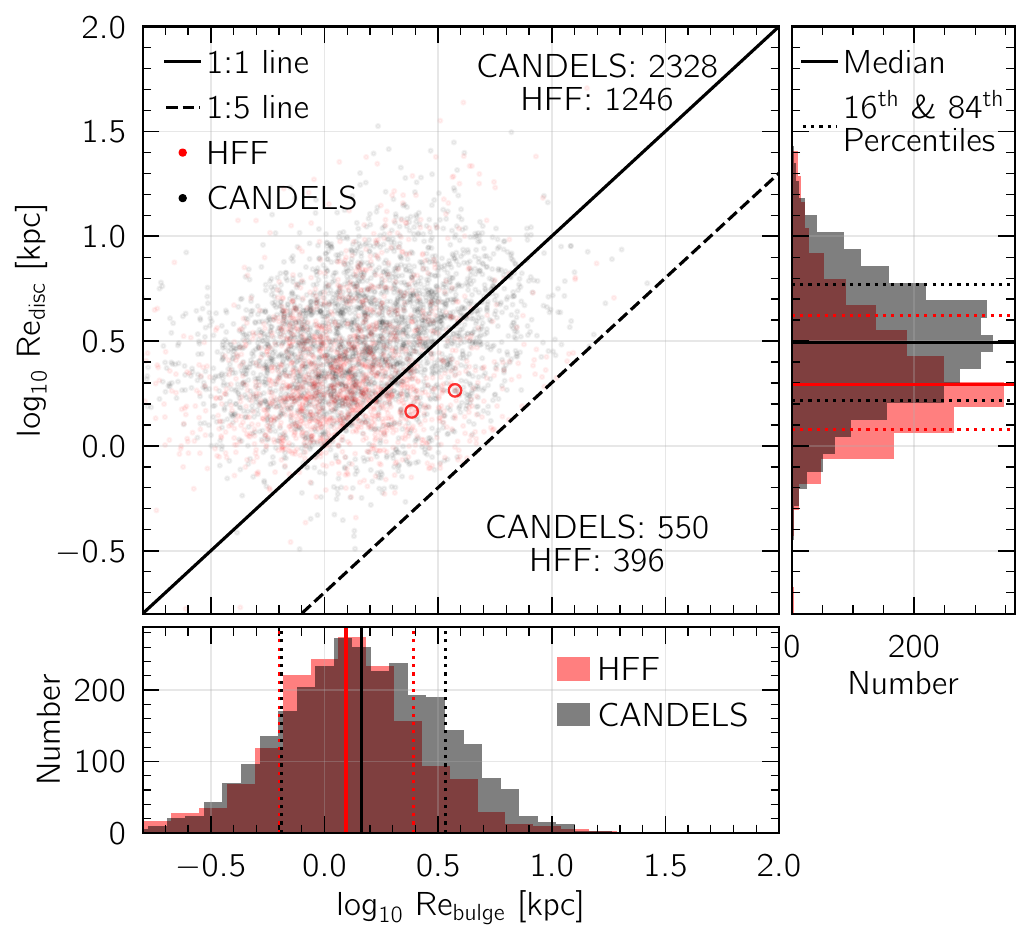}
    \vspace{-0.6cm}
    \caption{Comparison of the effective radius of the bulge and disc components, where both components make it into the final samples (i.e., objects from group (I) or (II) for which the bulge satisfies all criteria to be included in the bulge sample and the disc component is included in the final disc sample). The one-to-one line indicates where the disc and bulge sizes are equal while the one-to-five line indicates a cut that we apply -- namely, Re$_\mathrm{bulge} \leq5\times$ Re$_\mathrm{disc}$. The number of CANDELS and HFF galaxies shown at the top right are the number of galaxies above the one-to-one line, while the numbers at the bottom, right-hand side are the number of galaxies below the one-to-one line, i.e. for which the bulge is the larger component. 
    %The majority of objects lie above the black line, indicating that the disc is typically fit as the larger component, as expected. 
    In Figure \ref{fig:examples_flipped_comp}, we show the fitting results for the two galaxies indicated with larger red circles. The histograms show the disc and bulge size distributions for the HFF and CANDELS samples, with the median and 16$^\mathrm{th}$ and 84$^\mathrm{th}$ percentiles indicated with solid and dotted lines, respectively. }
    \label{fig:flipped_comp}
\end{figure}

\begin{figure*}
    \centering
    \includegraphics[width=0.96\textwidth]{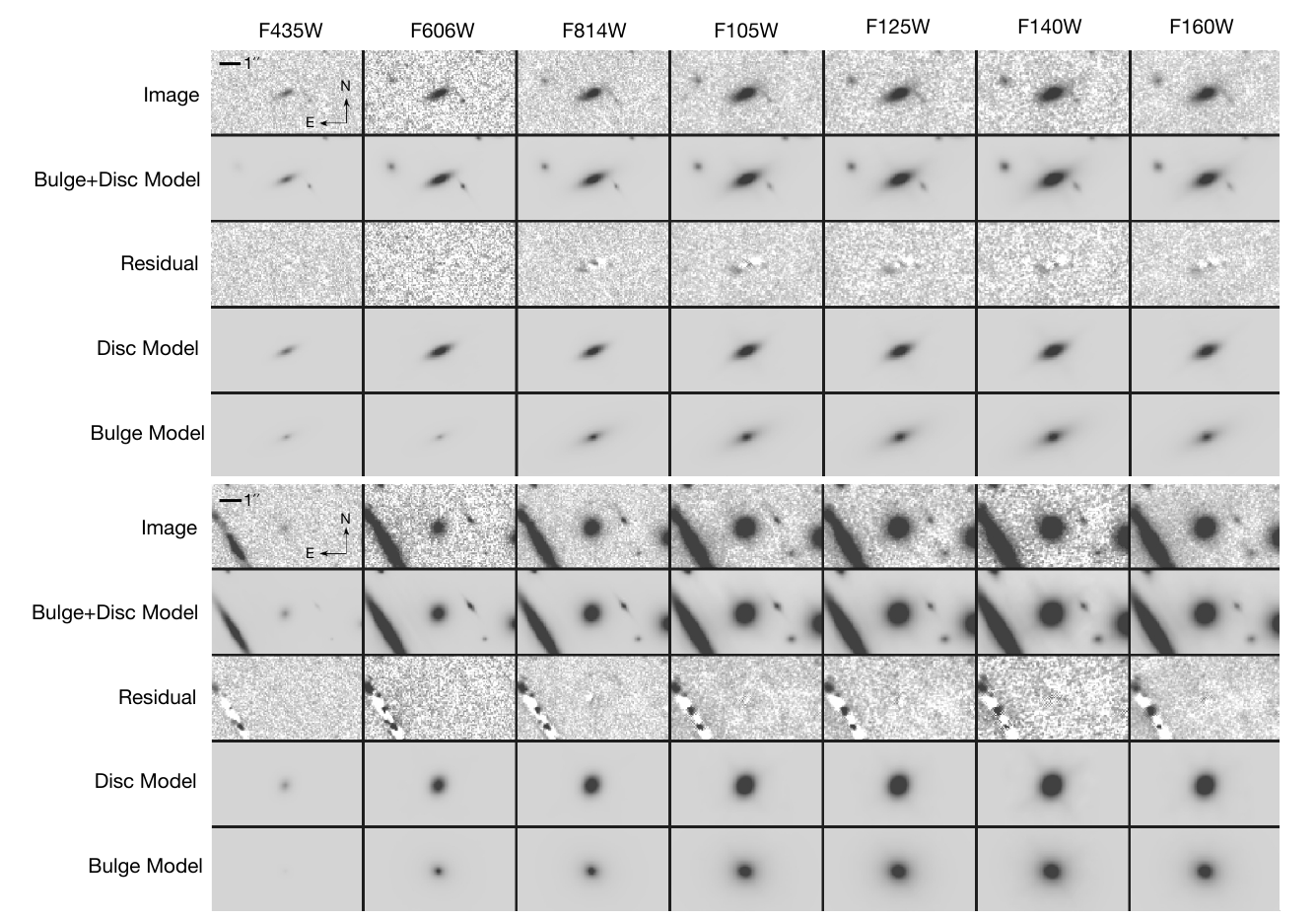}
    \vspace{-0.2cm}\caption{Images, total bulge+disc models, residuals, and individual disc models and bulge models for two example galaxies where the bulge is modelled as the larger component. In both randomly selected galaxies, the bulge is dominating the light profile in the centre, as high Sérsic index fits always do, even though it is the larger component. }
    \label{fig:examples_flipped_comp}
\end{figure*}

In \S\ref{sec:flippedcomps313}, we briefly discuss that in the modelling, the bulge and disc models may be reversed such that the bulge of the fit models the disc structure of the galaxy while the disc component dominates the light profile in the inner regions.
Several previous studies that have performed bulge+disc decompositions have found it necessary to flip the models if they satisfy certain criteria that suggest the components have been reversed \citep[e.g.][]{Lange2016, Fischer2019MNRAS, BUDDI_MaNGA2022MNRAS, Jegatheesan2024}. We argue that our sample is robust against this effect primarily because we have constrained the Sérsic indices of the bulges and discs to be $n=4$ and $n=1$, respectively. We find no strong evidence suggesting that certain disc and bulge models need to be flipped but we discuss this problem and the series of checks we have made here.

Following the idea from \cite{Lange2016} that the bulge component should be the smaller component, we compare the sizes of the discs to the sizes of the bulges for galaxies from groups (I) and (II) for which the two-component model is preferred and both the bulge and disc components satisfy all of the conditions to make it into the final bulge and disc samples, respectively. This sample consists of 1642 HFF and 2878 CANDELS galaxies. In Figure \ref{fig:flipped_comp}, we plot the sizes of the bulges against the sizes of the discs for these objects. Roughly 75\% of galaxies lie above the black, one-to-one line, 
%with 2328 CANDELS and 1246 HFF galaxies having a larger disc than bulge, and 550 CANDELS and 396 HFF galaxies having a larger bulge, 
showing that for the majority of the objects, the bulge is already the smaller component, as expected. We exclude galaxies for which the effective radius of the bulge is five times larger than the effective radius of the disc as we find that these are typically objects where the bulge is fitting flux from a neighbouring object. This criterion is indicated as a dashed line in Figure~\ref{fig:flipped_comp}. %We reiterate here that we do not flip components or discard bulges which are larger than discs for that reason alone. Instead we keep them as bulges, and are aware that at least some of them will not resemble true bulge components of galaxies.

We find that discs and bulges above and below the one-to-one line in Figure~\ref{fig:flipped_comp} have consistent g$-$r colours, which we obtain directly from the \textsc{GalfitM} modelling. This provides additional support for keeping galaxies with larger bulges than discs in our final sample. We have also visually inspected $\sim 25\%$ of the galaxies that have a bulge which is larger than the disc component. In general, we find that these objects have reliable fits without any obvious reason to exclude them from the final sample (e.g.~neighbours, bad residuals, etc). In Figure \ref{fig:examples_flipped_comp}, the models for two example galaxies -- one from the MACS1149 cluster and one from the MACS1149 parallel field -- are shown. These objects were randomly chosen in order to highlight that the bulge and disc models appear reliable upon visual inspection and that even when the bulges are modelled as the larger components, they still appear bulge-like. For both example galaxies, the bulge is modelled with a larger radius than the disc, but it is still dominating the light profile in the centre and is faint in the outskirts. This is by design since we fit all bulges with $n=4$ Sérsic profiles, which are more centrally concentrated than disc light profiles, which we have constrained to be exponential. Based on these results, we find no strong reasons to flip the models, or exclude galaxies for which the bulge component is modelled with a larger radius than the disc.

\begin{figure*}
    \centering
    \includegraphics[width=0.9\textwidth]{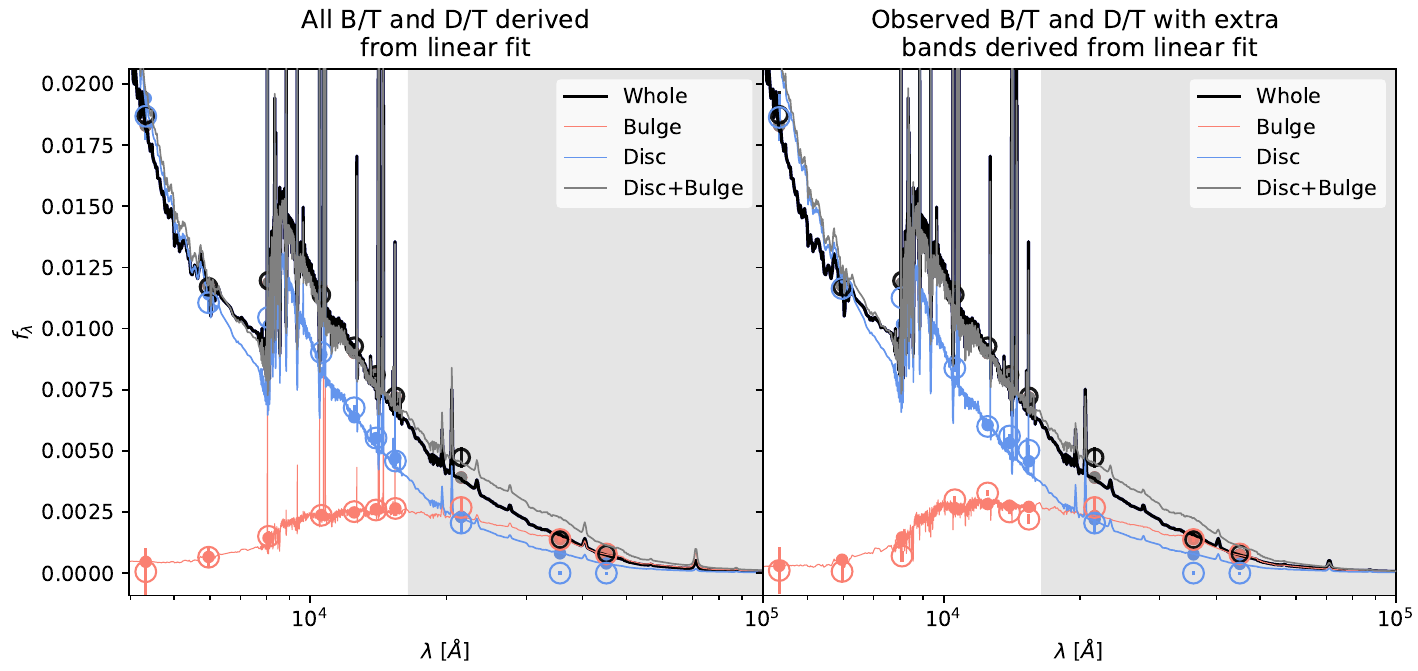}
    \vspace{-0.2cm}
    \caption{SEDs for the bulge (red) and disc (blue) components derived using the two methods described in this appendix. The sum of the bulge and disc SEDs is plotted in grey and compared to the SED of the whole galaxy, shown in black. In order to obtain B/T and D/T measurements for bands that were not used in the \textsc{GalfitM} modelling, indicated by the shaded grey region, we fit a linear relation to the B/T and D/T and extrapolate it. For the SEDs on the left, we derive all luminosity ratios using that linear fit. On the right, we use the same linear fit for the bands without measurements and we use the measured B/T and D/T from the fitting directly for bands where they exist. Both approaches reproduce the total SED well but the one of the left better reproduces the blue end of the spectrum. }
    \label{fig:SED_comparisons}
\end{figure*}

Lastly, from Figure \ref{fig:flipped_comp}, we note that the histograms indicate that bulges and discs from the HFF sample are on average smaller than those from the CANDELS fields. This is expected as we model fainter objects in the HFF fields because the images are deeper. When we exclude HFF galaxies that are fainter than the CANDELS magnitude limit that we apply in \S \ref{sec:phot_cuts}, the bulge and disc size distributions from the HFF sample align with those from the CANDELS sample. As a result, the difference in size between the two samples is primarily caused by the different magnitude limits applied, although other effects such as the different pixel scales of the images and the density of the fields could play a role.

\section{Galaxy Component Masses}\label{appendix_mass}

In this appendix, we discuss a series of tests that we have performed to assess our ability to obtain reliable stellar masses for bulge and disc components. As discussed in \S\ref{sec:mass}, there are a number of possible methods for deriving component stellar masses. One option is to generate photometric catalogues for the bulge and disc components separately and fit their SEDs. An alternative is to use a relation between mass-to-light ratio ($\Upsilon_\star$) and colour. Here, we compare stellar mass estimates from both methods. We also compare these masses to the global stellar masses reported in the HFF-DeepSpace and 3D-HST catalogues. Ideally, the bulge mass plus the disc mass that we derive should roughly equal the total mass from these catalogues.

We begin by providing additional details on how our bulge and disc photometric catalogues are generated. We adopt two methods to split objects' fluxes into their bulge and disc flux. The first method where we fit a line to the B/T and D/T measurements in order to obtain these values for filters redder than the H-band, is described in full detail in \S\ref{sec:results}. In a second approach, we instead use the measured B/T and D/T in bands where we have such measurements and only use the B/T and D/T measured from the linear fit at wavelengths where we do not have these measurements directly from \textsc{GalfitM}. By design, the first approach results in smoother SEDs.

Figure~\ref{fig:SED_comparisons} shows SEDs from both methods, where the ones on the left are obtained using the first method (i.e.~all B/T and D/T are obtained from the best-fitting line that we derive), while those in right panel are obtained via our second method described above. The red, blue, and grey SEDs are for the bulge components, disc components, and the sum of the disc and bulge (i.e.~the sum of the red and blue spectra), respectively. The SED of the whole galaxy from the photometric catalogues is shown in black. For all SEDs, the circles with error bars show the input fluxes and uncertainties, while the filled points show the flux obtained from the best-fitting SED from \textsc{eazy} \citep{eazy_brammer}. Bands that were not used in the \textsc{GalfitM} modelling are indicated with a grey shaded background.

In general, we find that both methods reproduce the SEDs of galaxies, as a whole, but the first method performs slightly better (i.e.~the bulge+disc SEDs from the first method better match the SED of the whole galaxy). This is also the case for the example shown in Figure~\ref{fig:SED_comparisons}, where the grey SED on the left better reproduces the SED shown in black at bluer wavelengths. Additionally, the recovered fluxes (filled circles) better match the input fluxes and their uncertainties from the bulge and disc catalogues (empty circles with error bars) in the left panel. 
%Particularly for the bulge, we note that the red circles, which indicate the flux obtained from the SED fit, are in better agreement with these empty circles, again suggesting that the method used to generate the SEDs shown on the left side of Figure~\ref{fig:SED_comparisons} is more secure. 
This can be most clearly seen for the bulge SEDs.
We have carefully visually inspected SEDs for over 1000 galaxies and their components, finding that the first method results in SEDs that are in somewhat better agreement with the SEDs from the 3D-HST and HFF-DeepSpace photometric catalogues. Thus, we choose to use the method where we obtain B/T and D/T luminosity ratios for all bands from the line that we fit to these ratios as a function of wavelength to derive the best fit spectra, and the galaxy masses.

%From these SED fits to our bulge and disc samples, we obtain rest-frame colours with \textsc{EAZY} \citep{eazy_brammer} which we use to derive the positions of individual components on the UVJ diagram, discussed in \S\ref{sec:comp_UVJ} as well as sSFR measurements for the discs and bulges with \textsc{FAST} \citep{Kriek2009} which are used in \S\ref{sec:results}. These measurements can also be used to derive stellar masses of individual components that can serve as a check of the stellar masses that we derive by using our empirically calibrated relation between mass-to-light ratio and colour.
%As we have made a number of approximations and assumptions while generating the bulge, disc, and total catalogues, we choose to use the relation between the mass-to-light ratio and colour to obtain stellar mass estimates for our galaxies in this paper. Nonetheless, we show the comparison between the methods, as well as a comparison to the stellar masses reported in the HFF-DeepSpace catalogues \citep{Shipley2018}, in Figure ~\ref{fig:Mass_comparisons}.

\begin{figure*}
    \centering
    \includegraphics[width=1.0\textwidth]{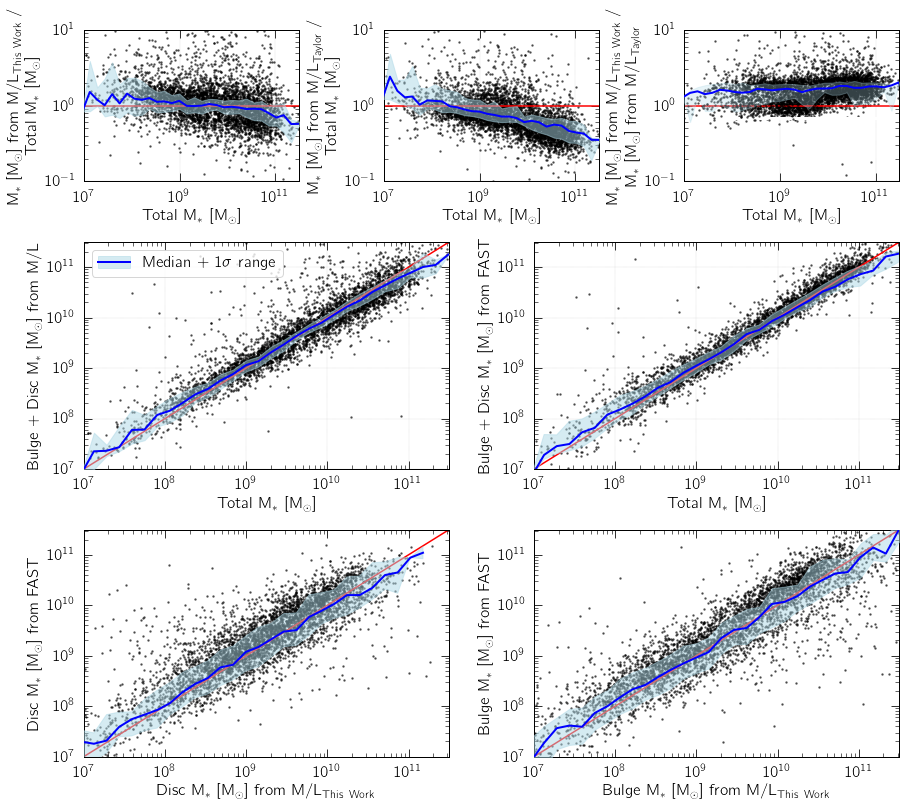}
    \vspace{-0.6cm}\caption{\textit{Top:} Ratio of bulge+disc mass derived by using relations between mass-to-light ratio and colour and stellar masses of galaxies from the 3D-HST and HFF-DeepSpace catalogues, which we term `total' masses. From left to right, we compare these `total' masses to masses derived using our empirically calibrated $\log_{10}(\Upsilon_\mathrm{V})$ -- colour relation, the  $\log_{10}(\Upsilon_\mathrm{i})$ -- colour relation from \protect\cite{Taylor2011}, and in the ratio between masses derived from these two relations. \textit{Middle:} Comparisons of the bulge and disc mass against the `total' mass of galaxies when bulge and disc masses are derived using two different approaches described in the text. \textit{Bottom:} Comparisons of the individual component masses obtained using these same two approaches.}\label{fig:Mass_comparisons}
\end{figure*}

In Figure \ref{fig:Mass_comparisons}, we compare stellar mass estimates obtained from several different approaches. First, we investigate the differences in stellar mass that are introduced by using different $\log_{10}(\Upsilon_\star)$ -- colour relations. Specifically, we compare the mass of the entire galaxy (i.e.~the mass of the bulge plus the mass of the disc) obtained from our relation and the \cite{Taylor2011} relation to stellar masses of galaxies from the HFF-DeepSpace and 3D-HST catalogues. These are shown in the top of Figure \ref{fig:Mass_comparisons}, where we term the stellar masses from the HFF-DeepSpace and 3D-HST catalogues as the `total' mass. 
From left to right, we show
%the mass of the HFF galaxies as a whole (i.e.~the total mass) derived from FAST and using the mass-to-light ratio and colour empirical relation described in \ref{sec:mass}. From left to right, we show, 
the ratios of the stellar masses derived from our $\log_{10}(\Upsilon_\mathrm{V})$ -- colour relation to the `total' mass, the ratios of the stellar masses obtained by using the \cite{Taylor2011} $\log_{10}(\Upsilon_\mathrm{i})$ -- colour relation to the `total' mass, and the ratio of stellar mass obtained by using each of the two relations. 
The running median and 1$\sigma$ scatter are indicated for each as blue lines and shaded regions, respectively. 
In general, we find decent agreement between all three. However, there appears to be a larger offset from unity, especially at high masses, when using the \cite{Taylor2011} relation, 
as shown in the top middle panel. This supports the need to derive our own empirically calibrated relation. Furthermore, in comparing stellar masses from our $\log_{10}(\Upsilon_\mathrm{V})$ -- colour relation against those from \cite{Taylor2011} in the top right panel, we find that our relation results in larger stellar masses, likely due to the use of mass-to-light ratio in different rest-frame bands and different assumptions.

Second, we compare the derived stellar masses of the individual bulge and disc components that are obtained using different approaches. This is illustrated in the bottom four panels of Figure \ref{fig:Mass_comparisons}, where we include galaxies from groups (I) and (II) (see \S\ref{sec:GroupI} and \S\ref{sec:all_good} for more details) where both the disc and bulge have been included in the final disc and bulge samples, respectively. 
%The sample of galaxies used here is similar to the one shown in Figure~\ref{fig:flipped_comp} but we remind the reader that we further exclude galaxies that have magnitude Chebyshev polynomials that are not monotonically increasing or decreasing (see \S\ref{sec:mass} for more details). 
There are a total of 4358 galaxies in the sample used here and for each, the masses of the discs plus the masses of the bulges are plotted against the `total' masses (i.e.~galaxy masses from the HFF-DeepSpace and 3D-HST photometric catalogues) in the middle row of Figure~\ref{fig:Mass_comparisons}. In the left panel of the middle row, the masses of the bulges and discs are obtained from our $\log_{10}(\Upsilon_\mathrm{V})$ -- colour relation. In the right panel of the middle row, we show the results when using the bulge and disc photometric catalogues that we construct as described above. These catalogues are then used as inputs to \textsc{FAST} to derive stellar masses of the individual components. For both, the mass of the disc plus the mass of the bulge agrees well with the `total' mass of the galaxy, as would be expected. The one-to-one line is indicated in red and the running median and scatter are again shown in blue.

Lastly in the bottom row of Figure~\ref{fig:Mass_comparisons}, we show direct comparisons of the disc and bulge stellar masses from both methods. Specifically, the stellar masses of the components on the y-axis are obtained from \textsc{FAST} while those on the x-axis are obtained using our empirical relation between mass-to-light ratio and colour. For both components, we find that the stellar masses from these approaches are consistent, suggesting that our results do not strongly depend on which method we choose. But, in the main analysis, we use the method that relies on the $\log_{10}(\Upsilon_\mathrm{V})$ -- colour relation because it has been widely used in the literature and it is, in a sense, the simpler approach. In order to use \textsc{FAST} to obtain component masses, we made a series of assumptions, particularly about extrapolating the B/T and D/T luminosity ratios to wavelengths that were not sampled in the \textsc{GalfitM} modelling, which may be biasing the mass estimates obtained with this method.

%%%%%%%%%%%%%%%%%%%%%%%%%%%%%%%%%%%%%%%%%%%%%%%%%%

% Don't change these lines
\bsp	% typesetting comment
\label{lastpage}
\end{document}